REQUEST: archive paper
PART: 1 of 1
PASSWORD: Dean
FORMAT: LATeX with embedded PS figures
KEYWORDS: tent map, Frobenius-Perron operator, spectral decomposition,
KEYWORDS: Ginzburg-Landau equation, asymptotic periodicity,
KEYWORDS: band-splitting points.
AUTHORS: S.Subbiah, D.J.Driebe.
TITLE:
Spectral Decomposition of the Tent map with Varying Height.
ABSTRACT:
The generalized spectral decomposition of the Frobenius-Perron
operator of the tent map with varying height is determined at the
band-splitting points. The decomposition includes both decay onto the
attracting set and the approach to the asymptotically periodic state on
the attractor. Explicit compact expressions for the polynomial eigenstates
are obtained using algebraic techniques.
PAPER: 
\documentclass[11pt]{article}

\def\endofps{EndOfTheIncludedPostscriptMagicCookie}
\chardef\other=12
\newwrite\psdumphandle 
\outer\def\psdump#1{\par\medbreak
  \immediate\openout\psdumphandle=#1
  \copytoblankline}
\def\copytoblankline{\begingroup\setupcopy\copypsline}
\def\setupcopy{\def\do##1{\catcode`##1=\other}\dospecials
  \catcode`\\=\other \obeylines}
{\obeylines \gdef\copypsline#1
  {\def\next{#1}%
  \ifx\next\endofps\let\next=\endgroup %
  \else\immediate\write\psdumphandle{\next} \let\next=\copypsline\fi\next}}
\outer\def\closepsdump{
  \immediate\closeout\psdumphandle}

\psdump{fig13.eps}
userdict begin /AltsysDict 245 dict def end
AltsysDict begin
/bdf{bind def}bind def
/xdf{exch def}bdf
/defed{where{pop true}{false}ifelse}bdf
/ndf{1 index where{pop pop pop}{dup xcheck{bind}if def}ifelse}bdf
/d{setdash}bdf
/h{closepath}bdf
/H{}bdf
/J{setlinecap}bdf
/j{setlinejoin}bdf
/M{setmiterlimit}bdf
/n{newpath}bdf
/N{newpath}bdf
/q{gsave}bdf
/Q{grestore}bdf
/w{setlinewidth}bdf
/sepdef{
 dup where not
 {
AltsysSepDict
 }
 if 
 3 1 roll exch put
}bdf
/st{settransfer}bdf
/colorimage defed /_rci xdf
/_NXLevel2 defed { 
 _NXLevel2 not {   
/colorimage where {
userdict eq {
/_rci false def 
} if
} if
 } if
} if
/md defed{ 
 md type /dicttype eq {  
/colorimage where { 
md eq { 
/_rci false def 
}if
}if
/settransfer where {
md eq {
/st systemdict /settransfer get def
}if
}if
 }if 
}if
/setstrokeadjust defed
{
 true setstrokeadjust
 /C{curveto}bdf
 /L{lineto}bdf
 /m{moveto}bdf
}
{
 /dr{transform .25 sub round .25 add 
exch .25 sub round .25 add exch itransform}bdf
 /C{dr curveto}bdf
 /L{dr lineto}bdf
 /m{dr moveto}bdf
 /setstrokeadjust{pop}bdf 
}ifelse
/rectstroke defed /xt xdf
xt {/yt save def} if
/privrectpath { 
 4 -2 roll m
 dtransform round exch round exch idtransform 
 2 copy 0 lt exch 0 lt xor
 {dup 0 exch rlineto exch 0 rlineto neg 0 exch rlineto}
 {exch dup 0 rlineto exch 0 exch rlineto neg 0 rlineto}
 ifelse
 closepath
}bdf
/rectclip{newpath privrectpath clip newpath}def
/rectfill{gsave newpath privrectpath fill grestore}def
/rectstroke{gsave newpath privrectpath stroke grestore}def
xt {yt restore} if
/_fonthacksave false def
/currentpacking defed 
{
 /_bfh {/_fonthacksave currentpacking def false setpacking} bdf
 /_efh {_fonthacksave setpacking} bdf
}
{
 /_bfh {} bdf
 /_efh {} bdf
}ifelse
/packedarray{array astore readonly}ndf
/` 
{ 
 false setoverprint

 /-save0- save def
 5 index concat
 pop
 storerect left bottom width height rectclip
 pop
 
 /dict_count countdictstack def
 /op_count count 1 sub def
 userdict begin
 
 /showpage {} def
 
 0 setgray 0 setlinecap 1 setlinewidth
 0 setlinejoin 10 setmiterlimit [] 0 setdash newpath
 
} bdf
/currentpacking defed{true setpacking}if
/min{2 copy gt{exch}if pop}bdf
/max{2 copy lt{exch}if pop}bdf
/xformfont { currentfont exch makefont setfont } bdf
/fhnumcolors 1 
 statusdict begin
/processcolors defed 
{
pop processcolors
}
{
/deviceinfo defed {
deviceinfo /Colors known {
pop deviceinfo /Colors get
} if
} if
} ifelse
 end 
def
/printerRes 
 gsave
 matrix defaultmatrix setmatrix
 72 72 dtransform
 abs exch abs
 max
 grestore
 def
/graycalcs
[
 {Angle Frequency}   
 {GrayAngle GrayFrequency} 
 {0 Width Height matrix defaultmatrix idtransform 
dup mul exch dup mul add sqrt 72 exch div} 
 {0 GrayWidth GrayHeight matrix defaultmatrix idtransform 
dup mul exch dup mul add sqrt 72 exch div} 
] def
/calcgraysteps {
 forcemaxsteps
 {
maxsteps
 }
 {
/currenthalftone defed
{currenthalftone /dicttype eq}{false}ifelse
{
currenthalftone begin
HalftoneType 4 le
{graycalcs HalftoneType 1 sub get exec}
{
HalftoneType 5 eq
{
Default begin
{graycalcs HalftoneType 1 sub get exec}
end
}
{0 60} 
ifelse
}
ifelse
end
}
{
currentscreen pop exch 
}
ifelse
 
printerRes 300 max exch div exch 
2 copy 
sin mul round dup mul 
3 1 roll 
cos mul round dup mul 
add 1 add 
dup maxsteps gt {pop maxsteps} if 
 }
 ifelse
} bdf
/nextrelease defed { 
 /languagelevel defed not {    
/framebuffer defed { 
0 40 string framebuffer 9 1 roll 8 {pop} repeat
dup 516 eq exch 520 eq or
{
/fhnumcolors 3 def
/currentscreen {60 0 {pop pop 1}}bdf
/calcgraysteps {maxsteps} bdf
}if
}if
 }if
}if
fhnumcolors 1 ne {
 /calcgraysteps {maxsteps} bdf
} if
/currentpagedevice defed {

 currentpagedevice /PreRenderingEnhance known
 {
currentpagedevice /PreRenderingEnhance get
{
/calcgraysteps 
{
forcemaxsteps 
{maxsteps}
{256 maxsteps min}
ifelse
} def
} if
 } if
} if
/gradfrequency 144 def
printerRes 1000 lt {
 /gradfrequency 72 def
} if
/adjnumsteps {
 
 dup dtransform abs exch abs max  
 
 printerRes div       
 
 gradfrequency mul      
 round        
 5 max       
 min        
}bdf
/goodsep {
 spots exch get 4 get dup sepname eq exch (_vc_Registration) eq or
}bdf
/BeginGradation defed
{/bb{BeginGradation}bdf}
{/bb{}bdf}
ifelse
/EndGradation defed
{/eb{EndGradation}bdf}
{/eb{}bdf}
ifelse
/bottom -0 def 
/delta -0 def 
/frac -0 def 
/height -0 def 
/left -0 def 
/numsteps1 -0 def 
/radius -0 def 
/right -0 def 
/top -0 def 
/width -0 def 
/xt -0 def 
/yt -0 def 
/df currentflat def 
/tempstr 1 string def 
/clipflatness currentflat def 
/inverted? 
 0 currenttransfer exec .5 ge def
/tc1 [0 0 0 1] def 
/tc2 [0 0 0 1] def 
/storerect{/top xdf /right xdf /bottom xdf /left xdf 
/width right left sub def /height top bottom sub def}bdf
/concatprocs{
 systemdict /packedarray known 
 {dup type /packedarraytype eq 2 index type /packedarraytype eq or}{false}ifelse
 { 
/proc2 exch cvlit def /proc1 exch cvlit def
proc1 aload pop proc2 aload pop
proc1 length proc2 length add packedarray cvx
 }
 { 
/proc2 exch cvlit def /proc1 exch cvlit def
/newproc proc1 length proc2 length add array def
newproc 0 proc1 putinterval newproc proc1 length proc2 putinterval
newproc cvx
 }ifelse
}bdf
/i{dup 0 eq
 {pop df dup} 
 {dup} ifelse 
 /clipflatness xdf setflat
}bdf
version cvr 38.0 le
{/setrgbcolor{
currenttransfer exec 3 1 roll
currenttransfer exec 3 1 roll
currenttransfer exec 3 1 roll
setrgbcolor}bdf}if
/vms {/vmsv save def} bdf
/vmr {vmsv restore} bdf
/vmrs{vmsv restore /vmsv save def}bdf
/eomode{ 
 {/filler /eofill load def /clipper /eoclip load def}
 {/filler /fill load def /clipper /clip load def}
 ifelse
}bdf
/normtaper{}bdf
/logtaper{9 mul 1 add log}bdf
/CD{
 /NF exch def 
 {    
exch dup 
/FID ne 1 index/UniqueID ne and
{exch NF 3 1 roll put}
{pop pop}
ifelse
 }forall 
 NF
}bdf
/MN{
 1 index length   
 /Len exch def 
 dup length Len add  
 string dup    
 Len     
 4 -1 roll    
 putinterval   
 dup     
 0      
 4 -1 roll   
 putinterval   
}bdf
/RC{4 -1 roll /ourvec xdf 256 string cvs(|______)anchorsearch
 {1 index MN cvn/NewN exch def cvn
 findfont dup maxlength dict CD dup/FontName NewN put dup
 /Encoding ourvec put NewN exch definefont pop}{pop}ifelse}bdf
/RF{ 
 dup      
 FontDirectory exch   
 known     
 {pop 3 -1 roll pop}  
 {RC}
 ifelse
}bdf
/FF{dup 256 string cvs(|______)exch MN cvn dup FontDirectory exch known
 {exch pop findfont 3 -1 roll pop}
 {pop dup findfont dup maxlength dict CD dup dup
 /Encoding exch /Encoding get 256 array copy 7 -1 roll 
 {3 -1 roll dup 4 -2 roll put}forall put definefont}
 ifelse}bdf
/RFJ{ 
 dup      
 FontDirectory exch   
 known     
 {pop 3 -1 roll pop  
 FontDirectory /Ryumin-Light-83pv-RKSJ-H known 
 {pop pop /Ryumin-Light-83pv-RKSJ-H dup}if  
 }      
 {RC}
 ifelse
}bdf
/FFJ{dup 256 string cvs(|______)exch MN cvn dup FontDirectory exch known
 {exch pop findfont 3 -1 roll pop}
 {pop
dup FontDirectory exch known not 
 {FontDirectory /Ryumin-Light-83pv-RKSJ-H known 
{pop /Ryumin-Light-83pv-RKSJ-H}if 
 }if            
 dup findfont dup maxlength dict CD dup dup
 /Encoding exch /Encoding get 256 array copy 7 -1 roll 
 {3 -1 roll dup 4 -2 roll put}forall put definefont}
 ifelse}bdf
/fps{
 currentflat   
 exch     
 dup 0 le{pop 1}if 
 {
dup setflat 3 index stopped
{1.3 mul dup 3 index gt{pop setflat pop pop stop}if} 
{exit} 
ifelse
 }loop 
 pop setflat pop pop
}bdf
/fp{100 currentflat fps}bdf
/clipper{clip}bdf 
/W{/clipper load 100 clipflatness dup setflat fps}bdf
userdict begin /BDFontDict 29 dict def end
BDFontDict begin
/bu{}def
/bn{}def
/setTxMode{av 70 ge{pop}if pop}def
/gm{m}def
/show{pop}def
/gr{pop}def
/fnt{pop pop pop}def
/fs{pop}def
/fz{pop}def
/lin{pop pop}def
/:M {pop pop} def
/sf {pop} def
/S {pop} def
/@b {pop pop pop pop pop pop pop pop} def
/_bdsave /save load def
/_bdrestore /restore load def
/save { dup /fontsave eq {null} {_bdsave} ifelse } def
/restore { dup null eq { pop } { _bdrestore } ifelse } def
/fontsave null def
end
/MacVec 256 array def 
MacVec 0 /Helvetica findfont
/Encoding get 0 128 getinterval putinterval
MacVec 127 /DEL put MacVec 16#27 /quotesingle put MacVec 16#60 /grave put
/NUL/SOH/STX/ETX/EOT/ENQ/ACK/BEL/BS/HT/LF/VT/FF/CR/SO/SI
/DLE/DC1/DC2/DC3/DC4/NAK/SYN/ETB/CAN/EM/SUB/ESC/FS/GS/RS/US
MacVec 0 32 getinterval astore pop
/Adieresis/Aring/Ccedilla/Eacute/Ntilde/Odieresis/Udieresis/aacute
/agrave/acircumflex/adieresis/atilde/aring/ccedilla/eacute/egrave
/ecircumflex/edieresis/iacute/igrave/icircumflex/idieresis/ntilde/oacute
/ograve/ocircumflex/odieresis/otilde/uacute/ugrave/ucircumflex/udieresis
/dagger/degree/cent/sterling/section/bullet/paragraph/germandbls
/registered/copyright/trademark/acute/dieresis/notequal/AE/Oslash
/infinity/plusminus/lessequal/greaterequal/yen/mu/partialdiff/summation
/product/pi/integral/ordfeminine/ordmasculine/Omega/ae/oslash 
/questiondown/exclamdown/logicalnot/radical/florin/approxequal/Delta/guillemotleft
/guillemotright/ellipsis/nbspace/Agrave/Atilde/Otilde/OE/oe
/endash/emdash/quotedblleft/quotedblright/quoteleft/quoteright/divide/lozenge
/ydieresis/Ydieresis/fraction/currency/guilsinglleft/guilsinglright/fi/fl
/daggerdbl/periodcentered/quotesinglbase/quotedblbase
/perthousand/Acircumflex/Ecircumflex/Aacute
/Edieresis/Egrave/Iacute/Icircumflex/Idieresis/Igrave/Oacute/Ocircumflex
/apple/Ograve/Uacute/Ucircumflex/Ugrave/dotlessi/circumflex/tilde
/macron/breve/dotaccent/ring/cedilla/hungarumlaut/ogonek/caron
MacVec 128 128 getinterval astore pop
end 
AltsysDict begin
_bfh
_efh
0 dict dup begin
end 
/f0 /Symbol FF def
_bfh
_efh
MacVec 256 array copy
/f1 /|______Times-Roman dup RF findfont def
end 
AltsysDict begin 
/onlyk4{false}ndf
/ccmyk{dup 5 -1 roll sub 0 max exch}ndf
/cmyk2gray{
 4 -1 roll 0.3 mul 4 -1 roll 0.59 mul 4 -1 roll 0.11 mul
 add add add 1 min neg 1 add
}bdf
/setcmykcolor{1 exch sub ccmyk ccmyk ccmyk pop setrgbcolor}ndf
/maxcolor { 
 max max max  
} ndf
/maxspot {
 pop
} ndf
/setcmykcoloroverprint{4{dup -1 eq{pop 0}if 4 1 roll}repeat setcmykcolor}ndf
/findcmykcustomcolor{5 packedarray}ndf
/setcustomcolor{exch aload pop pop 4{4 index mul 4 1 roll}repeat setcmykcolor pop}ndf
/setseparationgray{setgray}ndf
/setoverprint{pop}ndf 
/currentoverprint false ndf
/cmykbufs2gray{
 0 1 2 index length 1 sub
 { 
4 index 1 index get 0.3 mul 
4 index 2 index get 0.59 mul 
4 index 3 index get 0.11 mul 
4 index 4 index get 
add add add cvi 255 min
255 exch sub
2 index 3 1 roll put
 }for
 4 1 roll pop pop pop
}bdf
/colorimage{
 pop pop
 [
5 -1 roll/exec cvx 
6 -1 roll/exec cvx 
7 -1 roll/exec cvx 
8 -1 roll/exec cvx
/cmykbufs2gray cvx
 ]cvx 
 image
}
version cvr 47.1 le 
statusdict /product get (Lino) anchorsearch{pop pop true}{pop false}ifelse
and{userdict begin bdf end}{ndf}ifelse
fhnumcolors 1 ne {/yt save def} if
/customcolorimage{
 aload pop
 (_vc_Registration) eq 
 {
pop pop pop pop separationimage
 }
 {
/ik xdf /iy xdf /im xdf /ic xdf
ic im iy ik cmyk2gray /xt xdf
currenttransfer
{dup 1.0 exch sub xt mul add}concatprocs
st 
image
 }
 ifelse
}ndf
fhnumcolors 1 ne {yt restore} if
fhnumcolors 3 ne {/yt save def} if
/customcolorimage{
 aload pop 
 (_vc_Registration) eq 
 {
pop pop pop pop separationimage
 }
 {
/ik xdf /iy xdf /im xdf /ic xdf
1.0 dup ic ik add min sub 
1.0 dup im ik add min sub 
1.0 dup iy ik add min sub 
/ic xdf /iy xdf /im xdf
currentcolortransfer
4 1 roll 
{dup 1.0 exch sub ic mul add}concatprocs 4 1 roll 
{dup 1.0 exch sub iy mul add}concatprocs 4 1 roll 
{dup 1.0 exch sub im mul add}concatprocs 4 1 roll 
setcolortransfer
{/dummy xdf dummy}concatprocs{dummy}{dummy}true 3 colorimage
 }
 ifelse
}ndf
fhnumcolors 3 ne {yt restore} if
fhnumcolors 4 ne {/yt save def} if
/customcolorimage{
 aload pop
 (_vc_Registration) eq 
 {
pop pop pop pop separationimage
 }
 {
/ik xdf /iy xdf /im xdf /ic xdf
currentcolortransfer
{1.0 exch sub ik mul ik sub 1 add}concatprocs 4 1 roll
{1.0 exch sub iy mul iy sub 1 add}concatprocs 4 1 roll
{1.0 exch sub im mul im sub 1 add}concatprocs 4 1 roll
{1.0 exch sub ic mul ic sub 1 add}concatprocs 4 1 roll
setcolortransfer
{/dummy xdf dummy}concatprocs{dummy}{dummy}{dummy}
true 4 colorimage
 }
 ifelse
}ndf
fhnumcolors 4 ne {yt restore} if
/separationimage{image}ndf
/newcmykcustomcolor{6 packedarray}ndf
/inkoverprint false ndf
/setinkoverprint{pop}ndf 
/setspotcolor { 
 spots exch get
 dup 4 get (_vc_Registration) eq
 {pop 1 exch sub setseparationgray}
 {0 5 getinterval exch setcustomcolor}
 ifelse
}ndf
/currentcolortransfer{currenttransfer dup dup dup}ndf
/setcolortransfer{st pop pop pop}ndf
/fas{}ndf
/sas{}ndf
/fhsetspreadsize{pop}ndf
/filler{fill}bdf 
/F{gsave {filler}fp grestore}bdf
/f{closepath F}bdf
/S{gsave {stroke}fp grestore}bdf
/s{closepath S}bdf
/bc4 [0 0 0 0] def 
/_lfp4 {
 /iosv inkoverprint def
 /cosv currentoverprint def
 /yt xdf       
 /xt xdf       
 /ang xdf      
 storerect
 /taperfcn xdf
 /k2 xdf /y2 xdf /m2 xdf /c2 xdf
 /k1 xdf /y1 xdf /m1 xdf /c1 xdf
 c1 c2 sub abs
 m1 m2 sub abs
 y1 y2 sub abs
 k1 k2 sub abs
 maxcolor      
 calcgraysteps mul abs round  
 height abs adjnumsteps   
 dup 2 lt {pop 1} if    
 1 sub /numsteps1 xdf
 currentflat mark    
 currentflat clipflatness  
 /delta top bottom sub numsteps1 1 add div def 
 /right right left sub def  
 /botsv top delta sub def  
 {
{
W
xt yt translate 
ang rotate
xt neg yt neg translate 
dup setflat 
/bottom botsv def
0 1 numsteps1 
{
numsteps1 dup 0 eq {pop 0.5 } { div } ifelse 
taperfcn /frac xdf
bc4 0 c2 c1 sub frac mul c1 add put
bc4 1 m2 m1 sub frac mul m1 add put
bc4 2 y2 y1 sub frac mul y1 add put
bc4 3 k2 k1 sub frac mul k1 add put
bc4 vc
1 index setflat 
{ 
mark {newpath left bottom right delta rectfill}stopped
{cleartomark exch 1.3 mul dup setflat exch 2 copy gt{stop}if}
{cleartomark exit}ifelse
}loop
/bottom bottom delta sub def
}for
}
gsave stopped grestore
{exch pop 2 index exch 1.3 mul dup 100 gt{cleartomark setflat stop}if}
{exit}ifelse
 }loop
 cleartomark setflat
 iosv setinkoverprint
 cosv setoverprint
}bdf
/bcs [0 0] def 
/_lfs4 {
 /iosv inkoverprint def
 /cosv currentoverprint def
 /yt xdf       
 /xt xdf       
 /ang xdf      
 storerect
 /taperfcn xdf
 /tint2 xdf      
 /tint1 xdf      
 bcs exch 1 exch put    
 tint1 tint2 sub abs    
 bcs 1 get maxspot    
 calcgraysteps mul abs round  
 height abs adjnumsteps   
 dup 2 lt {pop 2} if    
 1 sub /numsteps1 xdf
 currentflat mark    
 currentflat clipflatness  
 /delta top bottom sub numsteps1 1 add div def 
 /right right left sub def  
 /botsv top delta sub def  
 {
{
W
xt yt translate 
ang rotate
xt neg yt neg translate 
dup setflat 
/bottom botsv def
0 1 numsteps1 
{
numsteps1 div taperfcn /frac xdf
bcs 0
1.0 tint2 tint1 sub frac mul tint1 add sub
put bcs vc
1 index setflat 
{ 
mark {newpath left bottom right delta rectfill}stopped
{cleartomark exch 1.3 mul dup setflat exch 2 copy gt{stop}if}
{cleartomark exit}ifelse
}loop
/bottom bottom delta sub def
}for
}
gsave stopped grestore
{exch pop 2 index exch 1.3 mul dup 100 gt{cleartomark setflat stop}if}
{exit}ifelse
 }loop
 cleartomark setflat
 iosv setinkoverprint
 cosv setoverprint
}bdf
/_rfs4 {
 /iosv inkoverprint def
 /cosv currentoverprint def
 /tint2 xdf      
 /tint1 xdf      
 bcs exch 1 exch put    
 /radius xdf      
 /yt xdf       
 /xt xdf       
 tint1 tint2 sub abs    
 bcs 1 get maxspot    
 calcgraysteps mul abs round  
 radius abs adjnumsteps   
 dup 2 lt {pop 2} if    
 1 sub /numsteps1 xdf
 radius numsteps1 div 2 div /halfstep xdf 
 currentflat mark    
 currentflat clipflatness  
 {
{
dup setflat 
W 
0 1 numsteps1 
{
dup /radindex xdf
numsteps1 div /frac xdf
bcs 0
tint2 tint1 sub frac mul tint1 add
put bcs vc
1 index setflat 
{ 
newpath mark xt yt radius 1 frac sub mul halfstep add 0 360
{ arc
radindex numsteps1 ne 
{
xt yt 
radindex 1 add numsteps1 
div 1 exch sub 
radius mul halfstep add
dup xt add yt moveto 
360 0 arcn 
} if
fill
}stopped
{cleartomark exch 1.3 mul dup setflat exch 2 copy gt{stop}if}
{cleartomark exit}ifelse
}loop
}for
}
gsave stopped grestore
{exch pop 2 index exch 1.3 mul dup 100 gt{cleartomark setflat stop}if}
{exit}ifelse
 }loop
 cleartomark setflat
 iosv setinkoverprint
 cosv setoverprint
}bdf
/_rfp4 {
 /iosv inkoverprint def
 /cosv currentoverprint def
 /k2 xdf /y2 xdf /m2 xdf /c2 xdf
 /k1 xdf /y1 xdf /m1 xdf /c1 xdf
 /radius xdf      
 /yt xdf       
 /xt xdf       
 c1 c2 sub abs
 m1 m2 sub abs
 y1 y2 sub abs
 k1 k2 sub abs
 maxcolor      
 calcgraysteps mul abs round  
 radius abs adjnumsteps   
 dup 2 lt {pop 1} if    
 1 sub /numsteps1 xdf
 radius numsteps1 dup 0 eq {pop} {div} ifelse 
 2 div /halfstep xdf 
 currentflat mark    
 currentflat clipflatness  
 {
{
dup setflat 
W 
0 1 numsteps1 
{
dup /radindex xdf
numsteps1 dup 0 eq {pop 0.5 } { div } ifelse 
/frac xdf
bc4 0 c2 c1 sub frac mul c1 add put
bc4 1 m2 m1 sub frac mul m1 add put
bc4 2 y2 y1 sub frac mul y1 add put
bc4 3 k2 k1 sub frac mul k1 add put
bc4 vc
1 index setflat 
{ 
newpath mark xt yt radius 1 frac sub mul halfstep add 0 360
{ arc
radindex numsteps1 ne 
{
xt yt 
radindex 1 add 
numsteps1 dup 0 eq {pop} {div} ifelse 
1 exch sub 
radius mul halfstep add
dup xt add yt moveto 
360 0 arcn 
} if
fill
}stopped
{cleartomark exch 1.3 mul dup setflat exch 2 copy gt{stop}if}
{cleartomark exit}ifelse
}loop
}for
}
gsave stopped grestore
{exch pop 2 index exch 1.3 mul dup 100 gt{cleartomark setflat stop}if}
{exit}ifelse
 }loop
 cleartomark setflat
 iosv setinkoverprint
 cosv setoverprint
}bdf
/lfp4{_lfp4}ndf
/lfs4{_lfs4}ndf
/rfs4{_rfs4}ndf
/rfp4{_rfp4}ndf
/cvc [0 0 0 1] def 
/vc{
 AltsysDict /cvc 2 index put 
 aload length 4 eq
 {setcmykcolor}
 {setspotcolor}
 ifelse
}bdf 
/origmtx matrix currentmatrix def
/ImMatrix matrix currentmatrix def
0 setseparationgray
/imgr {1692 1584 2304 2376 } def 
/bleed 0 def 
/clpr {1692 1584 2304 2376 } def 
/xs 1 def 
/ys 1 def 
/botx 0 def 
/overlap 0 def 
/wdist 18 def 
0 2 mul fhsetspreadsize 
0 0 ne {/df 0 def /clipflatness 0 def} if 
/maxsteps 256 def 
/forcemaxsteps false def 
vms
-1698 -1652 translate
/currentpacking defed{false setpacking}if 
/spots[
1 0 0 0 (Process Cyan) false newcmykcustomcolor
0 1 0 0 (Process Magenta) false newcmykcustomcolor
0 0 1 0 (Process Yellow) false newcmykcustomcolor
0 0 0 1 (Process Black) false newcmykcustomcolor
]def
/textopf false def
/curtextmtx{}def
/otw .25 def
/msf{dup/curtextmtx xdf makefont setfont}bdf
/makesetfont/msf load def
/curtextheight{.707104 .707104 curtextmtx dtransform
 dup mul exch dup mul add sqrt}bdf
/ta2{ 
tempstr 2 index gsave exec grestore 
cwidth cheight rmoveto 
4 index eq{5 index 5 index rmoveto}if 
2 index 2 index rmoveto 
}bdf
/ta{exch systemdict/cshow known
{{/cheight xdf/cwidth xdf tempstr 0 2 index put ta2}exch cshow} 
{{tempstr 0 2 index put tempstr stringwidth/cheight xdf/cwidth xdf ta2}forall} 
ifelse 6{pop}repeat}bdf
/sts{/textopf currentoverprint def vc setoverprint
/ts{awidthshow}def exec textopf setoverprint}bdf
/stol{/xt currentlinewidth def 
 setlinewidth vc newpath 
 /ts{{false charpath stroke}ta}def exec 
 xt setlinewidth}bdf 
 
/strk{/textopf currentoverprint def vc setoverprint
 /ts{{false charpath stroke}ta}def exec 
 textopf setoverprint
 }bdf 
n
[] 0 d
3.863708 M
1 w
0 j
0 J
false setoverprint
0 i
false eomode
[0 0 0 1] vc
vms
1765.898 2194.8656 m
2205.7915 2194.8656 L
2205.7915 1754.9722 L
1765.898 1754.9722 L
1765.898 2194.8656 L
3.863693 M
s 
n
1765.898 1754.9722 m
2205.7915 2194.8656 L
0.5 w
3.863708 M
S 
n
1986.4824 2066.8651 m
1765.898 1754.9722 L
1 w
3.863693 M
S 
n
1986.4824 2066.8651 m
2205.7915 1754.9722 L
S 
n
1986.4824 2066.8651 m
2077.791 2066.8651 L
0.5 w
3.863708 M
S 
n
2077.791 2066.8651 m
2077.791 1938.8646 L
S 
n
2077.791 1938.8646 m
2077.791 1754.9722 L
3.863693 M
[4 4 ] 0 d
S 
[] 0 d
n
2024.8826 2015.6649 m
2024.8826 1754.9722 L
[4 4 ] 0 d
S 
[] 0 d
n
2024.1908 2013.7731 m
1948.1905 2013.7731 L
S 
n
1948.1905 2013.7731 m
1948.1905 1937.3728 L
S 
n
1948.1905 1937.3728 m
2078.191 1937.3728 L
S 
n
1948.1905 1937.3728 m
1948.1905 1754.9722 L
[4 4 ] 0 d
S 
[] 0 d
n
3.863708 M
S 
n
q
_bfh
_efh
{
f1 [19.200073 0 0 19.200073 0 0] makesetfont
1768.881607 1726.663788 m
0 0 32 0 0 (0) ts
}
true
[0 0 0 1]sts
Q
false setoverprint
S 
n
q
_bfh
_efh
{
f1 [19.200073 0 0 19.200073 0 0] makesetfont
2024.882599 1727.663788 m
0 0 32 0 0 (x*) ts
}
true
[0 0 0 1]sts
Q
false setoverprint
S 
n
q
_bfh
_efh
{
f1 [19.200073 0 0 19.200073 0 0] makesetfont
2205.791504 1729.372086 m
0 0 32 0 0 (1) ts
}
true
[0 0 0 1]sts
Q
false setoverprint
S 
n
q
_bfh
_efh
{
f1 [19.200073 0 0 19.200073 0 0] makesetfont
1741.789703 2177.373779 m
0 0 32 0 0 (1) ts
}
true
[0 0 0 1]sts
Q
1770 2067 m
1760 2067 L
1 w
3.863693 M
false setoverprint
S 
n
1762.6845 1991.1997 m
1720 1991.1997 L
1720 1966 L
1762.6845 1966 L
1762.6845 1991.1997 L
n
q
_bfh
_efh
{
f1 [24 0 0 24 0 0] makesetfont
1720 1972 m
0 0 32 0 0 (T\(x\)) ts
}
true
[0 0 0 1]sts
Q
1756.844 1774.6798 m
1747.204 1774.6798 L
1747.204 1754.4799 L
1756.844 1754.4799 L
1756.844 1774.6798 L
n
q
_bfh
_efh
{
f1 [19.200073 0 0 19.200073 0 0] makesetfont
1747.203979 1759.319946 m
0 0 32 0 0 (0) ts
}
true
[0 0 0 1]sts
Q
1986 1759 m
1986 1751 L
false setoverprint
S 
n
vmrs
2170.3642 1707.1997 m
2145 1707.1997 L
2145 1682 L
2170.3642 1682 L
2170.3642 1707.1997 L
n
q
_bfh
_efh
{
f1 [24 0 0 24 0 0] makesetfont
2145 1688 m
0 0 32 0 0 (IV) ts
}
true
[0 0 0 1]sts
Q
1995.0244 1707.1997 m
1979 1707.1997 L
1979 1682 L
1995.0244 1682 L
1995.0244 1707.1997 L
n
q
_bfh
_efh
{
f1 [24 0 0 24 0 0] makesetfont
1979 1688 m
0 0 32 0 0 (II) ts
}
true
[0 0 0 1]sts
Q
1882.0322 1709.1997 m
1874 1709.1997 L
1874 1684 L
1882.0322 1684 L
1882.0322 1709.1997 L
n
q
_bfh
_efh
{
f1 [24 0 0 24 0 0] makesetfont
1874 1690 m
0 0 32 0 0 (I) ts
}
true
[0 0 0 1]sts
Q
2010.04 1677.1997 m
1998 1677.1997 L
1998 1652 L
2010.04 1652 L
2010.04 1677.1997 L
n
q
_bfh
_efh
{
f1 [24 0 0 24 0 0] makesetfont
1998 1658 m
0 0 32 0 0 (x) ts
}
true
[0 0 0 1]sts
Q
1991.64 1745.1997 m
1982 1745.1997 L
1982 1724.9999 L
1991.64 1724.9999 L
1991.64 1745.1997 L
n
q
_bfh
_efh
{
f1 [19.200073 0 0 19.200073 0 0] makesetfont
1982 1729.83989 m
0 0 32 0 0 (x) ts
}
true
[0 0 0 1]sts
Q
1999.4314 1736.1997 m
1993 1736.1997 L
1993 1720.7998 L
1999.4314 1720.7998 L
1999.4314 1736.1997 L
n
q
_bfh
_efh
{
f1 [14.400146 0 0 14.400146 0 0] makesetfont
1993 1724.679779 m
0 0 32 0 0 (c) ts
}
true
[0 0 0 1]sts
Q
2128 1747 m
2076 1747 L
2076 1727 L
2128 1727 L
2128 1747 L
n
q
_bfh
_efh
{
f1 [19.200073 0 0 19.200073 0 0] makesetfont
2076 1731.640182 m
0 -2 rmoveto 0 0 32 0 0 (T) ts
0 2 rmoveto }
true
[0 0 0 1]sts
{
f1 [12 0 0 12 0 0] makesetfont
0 12 rmoveto 0 0 32 0 0 (\(1\)) ts
0 -12 rmoveto }
true
[0 0 0 1]sts
{
f1 [12 0 0 12 0 0] makesetfont
0 0 32 0 0 (\(x) ts
}
true
[0 0 0 1]sts
{
f1 [12 0 0 12 0 0] makesetfont
0 -3.5 rmoveto 0 0 32 0 0 (c) ts
0 3.5 rmoveto }
true
[0 0 0 1]sts
{
f1 [12 0 0 12 0 0] makesetfont
0 0 32 0 0 (\)) ts
}
true
[0 0 0 1]sts
Q
1960.0787 1746.5999 m
1915 1746.5999 L
1915 1726.4 L
1960.0787 1726.4 L
1960.0787 1746.5999 L
n
q
_bfh
_efh
{
f1 [19.200073 0 0 19.200073 0 0] makesetfont
1915 1731.240036 m
0 -2 rmoveto 0 0 32 0 0 (T) ts
0 2 rmoveto }
true
[0 0 0 1]sts
{
f1 [12 0 0 12 0 0] makesetfont
0 12 rmoveto 0 0 32 0 0 (\(2\)) ts
0 -12 rmoveto }
true
[0 0 0 1]sts
{
f1 [12 0 0 12 0 0] makesetfont
0 0 32 0 0 (\(x) ts
}
true
[0 0 0 1]sts
{
f1 [12 0 0 12 0 0] makesetfont
0 -3.5 rmoveto 0 0 32 0 0 (c) ts
0 3.5 rmoveto }
true
[0 0 0 1]sts
{
f1 [12 0 0 12 0 0] makesetfont
0 0 32 0 0 (\)) ts
}
true
[0 0 0 1]sts
Q
1776.5 1679 m
1937.5 1679 L
3.863693 M
false setoverprint
S 
n
true eomode
1776.9166 1679 m
1776.9166 1676.5 L
1770.6667 1679 L
1776.9166 1681.5 L
1776.9166 1679 L
f 
n
1937.0834 1679 m
1937.0834 1681.5 L
1943.3333 1679 L
1937.0834 1676.5 L
1937.0834 1679 L
f 
n
1958 1679 m
2016 1679 L
S 
n
1958.4166 1679 m
1958.4166 1676.5 L
1952.1667 1679 L
1958.4166 1681.5 L
1958.4166 1679 L
f 
n
2015.5834 1679 m
2015.5834 1681.5 L
2021.8333 1679 L
2015.5834 1676.5 L
2015.5834 1679 L
f 
n
2034.5 1679 m
2070 1679 L
S 
n
2034.9166 1679 m
2034.9166 1676.5 L
2028.6667 1679 L
2034.9166 1681.5 L
2034.9166 1679 L
f 
n
2069.5834 1679 m
2069.5834 1681.5 L
2075.8333 1679 L
2069.5834 1676.5 L
2069.5834 1679 L
f 
n
2086.5 1679 m
2195.5 1679 L
S 
n
2086.9166 1679 m
2086.9166 1676.5 L
2080.6667 1679 L
2086.9166 1681.5 L
2086.9166 1679 L
f 
n
2195.0834 1679 m
2195.0834 1681.5 L
2201.3333 1679 L
2195.0834 1676.5 L
2195.0834 1679 L
f 
n
2068.0165 1709.6997 m
2044 1709.6997 L
2044 1684.5 L
2068.0165 1684.5 L
2068.0165 1709.6997 L
n
q
_bfh
_efh
{
f1 [24 0 0 24 0 0] makesetfont
2044 1690.5 m
0 -3 rmoveto 0 0 32 0 0 (III) ts
0 3 rmoveto }
true
[0 0 0 1]sts
Q
1756.0666 2084.0999 m
1708.5 2084.0999 L
1708.5 2043.567 L
1756.0666 2043.567 L
1756.0666 2084.0999 L
n
q
q
[1 0 0 1 1708.5 2054.674057] concat
vms
q
[1 0 0 1 -1704 -2055.5] concat
vms
1734.1866 2085.8998 m
1717.5 2085.8998 L
1717.5 2060.7001 L
1734.1866 2060.7001 L
1734.1866 2085.8998 L
n
q
_bfh
_efh
{
f1 [24 0 0 24 0 0] makesetfont
1717.5 2066.700073 m
0 -3 rmoveto 0 0 32 4.666626 0 (2) ts
0 3 rmoveto }
true
[0 0 0 1]sts
Q
1717.5 2082.5 m
1732.5 2082.5 L
false setoverprint
S 
n
1717.5 2082.5 m
1710.5 2060.5 L
0.5 w
S 
n
1710.5 2060.5 m
1707.5 2071 L
2 w
S 
n
1707.5 2071 m
1704 2066.5 L
1 w
S 
n
1737.5 2083.5 m
1729.5 2055.5 L
S 
n
vmr
Q
vmr
Q
_bfh
_efh
{
f1 [24 0 0 24 0 0] makesetfont
1744.026642 2051.674057 m
0 11 rmoveto 0 0 32 0 0 (2) ts
0 -11 rmoveto }
true
[0 0 0 1]sts
Q
vmr
vmr
end
EndOfTheIncludedPostscriptMagicCookie

\closepsdump

\psdump{bif.eps}
userdict begin /AltsysDict 245 dict def end
AltsysDict begin
/bdf{bind def}bind def
/xdf{exch def}bdf
/defed{where{pop true}{false}ifelse}bdf
/ndf{1 index where{pop pop pop}{dup xcheck{bind}if def}ifelse}bdf
/d{setdash}bdf
/h{closepath}bdf
/H{}bdf
/J{setlinecap}bdf
/j{setlinejoin}bdf
/M{setmiterlimit}bdf
/n{newpath}bdf
/N{newpath}bdf
/q{gsave}bdf
/Q{grestore}bdf
/w{setlinewidth}bdf
/sepdef{
 dup where not
 {
AltsysSepDict
 }
 if 
 3 1 roll exch put
}bdf
/st{settransfer}bdf
/colorimage defed /_rci xdf
/_NXLevel2 defed { 
 _NXLevel2 not {   
/colorimage where {
userdict eq {
/_rci false def 
} if
} if
 } if
} if
/md defed{ 
 md type /dicttype eq {  
/colorimage where { 
md eq { 
/_rci false def 
}if
}if
/settransfer where {
md eq {
/st systemdict /settransfer get def
}if
}if
 }if 
}if
/setstrokeadjust defed
{
 true setstrokeadjust
 /C{curveto}bdf
 /L{lineto}bdf
 /m{moveto}bdf
}
{
 /dr{transform .25 sub round .25 add 
exch .25 sub round .25 add exch itransform}bdf
 /C{dr curveto}bdf
 /L{dr lineto}bdf
 /m{dr moveto}bdf
 /setstrokeadjust{pop}bdf 
}ifelse
/rectstroke defed /xt xdf
xt {/yt save def} if
/privrectpath { 
 4 -2 roll m
 dtransform round exch round exch idtransform 
 2 copy 0 lt exch 0 lt xor
 {dup 0 exch rlineto exch 0 rlineto neg 0 exch rlineto}
 {exch dup 0 rlineto exch 0 exch rlineto neg 0 rlineto}
 ifelse
 closepath
}bdf
/rectclip{newpath privrectpath clip newpath}def
/rectfill{gsave newpath privrectpath fill grestore}def
/rectstroke{gsave newpath privrectpath stroke grestore}def
xt {yt restore} if
/_fonthacksave false def
/currentpacking defed 
{
 /_bfh {/_fonthacksave currentpacking def false setpacking} bdf
 /_efh {_fonthacksave setpacking} bdf
}
{
 /_bfh {} bdf
 /_efh {} bdf
}ifelse
/packedarray{array astore readonly}ndf
/` 
{ 
 false setoverprint

 /-save0- save def
 5 index concat
 pop
 storerect left bottom width height rectclip
 pop
 
 /dict_count countdictstack def
 /op_count count 1 sub def
 userdict begin
 
 /showpage {} def
 
 0 setgray 0 setlinecap 1 setlinewidth
 0 setlinejoin 10 setmiterlimit [] 0 setdash newpath
 
} bdf
/currentpacking defed{true setpacking}if
/min{2 copy gt{exch}if pop}bdf
/max{2 copy lt{exch}if pop}bdf
/xformfont { currentfont exch makefont setfont } bdf
/fhnumcolors 1 
 statusdict begin
/processcolors defed 
{
pop processcolors
}
{
/deviceinfo defed {
deviceinfo /Colors known {
pop deviceinfo /Colors get
} if
} if
} ifelse
 end 
def
/printerRes 
 gsave
 matrix defaultmatrix setmatrix
 72 72 dtransform
 abs exch abs
 max
 grestore
 def
/graycalcs
[
 {Angle Frequency}   
 {GrayAngle GrayFrequency} 
 {0 Width Height matrix defaultmatrix idtransform 
dup mul exch dup mul add sqrt 72 exch div} 
 {0 GrayWidth GrayHeight matrix defaultmatrix idtransform 
dup mul exch dup mul add sqrt 72 exch div} 
] def
/calcgraysteps {
 forcemaxsteps
 {
maxsteps
 }
 {
/currenthalftone defed
{currenthalftone /dicttype eq}{false}ifelse
{
currenthalftone begin
HalftoneType 4 le
{graycalcs HalftoneType 1 sub get exec}
{
HalftoneType 5 eq
{
Default begin
{graycalcs HalftoneType 1 sub get exec}
end
}
{0 60} 
ifelse
}
ifelse
end
}
{
currentscreen pop exch 
}
ifelse
 
printerRes 300 max exch div exch 
2 copy 
sin mul round dup mul 
3 1 roll 
cos mul round dup mul 
add 1 add 
dup maxsteps gt {pop maxsteps} if 
 }
 ifelse
} bdf
/nextrelease defed { 
 /languagelevel defed not {    
/framebuffer defed { 
0 40 string framebuffer 9 1 roll 8 {pop} repeat
dup 516 eq exch 520 eq or
{
/fhnumcolors 3 def
/currentscreen {60 0 {pop pop 1}}bdf
/calcgraysteps {maxsteps} bdf
}if
}if
 }if
}if
fhnumcolors 1 ne {
 /calcgraysteps {maxsteps} bdf
} if
/currentpagedevice defed {

 currentpagedevice /PreRenderingEnhance known
 {
currentpagedevice /PreRenderingEnhance get
{
/calcgraysteps 
{
forcemaxsteps 
{maxsteps}
{256 maxsteps min}
ifelse
} def
} if
 } if
} if
/gradfrequency 144 def
printerRes 1000 lt {
 /gradfrequency 72 def
} if
/adjnumsteps {
 
 dup dtransform abs exch abs max  
 
 printerRes div       
 
 gradfrequency mul      
 round        
 5 max       
 min        
}bdf
/goodsep {
 spots exch get 4 get dup sepname eq exch (_vc_Registration) eq or
}bdf
/BeginGradation defed
{/bb{BeginGradation}bdf}
{/bb{}bdf}
ifelse
/EndGradation defed
{/eb{EndGradation}bdf}
{/eb{}bdf}
ifelse
/bottom -0 def 
/delta -0 def 
/frac -0 def 
/height -0 def 
/left -0 def 
/numsteps1 -0 def 
/radius -0 def 
/right -0 def 
/top -0 def 
/width -0 def 
/xt -0 def 
/yt -0 def 
/df currentflat def 
/tempstr 1 string def 
/clipflatness currentflat def 
/inverted? 
 0 currenttransfer exec .5 ge def
/tc1 [0 0 0 1] def 
/tc2 [0 0 0 1] def 
/storerect{/top xdf /right xdf /bottom xdf /left xdf 
/width right left sub def /height top bottom sub def}bdf
/concatprocs{
 systemdict /packedarray known 
 {dup type /packedarraytype eq 2 index type /packedarraytype eq or}{false}ifelse
 { 
/proc2 exch cvlit def /proc1 exch cvlit def
proc1 aload pop proc2 aload pop
proc1 length proc2 length add packedarray cvx
 }
 { 
/proc2 exch cvlit def /proc1 exch cvlit def
/newproc proc1 length proc2 length add array def
newproc 0 proc1 putinterval newproc proc1 length proc2 putinterval
newproc cvx
 }ifelse
}bdf
/i{dup 0 eq
 {pop df dup} 
 {dup} ifelse 
 /clipflatness xdf setflat
}bdf
version cvr 38.0 le
{/setrgbcolor{
currenttransfer exec 3 1 roll
currenttransfer exec 3 1 roll
currenttransfer exec 3 1 roll
setrgbcolor}bdf}if
/vms {/vmsv save def} bdf
/vmr {vmsv restore} bdf
/vmrs{vmsv restore /vmsv save def}bdf
/eomode{ 
 {/filler /eofill load def /clipper /eoclip load def}
 {/filler /fill load def /clipper /clip load def}
 ifelse
}bdf
/normtaper{}bdf
/logtaper{9 mul 1 add log}bdf
/CD{
 /NF exch def 
 {    
exch dup 
/FID ne 1 index/UniqueID ne and
{exch NF 3 1 roll put}
{pop pop}
ifelse
 }forall 
 NF
}bdf
/MN{
 1 index length   
 /Len exch def 
 dup length Len add  
 string dup    
 Len     
 4 -1 roll    
 putinterval   
 dup     
 0      
 4 -1 roll   
 putinterval   
}bdf
/RC{4 -1 roll /ourvec xdf 256 string cvs(|______)anchorsearch
 {1 index MN cvn/NewN exch def cvn
 findfont dup maxlength dict CD dup/FontName NewN put dup
 /Encoding ourvec put NewN exch definefont pop}{pop}ifelse}bdf
/RF{ 
 dup      
 FontDirectory exch   
 known     
 {pop 3 -1 roll pop}  
 {RC}
 ifelse
}bdf
/FF{dup 256 string cvs(|______)exch MN cvn dup FontDirectory exch known
 {exch pop findfont 3 -1 roll pop}
 {pop dup findfont dup maxlength dict CD dup dup
 /Encoding exch /Encoding get 256 array copy 7 -1 roll 
 {3 -1 roll dup 4 -2 roll put}forall put definefont}
 ifelse}bdf
/RFJ{ 
 dup      
 FontDirectory exch   
 known     
 {pop 3 -1 roll pop  
 FontDirectory /Ryumin-Light-83pv-RKSJ-H known 
 {pop pop /Ryumin-Light-83pv-RKSJ-H dup}if  
 }      
 {RC}
 ifelse
}bdf
/FFJ{dup 256 string cvs(|______)exch MN cvn dup FontDirectory exch known
 {exch pop findfont 3 -1 roll pop}
 {pop
dup FontDirectory exch known not 
 {FontDirectory /Ryumin-Light-83pv-RKSJ-H known 
{pop /Ryumin-Light-83pv-RKSJ-H}if 
 }if            
 dup findfont dup maxlength dict CD dup dup
 /Encoding exch /Encoding get 256 array copy 7 -1 roll 
 {3 -1 roll dup 4 -2 roll put}forall put definefont}
 ifelse}bdf
/fps{
 currentflat   
 exch     
 dup 0 le{pop 1}if 
 {
dup setflat 3 index stopped
{1.3 mul dup 3 index gt{pop setflat pop pop stop}if} 
{exit} 
ifelse
 }loop 
 pop setflat pop pop
}bdf
/fp{100 currentflat fps}bdf
/clipper{clip}bdf 
/W{/clipper load 100 clipflatness dup setflat fps}bdf
userdict begin /BDFontDict 29 dict def end
BDFontDict begin
/bu{}def
/bn{}def
/setTxMode{av 70 ge{pop}if pop}def
/gm{m}def
/show{pop}def
/gr{pop}def
/fnt{pop pop pop}def
/fs{pop}def
/fz{pop}def
/lin{pop pop}def
/:M {pop pop} def
/sf {pop} def
/S {pop} def
/@b {pop pop pop pop pop pop pop pop} def
/_bdsave /save load def
/_bdrestore /restore load def
/save { dup /fontsave eq {null} {_bdsave} ifelse } def
/restore { dup null eq { pop } { _bdrestore } ifelse } def
/fontsave null def
end
/MacVec 256 array def 
MacVec 0 /Helvetica findfont
/Encoding get 0 128 getinterval putinterval
MacVec 127 /DEL put MacVec 16#27 /quotesingle put MacVec 16#60 /grave put
/NUL/SOH/STX/ETX/EOT/ENQ/ACK/BEL/BS/HT/LF/VT/FF/CR/SO/SI
/DLE/DC1/DC2/DC3/DC4/NAK/SYN/ETB/CAN/EM/SUB/ESC/FS/GS/RS/US
MacVec 0 32 getinterval astore pop
/Adieresis/Aring/Ccedilla/Eacute/Ntilde/Odieresis/Udieresis/aacute
/agrave/acircumflex/adieresis/atilde/aring/ccedilla/eacute/egrave
/ecircumflex/edieresis/iacute/igrave/icircumflex/idieresis/ntilde/oacute
/ograve/ocircumflex/odieresis/otilde/uacute/ugrave/ucircumflex/udieresis
/dagger/degree/cent/sterling/section/bullet/paragraph/germandbls
/registered/copyright/trademark/acute/dieresis/notequal/AE/Oslash
/infinity/plusminus/lessequal/greaterequal/yen/mu/partialdiff/summation
/product/pi/integral/ordfeminine/ordmasculine/Omega/ae/oslash 
/questiondown/exclamdown/logicalnot/radical/florin/approxequal/Delta/guillemotleft
/guillemotright/ellipsis/nbspace/Agrave/Atilde/Otilde/OE/oe
/endash/emdash/quotedblleft/quotedblright/quoteleft/quoteright/divide/lozenge
/ydieresis/Ydieresis/fraction/currency/guilsinglleft/guilsinglright/fi/fl
/daggerdbl/periodcentered/quotesinglbase/quotedblbase
/perthousand/Acircumflex/Ecircumflex/Aacute
/Edieresis/Egrave/Iacute/Icircumflex/Idieresis/Igrave/Oacute/Ocircumflex
/apple/Ograve/Uacute/Ucircumflex/Ugrave/dotlessi/circumflex/tilde
/macron/breve/dotaccent/ring/cedilla/hungarumlaut/ogonek/caron
MacVec 128 128 getinterval astore pop
end 
AltsysDict begin
_bfh
_efh
0 dict dup begin
end 
/f0 /Symbol FF def
_bfh
_efh
MacVec 256 array copy
/f1 /|______Times-Roman dup RF findfont def
_bfh
_efh
0 dict dup begin
end 
/f2 /Math1 FF def
end 
AltsysDict begin 
/onlyk4{false}ndf
/ccmyk{dup 5 -1 roll sub 0 max exch}ndf
/cmyk2gray{
 4 -1 roll 0.3 mul 4 -1 roll 0.59 mul 4 -1 roll 0.11 mul
 add add add 1 min neg 1 add
}bdf
/setcmykcolor{1 exch sub ccmyk ccmyk ccmyk pop setrgbcolor}ndf
/maxcolor { 
 max max max  
} ndf
/maxspot {
 pop
} ndf
/setcmykcoloroverprint{4{dup -1 eq{pop 0}if 4 1 roll}repeat setcmykcolor}ndf
/findcmykcustomcolor{5 packedarray}ndf
/setcustomcolor{exch aload pop pop 4{4 index mul 4 1 roll}repeat setcmykcolor pop}ndf
/setseparationgray{setgray}ndf
/setoverprint{pop}ndf 
/currentoverprint false ndf
/cmykbufs2gray{
 0 1 2 index length 1 sub
 { 
4 index 1 index get 0.3 mul 
4 index 2 index get 0.59 mul 
4 index 3 index get 0.11 mul 
4 index 4 index get 
add add add cvi 255 min
255 exch sub
2 index 3 1 roll put
 }for
 4 1 roll pop pop pop
}bdf
/colorimage{
 pop pop
 [
5 -1 roll/exec cvx 
6 -1 roll/exec cvx 
7 -1 roll/exec cvx 
8 -1 roll/exec cvx
/cmykbufs2gray cvx
 ]cvx 
 image
}
version cvr 47.1 le 
statusdict /product get (Lino) anchorsearch{pop pop true}{pop false}ifelse
and{userdict begin bdf end}{ndf}ifelse
fhnumcolors 1 ne {/yt save def} if
/customcolorimage{
 aload pop
 (_vc_Registration) eq 
 {
pop pop pop pop separationimage
 }
 {
/ik xdf /iy xdf /im xdf /ic xdf
ic im iy ik cmyk2gray /xt xdf
currenttransfer
{dup 1.0 exch sub xt mul add}concatprocs
st 
image
 }
 ifelse
}ndf
fhnumcolors 1 ne {yt restore} if
fhnumcolors 3 ne {/yt save def} if
/customcolorimage{
 aload pop 
 (_vc_Registration) eq 
 {
pop pop pop pop separationimage
 }
 {
/ik xdf /iy xdf /im xdf /ic xdf
1.0 dup ic ik add min sub 
1.0 dup im ik add min sub 
1.0 dup iy ik add min sub 
/ic xdf /iy xdf /im xdf
currentcolortransfer
4 1 roll 
{dup 1.0 exch sub ic mul add}concatprocs 4 1 roll 
{dup 1.0 exch sub iy mul add}concatprocs 4 1 roll 
{dup 1.0 exch sub im mul add}concatprocs 4 1 roll 
setcolortransfer
{/dummy xdf dummy}concatprocs{dummy}{dummy}true 3 colorimage
 }
 ifelse
}ndf
fhnumcolors 3 ne {yt restore} if
fhnumcolors 4 ne {/yt save def} if
/customcolorimage{
 aload pop
 (_vc_Registration) eq 
 {
pop pop pop pop separationimage
 }
 {
/ik xdf /iy xdf /im xdf /ic xdf
currentcolortransfer
{1.0 exch sub ik mul ik sub 1 add}concatprocs 4 1 roll
{1.0 exch sub iy mul iy sub 1 add}concatprocs 4 1 roll
{1.0 exch sub im mul im sub 1 add}concatprocs 4 1 roll
{1.0 exch sub ic mul ic sub 1 add}concatprocs 4 1 roll
setcolortransfer
{/dummy xdf dummy}concatprocs{dummy}{dummy}{dummy}
true 4 colorimage
 }
 ifelse
}ndf
fhnumcolors 4 ne {yt restore} if
/separationimage{image}ndf
/newcmykcustomcolor{6 packedarray}ndf
/inkoverprint false ndf
/setinkoverprint{pop}ndf 
/setspotcolor { 
 spots exch get
 dup 4 get (_vc_Registration) eq
 {pop 1 exch sub setseparationgray}
 {0 5 getinterval exch setcustomcolor}
 ifelse
}ndf
/currentcolortransfer{currenttransfer dup dup dup}ndf
/setcolortransfer{st pop pop pop}ndf
/fas{}ndf
/sas{}ndf
/fhsetspreadsize{pop}ndf
/filler{fill}bdf 
/F{gsave {filler}fp grestore}bdf
/f{closepath F}bdf
/S{gsave {stroke}fp grestore}bdf
/s{closepath S}bdf
/bc4 [0 0 0 0] def 
/_lfp4 {
 /iosv inkoverprint def
 /cosv currentoverprint def
 /yt xdf       
 /xt xdf       
 /ang xdf      
 storerect
 /taperfcn xdf
 /k2 xdf /y2 xdf /m2 xdf /c2 xdf
 /k1 xdf /y1 xdf /m1 xdf /c1 xdf
 c1 c2 sub abs
 m1 m2 sub abs
 y1 y2 sub abs
 k1 k2 sub abs
 maxcolor      
 calcgraysteps mul abs round  
 height abs adjnumsteps   
 dup 2 lt {pop 1} if    
 1 sub /numsteps1 xdf
 currentflat mark    
 currentflat clipflatness  
 /delta top bottom sub numsteps1 1 add div def 
 /right right left sub def  
 /botsv top delta sub def  
 {
{
W
xt yt translate 
ang rotate
xt neg yt neg translate 
dup setflat 
/bottom botsv def
0 1 numsteps1 
{
numsteps1 dup 0 eq {pop 0.5 } { div } ifelse 
taperfcn /frac xdf
bc4 0 c2 c1 sub frac mul c1 add put
bc4 1 m2 m1 sub frac mul m1 add put
bc4 2 y2 y1 sub frac mul y1 add put
bc4 3 k2 k1 sub frac mul k1 add put
bc4 vc
1 index setflat 
{ 
mark {newpath left bottom right delta rectfill}stopped
{cleartomark exch 1.3 mul dup setflat exch 2 copy gt{stop}if}
{cleartomark exit}ifelse
}loop
/bottom bottom delta sub def
}for
}
gsave stopped grestore
{exch pop 2 index exch 1.3 mul dup 100 gt{cleartomark setflat stop}if}
{exit}ifelse
 }loop
 cleartomark setflat
 iosv setinkoverprint
 cosv setoverprint
}bdf
/bcs [0 0] def 
/_lfs4 {
 /iosv inkoverprint def
 /cosv currentoverprint def
 /yt xdf       
 /xt xdf       
 /ang xdf      
 storerect
 /taperfcn xdf
 /tint2 xdf      
 /tint1 xdf      
 bcs exch 1 exch put    
 tint1 tint2 sub abs    
 bcs 1 get maxspot    
 calcgraysteps mul abs round  
 height abs adjnumsteps   
 dup 2 lt {pop 2} if    
 1 sub /numsteps1 xdf
 currentflat mark    
 currentflat clipflatness  
 /delta top bottom sub numsteps1 1 add div def 
 /right right left sub def  
 /botsv top delta sub def  
 {
{
W
xt yt translate 
ang rotate
xt neg yt neg translate 
dup setflat 
/bottom botsv def
0 1 numsteps1 
{
numsteps1 div taperfcn /frac xdf
bcs 0
1.0 tint2 tint1 sub frac mul tint1 add sub
put bcs vc
1 index setflat 
{ 
mark {newpath left bottom right delta rectfill}stopped
{cleartomark exch 1.3 mul dup setflat exch 2 copy gt{stop}if}
{cleartomark exit}ifelse
}loop
/bottom bottom delta sub def
}for
}
gsave stopped grestore
{exch pop 2 index exch 1.3 mul dup 100 gt{cleartomark setflat stop}if}
{exit}ifelse
 }loop
 cleartomark setflat
 iosv setinkoverprint
 cosv setoverprint
}bdf
/_rfs4 {
 /iosv inkoverprint def
 /cosv currentoverprint def
 /tint2 xdf      
 /tint1 xdf      
 bcs exch 1 exch put    
 /radius xdf      
 /yt xdf       
 /xt xdf       
 tint1 tint2 sub abs    
 bcs 1 get maxspot    
 calcgraysteps mul abs round  
 radius abs adjnumsteps   
 dup 2 lt {pop 2} if    
 1 sub /numsteps1 xdf
 radius numsteps1 div 2 div /halfstep xdf 
 currentflat mark    
 currentflat clipflatness  
 {
{
dup setflat 
W 
0 1 numsteps1 
{
dup /radindex xdf
numsteps1 div /frac xdf
bcs 0
tint2 tint1 sub frac mul tint1 add
put bcs vc
1 index setflat 
{ 
newpath mark xt yt radius 1 frac sub mul halfstep add 0 360
{ arc
radindex numsteps1 ne 
{
xt yt 
radindex 1 add numsteps1 
div 1 exch sub 
radius mul halfstep add
dup xt add yt moveto 
360 0 arcn 
} if
fill
}stopped
{cleartomark exch 1.3 mul dup setflat exch 2 copy gt{stop}if}
{cleartomark exit}ifelse
}loop
}for
}
gsave stopped grestore
{exch pop 2 index exch 1.3 mul dup 100 gt{cleartomark setflat stop}if}
{exit}ifelse
 }loop
 cleartomark setflat
 iosv setinkoverprint
 cosv setoverprint
}bdf
/_rfp4 {
 /iosv inkoverprint def
 /cosv currentoverprint def
 /k2 xdf /y2 xdf /m2 xdf /c2 xdf
 /k1 xdf /y1 xdf /m1 xdf /c1 xdf
 /radius xdf      
 /yt xdf       
 /xt xdf       
 c1 c2 sub abs
 m1 m2 sub abs
 y1 y2 sub abs
 k1 k2 sub abs
 maxcolor      
 calcgraysteps mul abs round  
 radius abs adjnumsteps   
 dup 2 lt {pop 1} if    
 1 sub /numsteps1 xdf
 radius numsteps1 dup 0 eq {pop} {div} ifelse 
 2 div /halfstep xdf 
 currentflat mark    
 currentflat clipflatness  
 {
{
dup setflat 
W 
0 1 numsteps1 
{
dup /radindex xdf
numsteps1 dup 0 eq {pop 0.5 } { div } ifelse 
/frac xdf
bc4 0 c2 c1 sub frac mul c1 add put
bc4 1 m2 m1 sub frac mul m1 add put
bc4 2 y2 y1 sub frac mul y1 add put
bc4 3 k2 k1 sub frac mul k1 add put
bc4 vc
1 index setflat 
{ 
newpath mark xt yt radius 1 frac sub mul halfstep add 0 360
{ arc
radindex numsteps1 ne 
{
xt yt 
radindex 1 add 
numsteps1 dup 0 eq {pop} {div} ifelse 
1 exch sub 
radius mul halfstep add
dup xt add yt moveto 
360 0 arcn 
} if
fill
}stopped
{cleartomark exch 1.3 mul dup setflat exch 2 copy gt{stop}if}
{cleartomark exit}ifelse
}loop
}for
}
gsave stopped grestore
{exch pop 2 index exch 1.3 mul dup 100 gt{cleartomark setflat stop}if}
{exit}ifelse
 }loop
 cleartomark setflat
 iosv setinkoverprint
 cosv setoverprint
}bdf
/lfp4{_lfp4}ndf
/lfs4{_lfs4}ndf
/rfs4{_rfs4}ndf
/rfp4{_rfp4}ndf
/cvc [0 0 0 1] def 
/vc{
 AltsysDict /cvc 2 index put 
 aload length 4 eq
 {setcmykcolor}
 {setspotcolor}
 ifelse
}bdf 
/origmtx matrix currentmatrix def
/ImMatrix matrix currentmatrix def
0 setseparationgray
/imgr {1692 1584 2304 2376 } def 
/bleed 0 def 
/clpr {1692 1584 2304 2376 } def 
/xs 1 def 
/ys 1 def 
/botx 0 def 
/overlap 0 def 
/wdist 18 def 
0 2 mul fhsetspreadsize 
0 0 ne {/df 0 def /clipflatness 0 def} if 
/maxsteps 256 def 
/forcemaxsteps false def 
vms
-1705 -1820 translate
/currentpacking defed{false setpacking}if 
/spots[
1 0 0 0 (Process Cyan) false newcmykcustomcolor
0 1 0 0 (Process Magenta) false newcmykcustomcolor
0 0 1 0 (Process Yellow) false newcmykcustomcolor
0 0 0 1 (Process Black) false newcmykcustomcolor
]def
/textopf false def
/curtextmtx{}def
/otw .25 def
/msf{dup/curtextmtx xdf makefont setfont}bdf
/makesetfont/msf load def
/curtextheight{.707104 .707104 curtextmtx dtransform
 dup mul exch dup mul add sqrt}bdf
/ta2{ 
tempstr 2 index gsave exec grestore 
cwidth cheight rmoveto 
4 index eq{5 index 5 index rmoveto}if 
2 index 2 index rmoveto 
}bdf
/ta{exch systemdict/cshow known
{{/cheight xdf/cwidth xdf tempstr 0 2 index put ta2}exch cshow} 
{{tempstr 0 2 index put tempstr stringwidth/cheight xdf/cwidth xdf ta2}forall} 
ifelse 6{pop}repeat}bdf
/sts{/textopf currentoverprint def vc setoverprint
/ts{awidthshow}def exec textopf setoverprint}bdf
/stol{/xt currentlinewidth def 
 setlinewidth vc newpath 
 /ts{{false charpath stroke}ta}def exec 
 xt setlinewidth}bdf 
 
/strk{/textopf currentoverprint def vc setoverprint
 /ts{{false charpath stroke}ta}def exec 
 textopf setoverprint
 }bdf 
n
[] 0 d
3.863708 M
1 w
0 j
0 J
false setoverprint
0 i
false eomode
[0 0 0 1] vc
vms
2021.2443 1943.7763 m
2021.2443 2133.7763 L
0.5 w
S 
n
2021.2443 1943.7763 m
2021.2443 1861.2763 L
S 
n
2241.2443 2133.7763 m
2241.2443 1936.2763 L
S 
n
2241.2443 1936.2763 m
2241.2443 1861.2763 L
S 
n
2021.2443 1861.2763 m
2241.2443 1861.2763 L
S 
n
2021.2443 2133.7763 m
2241.2443 2133.7763 L
S 
n
2028.2843 1856.8761 m
2022.2443 1856.8761 L
2022.2443 1843.8763 L
2028.2843 1843.8763 L
2028.2843 1856.8761 L
n
q
_bfh
_efh
{
f1 [12 0 0 12 0 0] makesetfont
2022.244308 1847.276276 m
0 0 32 0 0 (1) ts
}
true
[0 0 0 1]sts
Q
2227.7443 1863.7763 m
2227.7443 1856.7763 L
false setoverprint
S 
n
2018.2565 1951.4824 m
2007.5982 1951.4824 L
2007.5982 1939.2325 L
2018.2565 1939.2325 L
2018.2565 1951.4824 L
n
q
_bfh
_efh
{
f1 [11.25 0 0 11.25 0 0] makesetfont
2007.598221 1942.482498 m
0 0 32 0 0 (x) ts
}
true
[0 0 0 1]sts
{
f1 [11.25 0 0 11.25 0 0] makesetfont
0 -4.6875 rmoveto 0 0 32 0 0 (c) ts
0 4.6875 rmoveto }
true
[0 0 0 1]sts
Q
2024.7443 2120.7763 m
2017.2443 2120.7763 L
false setoverprint
S 
n
2018.2443 1871.2763 m
2024.7443 1871.2763 L
S 
n
2227.8299 1897.7592 m
2197.3612 1897.7592 L
2197.3612 1887.7593 L
2227.8299 1887.7593 L
2227.8299 1897.7592 L
n
q
_bfh
_efh
{
f1 [6 0 0 6 0 0] makesetfont
2197.361221 1890.559326 m
0 0 32 0 0 (   ) ts
}
true
[0 0 0 1]sts
{
f1 [9 0 0 9 0 0] makesetfont
0 0 32 0 0 (T) ts
}
true
[0 0 0 1]sts
{
f1 [7 0 0 7 0 0] makesetfont
0 5.5 rmoveto 0 0 32 0 0 (\(2\)) ts
0 -5.5 rmoveto }
true
[0 0 0 1]sts
{
f1 [7 0 0 7 0 0] makesetfont
0 0 32 0 0 (\() ts
}
true
[0 0 0 1]sts
{
f1 [9 0 0 9 0 0] makesetfont
0 0 32 0 0 (x) ts
}
true
[0 0 0 1]sts
{
f1 [7 0 0 7 0 0] makesetfont
0 -1.5 rmoveto 0 0 32 0 0 (c) ts
0 1.5 rmoveto }
true
[0 0 0 1]sts
{
f1 [7 0 0 7 0 0] makesetfont
0 0 32 0 0 (\)) ts
}
true
[0 0 0 1]sts
Q
2200.5795 2032.3761 m
2174.2443 2032.3761 L
2174.2443 2021.7762 L
2200.5795 2021.7762 L
2200.5795 2032.3761 L
n
q
_bfh
_efh
{
f1 [9.600037 0 0 9.600037 0 0] makesetfont
2174.244308 2024.696228 m
0 -1 rmoveto 0 0 32 0 0 (T) ts
0 1 rmoveto }
true
[0 0 0 1]sts
{
f1 [7 0 0 7 0 0] makesetfont
0 6 rmoveto 0 0 32 0 0 (\(3\)) ts
0 -6 rmoveto }
true
[0 0 0 1]sts
{
f1 [7 0 0 7 0 0] makesetfont
0 0 32 0 0 (\() ts
}
true
[0 0 0 1]sts
{
f1 [9 0 0 9 0 0] makesetfont
0 0 32 0 0 (x) ts
}
true
[0 0 0 1]sts
{
f1 [7 0 0 7 0 0] makesetfont
0 -1.5 rmoveto 0 0 32 0 0 (c) ts
0 1.5 rmoveto }
true
[0 0 0 1]sts
{
f1 [7 0 0 7 0 0] makesetfont
0 0 32 0 0 (\)) ts
}
true
[0 0 0 1]sts
Q
2182.5795 2046.8761 m
2156.2443 2046.8761 L
2156.2443 2036.2762 L
2182.5795 2036.2762 L
2182.5795 2046.8761 L
n
q
_bfh
_efh
{
f1 [9.600037 0 0 9.600037 0 0] makesetfont
2156.244308 2039.196228 m
0 -1 rmoveto 0 0 32 0 0 (T) ts
0 1 rmoveto }
true
[0 0 0 1]sts
{
f1 [7 0 0 7 0 0] makesetfont
0 6 rmoveto 0 0 32 0 0 (\(5\)) ts
0 -6 rmoveto }
true
[0 0 0 1]sts
{
f1 [7 0 0 7 0 0] makesetfont
0 0 32 0 0 (\() ts
}
true
[0 0 0 1]sts
{
f1 [9 0 0 9 0 0] makesetfont
0 0 32 0 0 (x) ts
}
true
[0 0 0 1]sts
{
f1 [7 0 0 7 0 0] makesetfont
0 -1.5 rmoveto 0 0 32 0 0 (c) ts
0 1.5 rmoveto }
true
[0 0 0 1]sts
{
f1 [7 0 0 7 0 0] makesetfont
0 0 32 0 0 (\)) ts
}
true
[0 0 0 1]sts
Q
2162.3095 1952.3761 m
2134.2443 1952.3761 L
2134.2443 1941.7762 L
2162.3095 1941.7762 L
2162.3095 1952.3761 L
n
q
_bfh
_efh
{
f1 [9.600037 0 0 9.600037 0 0] makesetfont
2134.244308 1944.696228 m
0 -1 rmoveto 0 0 32 0 0 (T) ts
0 1 rmoveto }
true
[0 0 0 1]sts
{
f1 [7 0 0 7 0 0] makesetfont
0 6 rmoveto 0 0 32 0 0 (\(6\)) ts
0 -6 rmoveto }
true
[0 0 0 1]sts
{
f1 [7 0 0 7 0 0] makesetfont
0 0 32 0 0 (\() ts
}
true
[0 0 0 1]sts
{
f1 [9 0 0 9 0 0] makesetfont
0 0 32 0 0 (x) ts
}
true
[0 0 0 1]sts
{
f1 [7 0 0 7 0 0] makesetfont
0 -1.5 rmoveto 0 0 32 0 0 (c) ts
0 1.5 rmoveto }
true
[0 0 0 1]sts
{
f1 [7 0 0 7 0 0] makesetfont
0 0 32 0 0 (\)) ts
}
true
[0 0 0 1]sts
Q
2210.0795 2071.3761 m
2183.7443 2071.3761 L
2183.7443 2060.7762 L
2210.0795 2060.7762 L
2210.0795 2071.3761 L
n
q
_bfh
_efh
{
f1 [9.600037 0 0 9.600037 0 0] makesetfont
2183.744308 2063.696228 m
0 0 32 0 0 (T) ts
}
true
[0 0 0 1]sts
{
f1 [7 0 0 7 0 0] makesetfont
0 6 rmoveto 0 0 32 0 0 (\(7\)) ts
0 -6 rmoveto }
true
[0 0 0 1]sts
{
f1 [7 0 0 7 0 0] makesetfont
0 0 32 0 0 (\() ts
}
true
[0 0 0 1]sts
{
f1 [9 0 0 9 0 0] makesetfont
0 0 32 0 0 (x) ts
}
true
[0 0 0 1]sts
{
f1 [7 0 0 7 0 0] makesetfont
0 -1.5 rmoveto 0 0 32 0 0 (c) ts
0 1.5 rmoveto }
true
[0 0 0 1]sts
{
f1 [7 0 0 7 0 0] makesetfont
0 0 32 0 0 (\)) ts
}
true
[0 0 0 1]sts
Q
2185.0795 1931.3761 m
2158.7443 1931.3761 L
2158.7443 1920.7762 L
2185.0795 1920.7762 L
2185.0795 1931.3761 L
n
q
_bfh
_efh
{
f1 [9.600037 0 0 9.600037 0 0] makesetfont
2158.744308 1923.696228 m
0 -1 rmoveto 0 0 32 0 0 (T) ts
0 1 rmoveto }
true
[0 0 0 1]sts
{
f1 [7 0 0 7 0 0] makesetfont
0 6 rmoveto 0 0 32 0 0 (\(8\)) ts
0 -6 rmoveto }
true
[0 0 0 1]sts
{
f1 [7 0 0 7 0 0] makesetfont
0 0 32 0 0 (\() ts
}
true
[0 0 0 1]sts
{
f1 [9 0 0 9 0 0] makesetfont
0 0 32 0 0 (x) ts
}
true
[0 0 0 1]sts
{
f1 [7 0 0 7 0 0] makesetfont
0 -1.5 rmoveto 0 0 32 0 0 (c) ts
0 1.5 rmoveto }
true
[0 0 0 1]sts
{
f1 [7 0 0 7 0 0] makesetfont
0 0 32 0 0 (\)) ts
}
true
[0 0 0 1]sts
Q
1998.2843 2006.8761 m
1992.2443 2006.8761 L
1992.2443 1993.8763 L
1998.2843 1993.8763 L
1998.2843 2006.8761 L
n
q
_bfh
_efh
{
f1 [12 0 0 12 0 0] makesetfont
1992.244308 1997.276276 m
0 -5 rmoveto 0 0 32 0 0 (x) ts
0 5 rmoveto }
true
[0 0 0 1]sts
Q
2132.4006 1825.8075 m
2132.1818 1825.37 L
2131.9006 1824.9013 L
2131.5568 1824.3388 L
2131.2756 1823.8388 L
2130.9943 1823.4638 L
2130.6818 1823.0263 L
2130.3381 1822.5888 L
2130.0568 1822.2138 L
2129.6818 1821.745 L
2129.3068 1821.4325 L
2128.7443 1821.0888 L
2128.1193 1820.9013 L
2127.5568 1821.0575 L
2127.1193 1821.4325 L
2126.8693 1822.1513 L
2126.7756 1822.9013 L
2126.8693 1823.5888 L
2127.0568 1824.2763 L
2127.3068 1824.8388 L
2127.8381 1825.5888 L
2128.4943 1825.9325 L
2129.0881 1826.1513 L
2129.7131 1826.12 L
2130.1506 1825.745 L
2130.4006 1825.245 L
2130.4006 1824.8075 L
2130.4006 1824.4638 L
2130.4318 1823.9638 L
2130.4006 1823.495 L
2130.3693 1823.12 L
2130.3381 1822.3075 L
2130.3693 1821.7138 L
2130.4318 1821.3388 L
2130.5881 1821.0888 L
2131.1193 1821.0263 L
2131.4943 1821.3388 L
2131.7131 1821.5888 L
2131.8693 1821.8388 L
false setoverprint
S 
n
2016.5051 2127.8761 m
1985.3698 2127.8761 L
1985.3698 2117.2762 L
2016.5051 2117.2762 L
2016.5051 2127.8761 L
n
q
_bfh
_efh
{
f1 [9.600037 0 0 9.600037 0 0] makesetfont
1985.369843 2120.196228 m
0 -1 rmoveto 0 0 32 0 0 (T) ts
0 1 rmoveto }
true
[0 0 0 1]sts
{
f1 [9.600037 0 0 9.600037 0 0] makesetfont
0 -5 rmoveto 0 0 32 0 0 (1) ts
0 5 rmoveto }
true
[0 0 0 1]sts
{
f1 [7 0 0 7 0 0] makesetfont
0 6 rmoveto 0 0 32 0 0 (\(1\)) ts
0 -6 rmoveto }
true
[0 0 0 1]sts
{
f1 [7 0 0 7 0 0] makesetfont
0 0 32 0 0 (\() ts
}
true
[0 0 0 1]sts
{
f1 [9 0 0 9 0 0] makesetfont
0 0 32 0 0 (x) ts
}
true
[0 0 0 1]sts
{
f1 [7 0 0 7 0 0] makesetfont
0 -1.5 rmoveto 0 0 32 0 0 (c) ts
0 1.5 rmoveto }
true
[0 0 0 1]sts
{
f1 [7 0 0 7 0 0] makesetfont
0 0 32 0 0 (\)) ts
}
true
[0 0 0 1]sts
Q
vmrs
2017.3795 1877.3761 m
1986.2443 1877.3761 L
1986.2443 1866.7762 L
2017.3795 1866.7762 L
2017.3795 1877.3761 L
n
q
_bfh
_efh
{
f1 [9.600037 0 0 9.600037 0 0] makesetfont
1986.244308 1869.696228 m
0 -1 rmoveto 0 0 32 0 0 (T) ts
0 1 rmoveto }
true
[0 0 0 1]sts
{
f1 [9.600037 0 0 9.600037 0 0] makesetfont
0 -5 rmoveto 0 0 32 0 0 (1) ts
0 5 rmoveto }
true
[0 0 0 1]sts
{
f1 [7 0 0 7 0 0] makesetfont
0 6 rmoveto 0 0 32 0 0 (\(2\)) ts
0 -6 rmoveto }
true
[0 0 0 1]sts
{
f1 [7 0 0 7 0 0] makesetfont
0 0 32 0 0 (\() ts
}
true
[0 0 0 1]sts
{
f1 [9 0 0 9 0 0] makesetfont
0 0 32 0 0 (x) ts
}
true
[0 0 0 1]sts
{
f1 [7 0 0 7 0 0] makesetfont
0 -1.5 rmoveto 0 0 32 0 0 (c) ts
0 1.5 rmoveto }
true
[0 0 0 1]sts
{
f1 [7 0 0 7 0 0] makesetfont
0 0 32 0 0 (\)) ts
}
true
[0 0 0 1]sts
Q
2309.7443 2061.2763 m
2201.7443 2061.2763 L
2201.7443 2055.7763 L
2309.7443 2055.7763 L
2309.7443 2061.2763 L
n
2066.2765 2109.3761 m
2052.2443 2109.3761 L
2052.2443 2096.3763 L
2066.2765 2096.3763 L
2066.2765 2109.3761 L
n
q
_bfh
_efh
{
f1 [12 0 0 12 0 0] makesetfont
2052.244308 2099.776276 m
0 -1.5 rmoveto 0 0 32 0 0 (\(b\)) ts
0 1.5 rmoveto }
true
[0 0 0 1]sts
Q
2016.973 1854.4631 m
2246.377 1854.4631 L
2246.377 2138.0201 L
2016.973 2138.0201 L
2016.973 1854.4631 L
h
q
W
n
q
[0.375 0 0 0.75 1935.924973 -9613.758392] concat
vms
230.6937 15409.989 m
254.3253 15420.132 L
280.1042 15431.192 L
304.3109 15441.586 L
327.594 15451.575 L
352.376 15462.214 L
376.228 15472.448 L
401.584 15483.331 L
426.011 15493.816 L
449.508 15503.898 L
474.51 15514.628 L
498.582 15524.96 L
521.725 15534.895 L
546.372 15545.472 L
570.089 15555.651 L
595.311 15566.473 L
619.598 15576.903 L
642.966 15586.93 L
667.828 15597.599 L
691.765 15607.876 L
717.201 15618.789 L
741.714 15629.311 L
765.297 15639.436 L
790.378 15650.196 L
813.306 15660.039 L
0.5 w
10 M
false setoverprint
[] 0 d
S 
[] 0 d
n
230.6937 15409.989 m
231.4155 15409.989 L
232.064 15409.989 L
232.8225 15409.983 L
233.5383 15409.983 L
234.8046 15409.983 L
236.1871 15409.977 L
237.7043 15409.971 L
239.3131 15409.965 L
242.2006 15409.946 L
245.1798 15409.922 L
247.9693 15409.891 L
254.3253 15409.805 L
260.8159 15409.695 L
266.7437 15409.567 L
280.1042 15409.194 L
291.7212 15408.778 L
304.3966 15408.227 L
329.258 15406.839 L
353.195 15405.12 L
378.638 15402.893 L
403.144 15400.348 L
426.727 15397.528 L
451.815 15394.139 L
475.966 15390.487 L
499.2 15386.62 L
523.927 15382.118 L
547.73 15377.408 L
573.032 15372 L
597.404 15366.396 L
620.852 15360.646 L
645.805 15354.131 L
669.822 15347.481 L
692.915 15340.733 L
717.513 15333.172 L
741.176 15325.519 L
766.343 15316.986 L
790.586 15308.378 L
813.306 15299.967 L
[] 0 d
S 
[] 0 d
n
230.6937 15409.989 m
254.3253 15419.948 L
280.1042 15430.372 L
304.3109 15439.726 L
327.594 15448.309 L
352.376 15456.971 L
376.228 15464.826 L
401.584 15472.644 L
426.011 15479.624 L
449.508 15485.815 L
474.51 15491.816 L
498.582 15496.997 L
521.725 15501.408 L
546.372 15505.47 L
570.089 15508.743 L
595.311 15511.52 L
607.907 15512.621 L
619.598 15513.478 L
631.631 15514.181 L
642.966 15514.683 L
648.998 15514.885 L
652.167 15514.97 L
655.519 15515.05 L
658.345 15515.105 L
661.447 15515.154 L
664.267 15515.191 L
665.546 15515.203 L
666.904 15515.209 L
667.656 15515.215 L
668.494 15515.221 L
669.259 15515.221 L
669.962 15515.227 L
670.366 15515.227 L
670.788 15515.227 L
671.217 15515.227 L
671.681 15515.227 L
672.085 15515.227 L
672.52 15515.227 L
673.297 15515.227 L
674.08 15515.227 L
674.52 15515.221 L
674.93 15515.221 L
676.471 15515.215 L
677.866 15515.203 L
679.353 15515.191 L
682.142 15515.16 L
684.761 15515.123 L
690.707 15515.001 L
697.069 15514.817 L
703.113 15514.597 L
[] 0 d
S 
[] 0 d
n
703.113 15514.597 m
714.455 15514.047 L
726.592 15513.257 L
737.848 15512.352 L
761.822 15509.826 L
787.289 15506.247 L
811.832 15501.897 L
813.306 15501.604 L
[] 0 d
S 
[] 0 d
n
230.6937 15409.989 m
231.4155 15409.989 L
232.064 15409.989 L
232.8225 15409.989 L
233.5383 15409.989 L
234.199 15409.989 L
234.8046 15409.989 L
235.4775 15409.989 L
236.1871 15409.989 L
236.8968 15409.989 L
237.6614 15409.989 L
239.044 15409.989 L
239.8882 15409.989 L
240.6957 15409.989 L
242.2006 15409.989 L
243.6565 15409.989 L
245.2042 15409.989 L
247.9999 15409.989 L
249.7067 15409.989 L
251.3156 15409.989 L
254.3253 15409.995 L
257.4636 15409.995 L
260.7792 15410.001 L
266.7132 15410.007 L
273.0936 15410.026 L
280.1042 15410.044 L
286.344 15410.075 L
292.0394 15410.105 L
304.8309 15410.191 L
316.509 15410.307 L
329.209 15410.472 L
341.303 15410.674 L
354.156 15410.949 L
377.249 15411.61 L
401.847 15412.601 L
425.51 15413.874 L
450.677 15415.629 L
474.92 15417.776 L
498.234 15420.309 L
523.046 15423.57 L
546.929 15427.32 L
572.316 15432.024 L
596.773 15437.328 L
620.307 15443.201 L
645.34 15450.352 L
669.443 15458.188 L
692.621 15466.661 L
717.299 15476.761 L
741.047 15487.607 L
766.3 15500.423 L
[] 0 d
S 
[] 0 d
n
766.3 15500.423 m
790.623 15514.102 L
813.306 15528.104 L
[] 0 d
S 
[] 0 d
n
230.6937 15409.989 m
254.3253 15420.126 L
280.1042 15431.131 L
304.3109 15441.371 L
327.594 15451.08 L
352.376 15461.204 L
376.228 15470.68 L
401.584 15480.395 L
426.011 15489.32 L
449.508 15497.432 L
474.51 15505.452 L
498.582 15512.499 L
521.725 15518.555 L
546.372 15524.128 L
570.089 15528.527 L
582.147 15530.362 L
595.311 15532.044 L
607.907 15533.31 L
619.598 15534.173 L
626.064 15534.516 L
629.239 15534.65 L
632.157 15534.748 L
634.812 15534.821 L
637.62 15534.883 L
639.082 15534.901 L
640.44 15534.919 L
641.223 15534.925 L
641.939 15534.932 L
642.33 15534.938 L
642.758 15534.938 L
643.535 15534.938 L
644.281 15534.944 L
644.967 15534.944 L
645.713 15534.938 L
646.502 15534.938 L
647.193 15534.932 L
647.836 15534.925 L
649.286 15534.913 L
650.894 15534.889 L
652.357 15534.864 L
655.635 15534.779 L
658.804 15534.675 L
662.242 15534.528 L
668.488 15534.185 L
674.275 15533.763 L
680.644 15533.194 L
693.515 15531.677 L
705.413 15529.823 L
718.192 15527.328 L
743.445 15520.77 L
[] 0 d
S 
[] 0 d
n
743.445 15520.77 m
767.768 15512.266 L
791.167 15501.891 L
813.306 15489.944 L
[] 0 d
S 
[] 0 d
n
230.6937 15409.989 m
231.4155 15409.989 L
232.064 15409.989 L
232.8225 15409.983 L
233.5383 15409.983 L
234.8046 15409.983 L
236.1871 15409.977 L
237.7043 15409.971 L
239.3131 15409.965 L
242.2006 15409.946 L
247.9693 15409.891 L
254.3253 15409.812 L
260.7425 15409.708 L
267.6369 15409.573 L
280.1042 15409.261 L
305.3203 15408.417 L
329.613 15407.383 L
352.981 15406.257 L
377.843 15405.003 L
401.78 15403.829 L
427.216 15402.734 L
439.983 15402.281 L
451.729 15401.951 L
457.657 15401.816 L
463.199 15401.718 L
468.974 15401.633 L
472.033 15401.602 L
473.623 15401.59 L
475.312 15401.578 L
476.847 15401.571 L
477.667 15401.565 L
478.542 15401.565 L
479.349 15401.559 L
480.089 15401.559 L
480.793 15401.559 L
481.533 15401.559 L
481.961 15401.559 L
482.347 15401.559 L
482.787 15401.559 L
483.209 15401.559 L
483.956 15401.559 L
484.769 15401.565 L
485.662 15401.565 L
486.611 15401.571 L
488.305 15401.584 L
489.767 15401.59 L
491.333 15401.608 L
494.166 15401.639 L
497.536 15401.688 L
500.613 15401.749 L
[] 0 d
S 
[] 0 d
n
500.613 15401.749 m
506.651 15401.889 L
512.249 15402.067 L
518.788 15402.33 L
524.936 15402.624 L
537.514 15403.413 L
548.905 15404.349 L
572.524 15407.065 L
584.74 15408.925 L
597.636 15411.261 L
620.901 15416.577 L
645.658 15423.973 L
669.491 15433.046 L
694.83 15445.091 L
719.232 15459.344 L
742.717 15475.813 L
767.695 15496.667 L
791.748 15520.348 L
813.306 15544.897 L
[] 0 d
S 
[] 0 d
n
230.6937 15409.989 m
254.3253 15419.954 L
280.1042 15430.44 L
304.3109 15439.964 L
327.594 15448.878 L
352.376 15458.17 L
376.228 15467.004 L
401.584 15476.388 L
426.011 15485.545 L
449.508 15494.599 L
474.51 15504.681 L
498.582 15515.007 L
521.725 15525.719 L
546.372 15538.211 L
570.089 15551.565 L
582.147 15558.949 L
588.515 15563.029 L
589.396 15563.604 L
590.35 15564.228 L
590.791 15564.515 L
591.249 15564.65 L
591.647 15564.73 L
592.081 15564.809 L
595.311 15565.427 L
607.907 15567.464 L
613.908 15568.223 L
619.598 15568.816 L
622.589 15569.067 L
625.844 15569.305 L
628.547 15569.465 L
631.533 15569.605 L
633.197 15569.666 L
634.757 15569.715 L
636.402 15569.752 L
637.332 15569.764 L
637.742 15569.77 L
638.182 15569.776 L
638.923 15569.783 L
639.614 15569.783 L
640.036 15569.783 L
640.434 15569.783 L
641.174 15569.783 L
641.963 15569.776 L
642.691 15569.77 L
643.486 15569.758 L
644.33 15569.746 L
645.878 15569.709 L
647.346 15569.666 L
650.099 15569.55 L
653.072 15569.385 L
[] 0 d
S 
[] 0 d
n
653.072 15569.385 m
656.302 15569.153 L
662.12 15568.59 L
667.509 15567.905 L
680.087 15565.659 L
691.478 15562.784 L
703.272 15558.912 L
716.021 15553.626 L
741.133 15539.526 L
765.321 15520.77 L
788.586 15497.303 L
813.306 15465.768 L
[] 0 d
S 
[] 0 d
n
230.6937 15409.989 m
231.4155 15409.989 L
232.064 15409.989 L
232.8225 15409.989 L
233.5383 15409.989 L
234.199 15409.989 L
234.8046 15409.989 L
235.4775 15409.989 L
236.1871 15409.989 L
236.8968 15409.989 L
237.6614 15409.989 L
238.3833 15409.989 L
239.044 15409.989 L
239.8209 15409.989 L
240.6651 15409.989 L
241.4604 15409.989 L
242.2006 15409.989 L
242.9591 15409.989 L
243.785 15409.989 L
245.2776 15409.989 L
246.7336 15409.989 L
248.0672 15409.989 L
249.5232 15409.989 L
251.1137 15409.989 L
252.7776 15409.989 L
254.3253 15409.989 L
255.9281 15409.989 L
257.6716 15409.983 L
260.8343 15409.983 L
263.9175 15409.983 L
266.7682 15409.983 L
269.8392 15409.983 L
273.2343 15409.983 L
276.7763 15409.977 L
280.1042 15409.977 L
286.1177 15409.971 L
292.4492 15409.965 L
297.857 15409.952 L
303.7726 15409.94 L
310.0552 15409.922 L
315.7689 15409.897 L
322.437 15409.867 L
328.585 15409.83 L
340.318 15409.732 L
346.497 15409.671 L
353.042 15409.591 L
365.014 15409.396 L
376.221 15409.163 L
388.628 15408.821 L
399.969 15408.423 L
[] 0 d
S 
[] 0 d
n
399.969 15408.423 m
412.59 15407.866 L
424.292 15407.218 L
447.691 15405.468 L
460.269 15404.22 L
471.66 15402.874 L
496.203 15399.124 L
519.816 15394.194 L
544.934 15387.146 L
569.123 15378.197 L
581.296 15372.74 L
586.606 15370.134 L
589.408 15368.702 L
590.093 15368.348 L
590.821 15367.968 L
591.629 15367.981 L
592.381 15368.029 L
595.574 15368.256 L
598.529 15368.507 L
604.341 15369.106 L
610.458 15369.901 L
617.145 15370.972 L
629.428 15373.554 L
642.342 15377.2 L
654.363 15381.537 L
667.185 15387.287 L
692.603 15402.581 L
717.097 15423.031 L
740.656 15448.994 L
765.725 15484.567 L
789.858 15527.854 L
813.068 15579.136 L
813.306 15579.717 L
[] 0 d
S 
[] 0 d
n
vmr
Q
Q
n
2170.5795 1977.8762 m
2144.2443 1977.8762 L
2144.2443 1967.2763 L
2170.5795 1967.2763 L
2170.5795 1977.8762 L
n
q
_bfh
_efh
{
f1 [9.600037 0 0 9.600037 0 0] makesetfont
2144.244308 1970.196289 m
0 -1 rmoveto 0 0 32 0 0 (T) ts
0 1 rmoveto }
true
[0 0 0 1]sts
{
f1 [7 0 0 7 0 0] makesetfont
0 6 rmoveto 0 0 32 0 0 (\(4\)) ts
0 -6 rmoveto }
true
[0 0 0 1]sts
{
f1 [7 0 0 7 0 0] makesetfont
0 0 32 0 0 (\() ts
}
true
[0 0 0 1]sts
{
f1 [9 0 0 9 0 0] makesetfont
0 0 32 0 0 (x) ts
}
true
[0 0 0 1]sts
{
f1 [7 0 0 7 0 0] makesetfont
0 -1.5 rmoveto 0 0 32 0 0 (c) ts
0 1.5 rmoveto }
true
[0 0 0 1]sts
{
f1 [7 0 0 7 0 0] makesetfont
0 0 32 0 0 (\)) ts
}
true
[0 0 0 1]sts
Q
2115.2443 1864.7763 m
2115.2443 1856.7763 L
0.5 w
false setoverprint
S 
n
2237.0025 1854.3762 m
2221.7443 1854.3762 L
2221.7443 1841.3763 L
2237.0025 1841.3763 L
2237.0025 1854.3762 L
n
q
_bfh
_efh
{
f2 [12 0 0 12 0 0] makesetfont
2221.744308 1844.776337 m
0 0 32 0 0 (2) ts
}
true
[0 0 0 1]sts
{
f2 [7 0 0 7 0 0] makesetfont
0 5 rmoveto 0 0 32 0 0 (1\/2) ts
0 -5 rmoveto }
true
[0 0 0 1]sts
Q
2125.0025 1854.3762 m
2109.7443 1854.3762 L
2109.7443 1841.3763 L
2125.0025 1841.3763 L
2125.0025 1854.3762 L
n
q
_bfh
_efh
{
f2 [12 0 0 12 0 0] makesetfont
2109.744308 1844.776337 m
0 0 32 0 0 (2) ts
}
true
[0 0 0 1]sts
{
f2 [7 0 0 7 0 0] makesetfont
0 5 rmoveto 0 0 32 0 0 (1\/4) ts
0 -5 rmoveto }
true
[0 0 0 1]sts
Q
2206.0795 2108.8762 m
2179.7443 2108.8762 L
2179.7443 2098.2763 L
2206.0795 2098.2763 L
2206.0795 2108.8762 L
n
q
_bfh
_efh
{
f1 [9.600037 0 0 9.600037 0 0] makesetfont
2179.744308 2101.196289 m
0 -1 rmoveto 0 0 32 0 0 (T) ts
0 1 rmoveto }
true
[0 0 0 1]sts
{
f1 [7 0 0 7 0 0] makesetfont
0 6 rmoveto 0 0 32 0 0 (\(1\)) ts
0 -6 rmoveto }
true
[0 0 0 1]sts
{
f1 [7 0 0 7 0 0] makesetfont
0 0 32 0 0 (\() ts
}
true
[0 0 0 1]sts
{
f1 [9 0 0 9 0 0] makesetfont
0 0 32 0 0 (x) ts
}
true
[0 0 0 1]sts
{
f1 [7 0 0 7 0 0] makesetfont
0 -1.5 rmoveto 0 0 32 0 0 (c) ts
0 1.5 rmoveto }
true
[0 0 0 1]sts
{
f1 [7 0 0 7 0 0] makesetfont
0 0 32 0 0 (\)) ts
}
true
[0 0 0 1]sts
Q
1737.4694 1854.9757 m
1966.8734 1854.9757 L
1966.8734 2138.5327 L
1737.4694 2138.5327 L
1737.4694 1854.9757 L
h
q
W
n
q
[0.375 0 0 0.75 1656.421402 -9613.245819] concat
vms
230.6937 15409.989 m
254.3253 15420.132 L
280.1042 15431.192 L
304.3109 15441.586 L
327.594 15451.575 L
352.376 15462.214 L
376.228 15472.448 L
401.584 15483.331 L
426.011 15493.816 L
449.508 15503.898 L
474.51 15514.628 L
498.582 15524.96 L
521.725 15534.895 L
546.372 15545.472 L
570.089 15555.651 L
595.311 15566.473 L
619.598 15576.903 L
642.966 15586.93 L
667.828 15597.599 L
691.765 15607.876 L
717.201 15618.789 L
741.714 15629.311 L
765.297 15639.436 L
790.378 15650.196 L
813.306 15660.039 L
10 M
false setoverprint
[] 0 d
S 
[] 0 d
n
230.6937 15409.989 m
231.4155 15409.989 L
232.064 15409.989 L
232.8225 15409.983 L
233.5383 15409.983 L
234.8046 15409.983 L
236.1871 15409.977 L
237.7043 15409.971 L
239.3131 15409.965 L
242.2006 15409.946 L
245.1798 15409.922 L
247.9693 15409.891 L
254.3253 15409.805 L
260.8159 15409.695 L
266.7437 15409.567 L
280.1042 15409.194 L
291.7212 15408.778 L
304.3966 15408.227 L
329.258 15406.839 L
353.195 15405.12 L
378.638 15402.893 L
403.144 15400.348 L
426.727 15397.528 L
451.815 15394.139 L
475.966 15390.487 L
499.2 15386.62 L
523.927 15382.118 L
547.73 15377.408 L
573.032 15372 L
597.404 15366.396 L
620.852 15360.646 L
645.805 15354.131 L
669.822 15347.481 L
692.915 15340.733 L
717.513 15333.172 L
741.176 15325.519 L
766.343 15316.986 L
790.586 15308.378 L
813.306 15299.967 L
[] 0 d
S 
[] 0 d
n
230.6937 15409.989 m
254.3253 15419.948 L
280.1042 15430.372 L
304.3109 15439.726 L
327.594 15448.309 L
352.376 15456.971 L
376.228 15464.826 L
401.584 15472.644 L
426.011 15479.624 L
449.508 15485.815 L
474.51 15491.816 L
498.582 15496.997 L
521.725 15501.408 L
546.372 15505.47 L
570.089 15508.743 L
595.311 15511.52 L
607.907 15512.621 L
619.598 15513.478 L
631.631 15514.181 L
642.966 15514.683 L
648.998 15514.885 L
652.167 15514.97 L
655.519 15515.05 L
658.345 15515.105 L
661.447 15515.154 L
664.267 15515.191 L
665.546 15515.203 L
666.904 15515.209 L
667.656 15515.215 L
668.494 15515.221 L
669.259 15515.221 L
669.962 15515.227 L
670.366 15515.227 L
670.788 15515.227 L
671.217 15515.227 L
671.681 15515.227 L
672.085 15515.227 L
672.52 15515.227 L
673.297 15515.227 L
674.08 15515.227 L
674.52 15515.221 L
674.93 15515.221 L
676.471 15515.215 L
677.866 15515.203 L
679.353 15515.191 L
682.142 15515.16 L
684.761 15515.123 L
690.707 15515.001 L
697.069 15514.817 L
703.113 15514.597 L
[] 0 d
S 
[] 0 d
n
703.113 15514.597 m
714.455 15514.047 L
726.592 15513.257 L
737.848 15512.352 L
761.822 15509.826 L
787.289 15506.247 L
811.832 15501.897 L
813.306 15501.604 L
[] 0 d
S 
[] 0 d
n
230.6937 15409.989 m
231.4155 15409.989 L
232.064 15409.989 L
232.8225 15409.989 L
233.5383 15409.989 L
234.199 15409.989 L
234.8046 15409.989 L
235.4775 15409.989 L
236.1871 15409.989 L
236.8968 15409.989 L
237.6614 15409.989 L
239.044 15409.989 L
239.8882 15409.989 L
240.6957 15409.989 L
242.2006 15409.989 L
243.6565 15409.989 L
245.2042 15409.989 L
247.9999 15409.989 L
249.7067 15409.989 L
251.3156 15409.989 L
254.3253 15409.995 L
257.4636 15409.995 L
260.7792 15410.001 L
266.7132 15410.007 L
273.0936 15410.026 L
280.1042 15410.044 L
286.344 15410.075 L
292.0394 15410.105 L
304.8309 15410.191 L
316.509 15410.307 L
329.209 15410.472 L
341.303 15410.674 L
354.156 15410.949 L
377.249 15411.61 L
401.847 15412.601 L
425.51 15413.874 L
450.677 15415.629 L
474.92 15417.776 L
498.234 15420.309 L
523.046 15423.57 L
546.929 15427.32 L
572.316 15432.024 L
596.773 15437.328 L
620.307 15443.201 L
645.34 15450.352 L
669.443 15458.188 L
692.621 15466.661 L
717.299 15476.761 L
741.047 15487.607 L
766.3 15500.423 L
[] 0 d
S 
[] 0 d
n
766.3 15500.423 m
790.623 15514.102 L
813.306 15528.104 L
[] 0 d
S 
[] 0 d
n
230.6937 15409.989 m
254.3253 15420.126 L
280.1042 15431.131 L
304.3109 15441.371 L
327.594 15451.08 L
352.376 15461.204 L
376.228 15470.68 L
401.584 15480.395 L
426.011 15489.32 L
449.508 15497.432 L
474.51 15505.452 L
498.582 15512.499 L
521.725 15518.555 L
546.372 15524.128 L
570.089 15528.527 L
582.147 15530.362 L
595.311 15532.044 L
607.907 15533.31 L
619.598 15534.173 L
626.064 15534.516 L
629.239 15534.65 L
632.157 15534.748 L
634.812 15534.821 L
637.62 15534.883 L
639.082 15534.901 L
640.44 15534.919 L
641.223 15534.925 L
641.939 15534.932 L
642.33 15534.938 L
642.758 15534.938 L
643.535 15534.938 L
644.281 15534.944 L
644.967 15534.944 L
645.713 15534.938 L
646.502 15534.938 L
647.193 15534.932 L
647.836 15534.925 L
649.286 15534.913 L
650.894 15534.889 L
652.357 15534.864 L
655.635 15534.779 L
658.804 15534.675 L
662.242 15534.528 L
668.488 15534.185 L
674.275 15533.763 L
680.644 15533.194 L
693.515 15531.677 L
705.413 15529.823 L
718.192 15527.328 L
743.445 15520.77 L
[] 0 d
S 
[] 0 d
n
743.445 15520.77 m
767.768 15512.266 L
791.167 15501.891 L
813.306 15489.944 L
[] 0 d
S 
[] 0 d
n
230.6937 15409.989 m
231.4155 15409.989 L
232.064 15409.989 L
232.8225 15409.983 L
233.5383 15409.983 L
234.8046 15409.983 L
236.1871 15409.977 L
237.7043 15409.971 L
239.3131 15409.965 L
242.2006 15409.946 L
247.9693 15409.891 L
254.3253 15409.812 L
260.7425 15409.708 L
267.6369 15409.573 L
280.1042 15409.261 L
305.3203 15408.417 L
329.613 15407.383 L
352.981 15406.257 L
377.843 15405.003 L
401.78 15403.829 L
427.216 15402.734 L
439.983 15402.281 L
451.729 15401.951 L
457.657 15401.816 L
463.199 15401.718 L
468.974 15401.633 L
472.033 15401.602 L
473.623 15401.59 L
475.312 15401.578 L
476.847 15401.571 L
477.667 15401.565 L
478.542 15401.565 L
479.349 15401.559 L
480.089 15401.559 L
480.793 15401.559 L
481.533 15401.559 L
481.961 15401.559 L
482.347 15401.559 L
482.787 15401.559 L
483.209 15401.559 L
483.956 15401.559 L
484.769 15401.565 L
485.662 15401.565 L
486.611 15401.571 L
488.305 15401.584 L
489.767 15401.59 L
491.333 15401.608 L
494.166 15401.639 L
497.536 15401.688 L
500.613 15401.749 L
[] 0 d
S 
[] 0 d
n
500.613 15401.749 m
506.651 15401.889 L
512.249 15402.067 L
518.788 15402.33 L
524.936 15402.624 L
537.514 15403.413 L
548.905 15404.349 L
572.524 15407.065 L
584.74 15408.925 L
597.636 15411.261 L
620.901 15416.577 L
645.658 15423.973 L
669.491 15433.046 L
694.83 15445.091 L
719.232 15459.344 L
742.717 15475.813 L
767.695 15496.667 L
791.748 15520.348 L
813.306 15544.897 L
[] 0 d
S 
[] 0 d
n
230.6937 15409.989 m
254.3253 15419.954 L
280.1042 15430.44 L
304.3109 15439.964 L
327.594 15448.878 L
352.376 15458.17 L
376.228 15467.004 L
401.584 15476.388 L
426.011 15485.545 L
449.508 15494.599 L
474.51 15504.681 L
498.582 15515.007 L
521.725 15525.719 L
546.372 15538.211 L
570.089 15551.565 L
582.147 15558.949 L
588.515 15563.029 L
589.396 15563.604 L
590.35 15564.228 L
590.791 15564.515 L
591.249 15564.65 L
591.647 15564.73 L
592.081 15564.809 L
595.311 15565.427 L
607.907 15567.464 L
613.908 15568.223 L
619.598 15568.816 L
622.589 15569.067 L
625.844 15569.305 L
628.547 15569.465 L
631.533 15569.605 L
633.197 15569.666 L
634.757 15569.715 L
636.402 15569.752 L
637.332 15569.764 L
637.742 15569.77 L
638.182 15569.776 L
638.923 15569.783 L
639.614 15569.783 L
640.036 15569.783 L
640.434 15569.783 L
641.174 15569.783 L
641.963 15569.776 L
642.691 15569.77 L
643.486 15569.758 L
644.33 15569.746 L
645.878 15569.709 L
647.346 15569.666 L
650.099 15569.55 L
653.072 15569.385 L
[] 0 d
S 
[] 0 d
n
653.072 15569.385 m
656.302 15569.153 L
662.12 15568.59 L
667.509 15567.905 L
680.087 15565.659 L
691.478 15562.784 L
703.272 15558.912 L
716.021 15553.626 L
741.133 15539.526 L
765.321 15520.77 L
788.586 15497.303 L
813.306 15465.768 L
[] 0 d
S 
[] 0 d
n
230.6937 15409.989 m
231.4155 15409.989 L
232.064 15409.989 L
232.8225 15409.989 L
233.5383 15409.989 L
234.199 15409.989 L
234.8046 15409.989 L
235.4775 15409.989 L
236.1871 15409.989 L
236.8968 15409.989 L
237.6614 15409.989 L
238.3833 15409.989 L
239.044 15409.989 L
239.8209 15409.989 L
240.6651 15409.989 L
241.4604 15409.989 L
242.2006 15409.989 L
242.9591 15409.989 L
243.785 15409.989 L
245.2776 15409.989 L
246.7336 15409.989 L
248.0672 15409.989 L
249.5232 15409.989 L
251.1137 15409.989 L
252.7776 15409.989 L
254.3253 15409.989 L
255.9281 15409.989 L
257.6716 15409.983 L
260.8343 15409.983 L
263.9175 15409.983 L
266.7682 15409.983 L
269.8392 15409.983 L
273.2343 15409.983 L
276.7763 15409.977 L
280.1042 15409.977 L
286.1177 15409.971 L
292.4492 15409.965 L
297.857 15409.952 L
303.7726 15409.94 L
310.0552 15409.922 L
315.7689 15409.897 L
322.437 15409.867 L
328.585 15409.83 L
340.318 15409.732 L
346.497 15409.671 L
353.042 15409.591 L
365.014 15409.396 L
376.221 15409.163 L
388.628 15408.821 L
399.969 15408.423 L
[] 0 d
S 
[] 0 d
n
399.969 15408.423 m
412.59 15407.866 L
424.292 15407.218 L
447.691 15405.468 L
460.269 15404.22 L
471.66 15402.874 L
496.203 15399.124 L
519.816 15394.194 L
544.934 15387.146 L
569.123 15378.197 L
581.296 15372.74 L
586.606 15370.134 L
589.408 15368.702 L
590.093 15368.348 L
590.821 15367.968 L
591.629 15367.981 L
592.381 15368.029 L
595.574 15368.256 L
598.529 15368.507 L
604.341 15369.106 L
610.458 15369.901 L
617.145 15370.972 L
629.428 15373.554 L
642.342 15377.2 L
654.363 15381.537 L
667.185 15387.287 L
692.603 15402.581 L
717.097 15423.031 L
740.656 15448.994 L
765.725 15484.567 L
789.858 15527.854 L
813.068 15579.136 L
813.306 15579.717 L
[] 0 d
S 
[] 0 d
n
vmr
Q
Q
n
1742.2408 1944.2888 m
1742.2408 2134.2888 L
false setoverprint
S 
n
1742.2408 1944.2888 m
1742.2408 1861.7888 L
S 
n
1962.2408 2134.2888 m
1962.2408 1936.7888 L
S 
n
1962.2408 1936.7888 m
1962.2408 1861.7888 L
S 
n
1742.2408 1861.7888 m
1962.2408 1861.7888 L
S 
n
1742.2408 2134.2888 m
1962.2408 2134.2888 L
S 
n
1749.2807 1858.3887 m
1743.2408 1858.3887 L
1743.2408 1845.3889 L
1749.2807 1845.3889 L
1749.2807 1858.3887 L
n
q
_bfh
_efh
{
f1 [12 0 0 12 0 0] makesetfont
1743.240753 1848.788849 m
0 0 32 0 0 (1) ts
}
true
[0 0 0 1]sts
Q
1948.7408 1864.2888 m
1948.7408 1857.2888 L
false setoverprint
S 
n
1739.2529 1951.9949 m
1728.5947 1951.9949 L
1728.5947 1939.7451 L
1739.2529 1939.7451 L
1739.2529 1951.9949 L
n
q
_bfh
_efh
{
f1 [11.25 0 0 11.25 0 0] makesetfont
1728.594666 1942.995071 m
0 0 32 0 0 (x) ts
}
true
[0 0 0 1]sts
{
f1 [11.25 0 0 11.25 0 0] makesetfont
0 -4.6875 rmoveto 0 0 32 0 0 (c) ts
0 4.6875 rmoveto }
true
[0 0 0 1]sts
Q
1745.7408 2121.2888 m
1738.2408 2121.2888 L
false setoverprint
S 
n
1739.2408 1871.7888 m
1745.7408 1871.7888 L
S 
n
1719.2807 2007.3887 m
1713.2408 2007.3887 L
1713.2408 1994.3889 L
1719.2807 1994.3889 L
1719.2807 2007.3887 L
n
q
_bfh
_efh
{
f1 [12 0 0 12 0 0] makesetfont
1713.240753 1997.788849 m
0 -5 rmoveto 0 0 32 0 0 (x) ts
0 5 rmoveto }
true
[0 0 0 1]sts
Q
true eomode
1961.4908 2131.2888 m
1789.2408 1984.0388 L
1786.9908 1981.7888 L
1802.2408 1993.5388 L
1817.9908 2005.0388 L
1828.2408 2012.2888 L
1834.9908 2016.2888 L
1828.4908 2010.0388 L
1822.9908 2005.5388 L
1818.9908 2002.7888 L
1815.4908 2000.0388 L
1811.4908 1997.0388 L
1806.9908 1993.7888 L
1803.2408 1991.0388 L
1799.2408 1987.7888 L
1795.2408 1985.0388 L
1789.7408 1981.0388 L
1785.9908 1978.2888 L
1782.9908 1976.0388 L
1797.9908 1985.7888 L
1804.9908 1990.2888 L
1811.9908 1994.0388 L
1819.4908 1998.5388 L
1826.7408 2002.5388 L
1833.4908 2005.5388 L
1844.9908 2010.2888 L
1855.2408 2014.5388 L
1863.2408 2016.7888 L
1874.2408 2019.5388 L
1887.9908 2022.2888 L
1895.2408 2023.0388 L
1901.7408 2023.2888 L
1914.2408 2023.5388 L
1921.7408 2023.0388 L
1928.2408 2022.0388 L
1934.9908 2021.2888 L
1938.7408 2020.2888 L
1941.4908 2019.7888 L
1944.9908 2018.7888 L
1948.2408 2017.5388 L
1946.2408 2014.7888 L
1941.9908 2010.0388 L
1934.4908 2002.5388 L
1926.4908 1994.5388 L
1917.9908 1987.5388 L
1909.4908 1981.2888 L
1899.9908 1975.2888 L
1889.9908 1969.2888 L
1875.2408 1962.2888 L
1863.2408 1957.2888 L
1850.2408 1953.5388 L
1839.4908 1950.5388 L
1830.2408 1948.7888 L
1824.7408 1948.0388 L
1818.2408 1947.0388 L
1811.2408 1946.2888 L
1805.9908 1945.7888 L
1800.2408 1945.2888 L
1791.9908 1944.7888 L
1787.4908 1944.5388 L
1784.2408 1944.2888 L
1794.7408 1944.0388 L
1799.2408 1944.0388 L
1807.2408 1943.5388 L
1813.9908 1942.7888 L
1819.4908 1942.0388 L
1824.7408 1941.2888 L
1829.4908 1940.0388 L
1833.4908 1939.0388 L
1836.7408 1938.2888 L
1831.9908 1937.5388 L
1824.7408 1937.7888 L
1818.4908 1938.2888 L
1812.2408 1938.7888 L
1806.2408 1939.2888 L
1799.9908 1939.7888 L
1795.7408 1940.2888 L
1790.7408 1941.0388 L
1786.9908 1941.5388 L
1783.7408 1941.7888 L
1794.7408 1940.0388 L
1801.9908 1938.5388 L
1808.7408 1937.0388 L
1819.2408 1934.5388 L
1832.2408 1930.7888 L
1846.2408 1926.2888 L
1852.7408 1924.0388 L
1860.7408 1920.7888 L
1866.9908 1918.0388 L
1873.7408 1915.0388 L
1878.4908 1913.0388 L
1884.9908 1909.7888 L
1894.7408 1904.7888 L
1900.4908 1901.7888 L
1908.4908 1897.2888 L
1916.4908 1892.7888 L
1922.2408 1889.2888 L
1929.7408 1884.2888 L
1936.2408 1880.0388 L
1942.9908 1875.5388 L
1948.9908 1871.2888 L
1953.7408 1868.2888 L
1957.2408 1865.5388 L
1959.7408 1863.2888 L
1962.2408 1861.7888 L
1961.4908 2131.2888 L
[0 0 0 0.45]  vc
false setoverprint
f 
[0 0 0 1]  vc
S 
n
1854.6302 1827.3887 m
1848.6152 1827.3887 L
1848.6152 1820.3887 L
1854.6302 1820.3887 L
1854.6302 1827.3887 L
n
q
q
[1 0 0 1 1848.615204 1821.788788] concat
[0.5 0 0 0.5 0 0] concat
vms
q
[1 0 0 1 -509.0625 -15135.25] concat
vms
520.3125 15145.0625 m
519.875 15144.1875 L
519.3125 15143.25 L
518.625 15142.125 L
518.0625 15141.125 L
517.5 15140.375 L
516.875 15139.5 L
516.1875 15138.625 L
515.625 15137.875 L
514.875 15136.9375 L
514.125 15136.3125 L
513 15135.625 L
511.75 15135.25 L
510.625 15135.5625 L
509.75 15136.3125 L
509.25 15137.75 L
509.0625 15139.25 L
509.25 15140.625 L
509.625 15142 L
510.125 15143.125 L
511.1875 15144.625 L
512.5 15145.3125 L
513.6875 15145.75 L
514.9375 15145.6875 L
515.8125 15144.9375 L
516.3125 15143.9375 L
516.3125 15143.0625 L
516.3125 15142.375 L
516.375 15141.375 L
516.3125 15140.4375 L
516.25 15139.6875 L
516.1875 15138.0625 L
516.25 15136.875 L
516.375 15136.125 L
516.6875 15135.625 L
517.75 15135.5 L
518.5 15136.125 L
518.9375 15136.625 L
519.25 15137.125 L
1 w
S 
n
vmr
Q
vmr
Q
Q
1960.499 1854.8887 m
1945.2408 1854.8887 L
1945.2408 1841.8889 L
1960.499 1841.8889 L
1960.499 1854.8887 L
n
q
_bfh
_efh
{
f2 [12 0 0 12 0 0] makesetfont
1945.240753 1845.288849 m
0 0 32 0 0 (2) ts
}
true
[0 0 0 1]sts
{
f2 [7 0 0 7 0 0] makesetfont
0 5 rmoveto 0 0 32 0 0 (1\/2) ts
0 -5 rmoveto }
true
[0 0 0 1]sts
Q
1836.2408 1865.2888 m
1836.2408 1858.2888 L
false setoverprint
S 
n
1846.499 1855.3888 m
1831.2408 1855.3888 L
1831.2408 1842.3889 L
1846.499 1842.3889 L
1846.499 1855.3888 L
n
q
_bfh
_efh
{
f2 [12 0 0 12 0 0] makesetfont
1831.240753 1845.78891 m
0 0 32 0 0 (2) ts
}
true
[0 0 0 1]sts
{
f2 [7 0 0 7 0 0] makesetfont
0 5 rmoveto 0 0 32 0 0 (1\/4) ts
0 -5 rmoveto }
true
[0 0 0 1]sts
Q
1738.0439 2126.3887 m
1708.2408 2126.3887 L
1708.2408 2115.7888 L
1738.0439 2115.7888 L
1738.0439 2126.3887 L
n
q
_bfh
_efh
{
f1 [9.600037 0 0 9.600037 0 0] makesetfont
1708.240753 2118.708801 m
0 -1 rmoveto 0 0 32 0 0 (T) ts
0 1 rmoveto }
true
[0 0 0 1]sts
{
f1 [9.600037 0 0 9.600037 0 0] makesetfont
0 -5 rmoveto 0 0 32 0 0 (1) ts
0 5 rmoveto }
true
[0 0 0 1]sts
{
f1 [6 0 0 6 0 0] makesetfont
0 6 rmoveto 0 0 32 0 0 (\() ts
0 -6 rmoveto }
true
[0 0 0 1]sts
{
f1 [7 0 0 7 0 0] makesetfont
0 6 rmoveto 0 0 32 0 0 (1) ts
0 -6 rmoveto }
true
[0 0 0 1]sts
{
f1 [6 0 0 6 0 0] makesetfont
0 6 rmoveto 0 0 32 0 0 (\)) ts
0 -6 rmoveto }
true
[0 0 0 1]sts
{
f1 [6 0 0 6 0 0] makesetfont
0 0 32 0 0 (\() ts
}
true
[0 0 0 1]sts
{
f1 [9 0 0 9 0 0] makesetfont
0 0 32 0 0 (x) ts
}
true
[0 0 0 1]sts
{
f1 [7 0 0 7 0 0] makesetfont
0 -1.5 rmoveto 0 0 32 0 0 (c) ts
0 1.5 rmoveto }
true
[0 0 0 1]sts
{
f1 [6 0 0 6 0 0] makesetfont
0 0 32 0 0 (\)) ts
}
true
[0 0 0 1]sts
Q
1739.5439 1877.3888 m
1709.7408 1877.3888 L
1709.7408 1866.7889 L
1739.5439 1866.7889 L
1739.5439 1877.3888 L
n
q
_bfh
_efh
{
f1 [9.600037 0 0 9.600037 0 0] makesetfont
1709.740753 1869.708862 m
0 -1 rmoveto 0 0 32 0 0 (T) ts
0 1 rmoveto }
true
[0 0 0 1]sts
{
f1 [9.600037 0 0 9.600037 0 0] makesetfont
0 -5 rmoveto 0 0 32 0 0 (1) ts
0 5 rmoveto }
true
[0 0 0 1]sts
{
f1 [6 0 0 6 0 0] makesetfont
0 6 rmoveto 0 0 32 0 0 (\() ts
0 -6 rmoveto }
true
[0 0 0 1]sts
{
f1 [7 0 0 7 0 0] makesetfont
0 6 rmoveto 0 0 32 0 0 (2) ts
0 -6 rmoveto }
true
[0 0 0 1]sts
{
f1 [6 0 0 6 0 0] makesetfont
0 6 rmoveto 0 0 32 0 0 (\)) ts
0 -6 rmoveto }
true
[0 0 0 1]sts
{
f1 [6 0 0 6 0 0] makesetfont
0 0 32 0 0 (\() ts
}
true
[0 0 0 1]sts
{
f1 [9 0 0 9 0 0] makesetfont
0 0 32 0 0 (x) ts
}
true
[0 0 0 1]sts
{
f1 [7 0 0 7 0 0] makesetfont
0 -1.5 rmoveto 0 0 32 0 0 (c) ts
0 1.5 rmoveto }
true
[0 0 0 1]sts
{
f1 [6 0 0 6 0 0] makesetfont
0 0 32 0 0 (\)) ts
}
true
[0 0 0 1]sts
Q
1788.0991 2109.8887 m
1774.7408 2109.8887 L
1774.7408 2096.8889 L
1788.0991 2096.8889 L
1788.0991 2109.8887 L
n
q
_bfh
_efh
{
f1 [12 0 0 12 0 0] makesetfont
1774.740753 2100.288849 m
0 0 32 0 0 (\(a\)) ts
}
true
[0 0 0 1]sts
Q
vmr
vmr
end
EndOfTheIncludedPostscriptMagicCookie

\closepsdump

\psdump{fig21.eps}
userdict begin /AltsysDict 245 dict def end
AltsysDict begin
/bdf{bind def}bind def
/xdf{exch def}bdf
/defed{where{pop true}{false}ifelse}bdf
/ndf{1 index where{pop pop pop}{dup xcheck{bind}if def}ifelse}bdf
/d{setdash}bdf
/h{closepath}bdf
/H{}bdf
/J{setlinecap}bdf
/j{setlinejoin}bdf
/M{setmiterlimit}bdf
/n{newpath}bdf
/N{newpath}bdf
/q{gsave}bdf
/Q{grestore}bdf
/w{setlinewidth}bdf
/sepdef{
 dup where not
 {
AltsysSepDict
 }
 if 
 3 1 roll exch put
}bdf
/st{settransfer}bdf
/colorimage defed /_rci xdf
/_NXLevel2 defed { 
 _NXLevel2 not {   
/colorimage where {
userdict eq {
/_rci false def 
} if
} if
 } if
} if
/md defed{ 
 md type /dicttype eq {  
/colorimage where { 
md eq { 
/_rci false def 
}if
}if
/settransfer where {
md eq {
/st systemdict /settransfer get def
}if
}if
 }if 
}if
/setstrokeadjust defed
{
 true setstrokeadjust
 /C{curveto}bdf
 /L{lineto}bdf
 /m{moveto}bdf
}
{
 /dr{transform .25 sub round .25 add 
exch .25 sub round .25 add exch itransform}bdf
 /C{dr curveto}bdf
 /L{dr lineto}bdf
 /m{dr moveto}bdf
 /setstrokeadjust{pop}bdf 
}ifelse
/rectstroke defed /xt xdf
xt {/yt save def} if
/privrectpath { 
 4 -2 roll m
 dtransform round exch round exch idtransform 
 2 copy 0 lt exch 0 lt xor
 {dup 0 exch rlineto exch 0 rlineto neg 0 exch rlineto}
 {exch dup 0 rlineto exch 0 exch rlineto neg 0 rlineto}
 ifelse
 closepath
}bdf
/rectclip{newpath privrectpath clip newpath}def
/rectfill{gsave newpath privrectpath fill grestore}def
/rectstroke{gsave newpath privrectpath stroke grestore}def
xt {yt restore} if
/_fonthacksave false def
/currentpacking defed 
{
 /_bfh {/_fonthacksave currentpacking def false setpacking} bdf
 /_efh {_fonthacksave setpacking} bdf
}
{
 /_bfh {} bdf
 /_efh {} bdf
}ifelse
/packedarray{array astore readonly}ndf
/` 
{ 
 false setoverprint

 /-save0- save def
 5 index concat
 pop
 storerect left bottom width height rectclip
 pop
 
 /dict_count countdictstack def
 /op_count count 1 sub def
 userdict begin
 
 /showpage {} def
 
 0 setgray 0 setlinecap 1 setlinewidth
 0 setlinejoin 10 setmiterlimit [] 0 setdash newpath
 
} bdf
/currentpacking defed{true setpacking}if
/min{2 copy gt{exch}if pop}bdf
/max{2 copy lt{exch}if pop}bdf
/xformfont { currentfont exch makefont setfont } bdf
/fhnumcolors 1 
 statusdict begin
/processcolors defed 
{
pop processcolors
}
{
/deviceinfo defed {
deviceinfo /Colors known {
pop deviceinfo /Colors get
} if
} if
} ifelse
 end 
def
/printerRes 
 gsave
 matrix defaultmatrix setmatrix
 72 72 dtransform
 abs exch abs
 max
 grestore
 def
/graycalcs
[
 {Angle Frequency}   
 {GrayAngle GrayFrequency} 
 {0 Width Height matrix defaultmatrix idtransform 
dup mul exch dup mul add sqrt 72 exch div} 
 {0 GrayWidth GrayHeight matrix defaultmatrix idtransform 
dup mul exch dup mul add sqrt 72 exch div} 
] def
/calcgraysteps {
 forcemaxsteps
 {
maxsteps
 }
 {
/currenthalftone defed
{currenthalftone /dicttype eq}{false}ifelse
{
currenthalftone begin
HalftoneType 4 le
{graycalcs HalftoneType 1 sub get exec}
{
HalftoneType 5 eq
{
Default begin
{graycalcs HalftoneType 1 sub get exec}
end
}
{0 60} 
ifelse
}
ifelse
end
}
{
currentscreen pop exch 
}
ifelse
 
printerRes 300 max exch div exch 
2 copy 
sin mul round dup mul 
3 1 roll 
cos mul round dup mul 
add 1 add 
dup maxsteps gt {pop maxsteps} if 
 }
 ifelse
} bdf
/nextrelease defed { 
 /languagelevel defed not {    
/framebuffer defed { 
0 40 string framebuffer 9 1 roll 8 {pop} repeat
dup 516 eq exch 520 eq or
{
/fhnumcolors 3 def
/currentscreen {60 0 {pop pop 1}}bdf
/calcgraysteps {maxsteps} bdf
}if
}if
 }if
}if
fhnumcolors 1 ne {
 /calcgraysteps {maxsteps} bdf
} if
/currentpagedevice defed {

 currentpagedevice /PreRenderingEnhance known
 {
currentpagedevice /PreRenderingEnhance get
{
/calcgraysteps 
{
forcemaxsteps 
{maxsteps}
{256 maxsteps min}
ifelse
} def
} if
 } if
} if
/gradfrequency 144 def
printerRes 1000 lt {
 /gradfrequency 72 def
} if
/adjnumsteps {
 
 dup dtransform abs exch abs max  
 
 printerRes div       
 
 gradfrequency mul      
 round        
 5 max       
 min        
}bdf
/goodsep {
 spots exch get 4 get dup sepname eq exch (_vc_Registration) eq or
}bdf
/BeginGradation defed
{/bb{BeginGradation}bdf}
{/bb{}bdf}
ifelse
/EndGradation defed
{/eb{EndGradation}bdf}
{/eb{}bdf}
ifelse
/bottom -0 def 
/delta -0 def 
/frac -0 def 
/height -0 def 
/left -0 def 
/numsteps1 -0 def 
/radius -0 def 
/right -0 def 
/top -0 def 
/width -0 def 
/xt -0 def 
/yt -0 def 
/df currentflat def 
/tempstr 1 string def 
/clipflatness currentflat def 
/inverted? 
 0 currenttransfer exec .5 ge def
/tc1 [0 0 0 1] def 
/tc2 [0 0 0 1] def 
/storerect{/top xdf /right xdf /bottom xdf /left xdf 
/width right left sub def /height top bottom sub def}bdf
/concatprocs{
 systemdict /packedarray known 
 {dup type /packedarraytype eq 2 index type /packedarraytype eq or}{false}ifelse
 { 
/proc2 exch cvlit def /proc1 exch cvlit def
proc1 aload pop proc2 aload pop
proc1 length proc2 length add packedarray cvx
 }
 { 
/proc2 exch cvlit def /proc1 exch cvlit def
/newproc proc1 length proc2 length add array def
newproc 0 proc1 putinterval newproc proc1 length proc2 putinterval
newproc cvx
 }ifelse
}bdf
/i{dup 0 eq
 {pop df dup} 
 {dup} ifelse 
 /clipflatness xdf setflat
}bdf
version cvr 38.0 le
{/setrgbcolor{
currenttransfer exec 3 1 roll
currenttransfer exec 3 1 roll
currenttransfer exec 3 1 roll
setrgbcolor}bdf}if
/vms {/vmsv save def} bdf
/vmr {vmsv restore} bdf
/vmrs{vmsv restore /vmsv save def}bdf
/eomode{ 
 {/filler /eofill load def /clipper /eoclip load def}
 {/filler /fill load def /clipper /clip load def}
 ifelse
}bdf
/normtaper{}bdf
/logtaper{9 mul 1 add log}bdf
/CD{
 /NF exch def 
 {    
exch dup 
/FID ne 1 index/UniqueID ne and
{exch NF 3 1 roll put}
{pop pop}
ifelse
 }forall 
 NF
}bdf
/MN{
 1 index length   
 /Len exch def 
 dup length Len add  
 string dup    
 Len     
 4 -1 roll    
 putinterval   
 dup     
 0      
 4 -1 roll   
 putinterval   
}bdf
/RC{4 -1 roll /ourvec xdf 256 string cvs(|______)anchorsearch
 {1 index MN cvn/NewN exch def cvn
 findfont dup maxlength dict CD dup/FontName NewN put dup
 /Encoding ourvec put NewN exch definefont pop}{pop}ifelse}bdf
/RF{ 
 dup      
 FontDirectory exch   
 known     
 {pop 3 -1 roll pop}  
 {RC}
 ifelse
}bdf
/FF{dup 256 string cvs(|______)exch MN cvn dup FontDirectory exch known
 {exch pop findfont 3 -1 roll pop}
 {pop dup findfont dup maxlength dict CD dup dup
 /Encoding exch /Encoding get 256 array copy 7 -1 roll 
 {3 -1 roll dup 4 -2 roll put}forall put definefont}
 ifelse}bdf
/RFJ{ 
 dup      
 FontDirectory exch   
 known     
 {pop 3 -1 roll pop  
 FontDirectory /Ryumin-Light-83pv-RKSJ-H known 
 {pop pop /Ryumin-Light-83pv-RKSJ-H dup}if  
 }      
 {RC}
 ifelse
}bdf
/FFJ{dup 256 string cvs(|______)exch MN cvn dup FontDirectory exch known
 {exch pop findfont 3 -1 roll pop}
 {pop
dup FontDirectory exch known not 
 {FontDirectory /Ryumin-Light-83pv-RKSJ-H known 
{pop /Ryumin-Light-83pv-RKSJ-H}if 
 }if            
 dup findfont dup maxlength dict CD dup dup
 /Encoding exch /Encoding get 256 array copy 7 -1 roll 
 {3 -1 roll dup 4 -2 roll put}forall put definefont}
 ifelse}bdf
/fps{
 currentflat   
 exch     
 dup 0 le{pop 1}if 
 {
dup setflat 3 index stopped
{1.3 mul dup 3 index gt{pop setflat pop pop stop}if} 
{exit} 
ifelse
 }loop 
 pop setflat pop pop
}bdf
/fp{100 currentflat fps}bdf
/clipper{clip}bdf 
/W{/clipper load 100 clipflatness dup setflat fps}bdf
userdict begin /BDFontDict 29 dict def end
BDFontDict begin
/bu{}def
/bn{}def
/setTxMode{av 70 ge{pop}if pop}def
/gm{m}def
/show{pop}def
/gr{pop}def
/fnt{pop pop pop}def
/fs{pop}def
/fz{pop}def
/lin{pop pop}def
/:M {pop pop} def
/sf {pop} def
/S {pop} def
/@b {pop pop pop pop pop pop pop pop} def
/_bdsave /save load def
/_bdrestore /restore load def
/save { dup /fontsave eq {null} {_bdsave} ifelse } def
/restore { dup null eq { pop } { _bdrestore } ifelse } def
/fontsave null def
end
/MacVec 256 array def 
MacVec 0 /Helvetica findfont
/Encoding get 0 128 getinterval putinterval
MacVec 127 /DEL put MacVec 16#27 /quotesingle put MacVec 16#60 /grave put
/NUL/SOH/STX/ETX/EOT/ENQ/ACK/BEL/BS/HT/LF/VT/FF/CR/SO/SI
/DLE/DC1/DC2/DC3/DC4/NAK/SYN/ETB/CAN/EM/SUB/ESC/FS/GS/RS/US
MacVec 0 32 getinterval astore pop
/Adieresis/Aring/Ccedilla/Eacute/Ntilde/Odieresis/Udieresis/aacute
/agrave/acircumflex/adieresis/atilde/aring/ccedilla/eacute/egrave
/ecircumflex/edieresis/iacute/igrave/icircumflex/idieresis/ntilde/oacute
/ograve/ocircumflex/odieresis/otilde/uacute/ugrave/ucircumflex/udieresis
/dagger/degree/cent/sterling/section/bullet/paragraph/germandbls
/registered/copyright/trademark/acute/dieresis/notequal/AE/Oslash
/infinity/plusminus/lessequal/greaterequal/yen/mu/partialdiff/summation
/product/pi/integral/ordfeminine/ordmasculine/Omega/ae/oslash 
/questiondown/exclamdown/logicalnot/radical/florin/approxequal/Delta/guillemotleft
/guillemotright/ellipsis/nbspace/Agrave/Atilde/Otilde/OE/oe
/endash/emdash/quotedblleft/quotedblright/quoteleft/quoteright/divide/lozenge
/ydieresis/Ydieresis/fraction/currency/guilsinglleft/guilsinglright/fi/fl
/daggerdbl/periodcentered/quotesinglbase/quotedblbase
/perthousand/Acircumflex/Ecircumflex/Aacute
/Edieresis/Egrave/Iacute/Icircumflex/Idieresis/Igrave/Oacute/Ocircumflex
/apple/Ograve/Uacute/Ucircumflex/Ugrave/dotlessi/circumflex/tilde
/macron/breve/dotaccent/ring/cedilla/hungarumlaut/ogonek/caron
MacVec 128 128 getinterval astore pop
end 
AltsysDict begin
_bfh
_efh
0 dict dup begin
end 
/f0 /Symbol FF def
_bfh
_efh
MacVec 256 array copy
/f1 /|______Times-Roman dup RF findfont def
end 
AltsysDict begin 
/onlyk4{false}ndf
/ccmyk{dup 5 -1 roll sub 0 max exch}ndf
/cmyk2gray{
 4 -1 roll 0.3 mul 4 -1 roll 0.59 mul 4 -1 roll 0.11 mul
 add add add 1 min neg 1 add
}bdf
/setcmykcolor{1 exch sub ccmyk ccmyk ccmyk pop setrgbcolor}ndf
/maxcolor { 
 max max max  
} ndf
/maxspot {
 pop
} ndf
/setcmykcoloroverprint{4{dup -1 eq{pop 0}if 4 1 roll}repeat setcmykcolor}ndf
/findcmykcustomcolor{5 packedarray}ndf
/setcustomcolor{exch aload pop pop 4{4 index mul 4 1 roll}repeat setcmykcolor pop}ndf
/setseparationgray{setgray}ndf
/setoverprint{pop}ndf 
/currentoverprint false ndf
/cmykbufs2gray{
 0 1 2 index length 1 sub
 { 
4 index 1 index get 0.3 mul 
4 index 2 index get 0.59 mul 
4 index 3 index get 0.11 mul 
4 index 4 index get 
add add add cvi 255 min
255 exch sub
2 index 3 1 roll put
 }for
 4 1 roll pop pop pop
}bdf
/colorimage{
 pop pop
 [
5 -1 roll/exec cvx 
6 -1 roll/exec cvx 
7 -1 roll/exec cvx 
8 -1 roll/exec cvx
/cmykbufs2gray cvx
 ]cvx 
 image
}
version cvr 47.1 le 
statusdict /product get (Lino) anchorsearch{pop pop true}{pop false}ifelse
and{userdict begin bdf end}{ndf}ifelse
fhnumcolors 1 ne {/yt save def} if
/customcolorimage{
 aload pop
 (_vc_Registration) eq 
 {
pop pop pop pop separationimage
 }
 {
/ik xdf /iy xdf /im xdf /ic xdf
ic im iy ik cmyk2gray /xt xdf
currenttransfer
{dup 1.0 exch sub xt mul add}concatprocs
st 
image
 }
 ifelse
}ndf
fhnumcolors 1 ne {yt restore} if
fhnumcolors 3 ne {/yt save def} if
/customcolorimage{
 aload pop 
 (_vc_Registration) eq 
 {
pop pop pop pop separationimage
 }
 {
/ik xdf /iy xdf /im xdf /ic xdf
1.0 dup ic ik add min sub 
1.0 dup im ik add min sub 
1.0 dup iy ik add min sub 
/ic xdf /iy xdf /im xdf
currentcolortransfer
4 1 roll 
{dup 1.0 exch sub ic mul add}concatprocs 4 1 roll 
{dup 1.0 exch sub iy mul add}concatprocs 4 1 roll 
{dup 1.0 exch sub im mul add}concatprocs 4 1 roll 
setcolortransfer
{/dummy xdf dummy}concatprocs{dummy}{dummy}true 3 colorimage
 }
 ifelse
}ndf
fhnumcolors 3 ne {yt restore} if
fhnumcolors 4 ne {/yt save def} if
/customcolorimage{
 aload pop
 (_vc_Registration) eq 
 {
pop pop pop pop separationimage
 }
 {
/ik xdf /iy xdf /im xdf /ic xdf
currentcolortransfer
{1.0 exch sub ik mul ik sub 1 add}concatprocs 4 1 roll
{1.0 exch sub iy mul iy sub 1 add}concatprocs 4 1 roll
{1.0 exch sub im mul im sub 1 add}concatprocs 4 1 roll
{1.0 exch sub ic mul ic sub 1 add}concatprocs 4 1 roll
setcolortransfer
{/dummy xdf dummy}concatprocs{dummy}{dummy}{dummy}
true 4 colorimage
 }
 ifelse
}ndf
fhnumcolors 4 ne {yt restore} if
/separationimage{image}ndf
/newcmykcustomcolor{6 packedarray}ndf
/inkoverprint false ndf
/setinkoverprint{pop}ndf 
/setspotcolor { 
 spots exch get
 dup 4 get (_vc_Registration) eq
 {pop 1 exch sub setseparationgray}
 {0 5 getinterval exch setcustomcolor}
 ifelse
}ndf
/currentcolortransfer{currenttransfer dup dup dup}ndf
/setcolortransfer{st pop pop pop}ndf
/fas{}ndf
/sas{}ndf
/fhsetspreadsize{pop}ndf
/filler{fill}bdf 
/F{gsave {filler}fp grestore}bdf
/f{closepath F}bdf
/S{gsave {stroke}fp grestore}bdf
/s{closepath S}bdf
/bc4 [0 0 0 0] def 
/_lfp4 {
 /iosv inkoverprint def
 /cosv currentoverprint def
 /yt xdf       
 /xt xdf       
 /ang xdf      
 storerect
 /taperfcn xdf
 /k2 xdf /y2 xdf /m2 xdf /c2 xdf
 /k1 xdf /y1 xdf /m1 xdf /c1 xdf
 c1 c2 sub abs
 m1 m2 sub abs
 y1 y2 sub abs
 k1 k2 sub abs
 maxcolor      
 calcgraysteps mul abs round  
 height abs adjnumsteps   
 dup 2 lt {pop 1} if    
 1 sub /numsteps1 xdf
 currentflat mark    
 currentflat clipflatness  
 /delta top bottom sub numsteps1 1 add div def 
 /right right left sub def  
 /botsv top delta sub def  
 {
{
W
xt yt translate 
ang rotate
xt neg yt neg translate 
dup setflat 
/bottom botsv def
0 1 numsteps1 
{
numsteps1 dup 0 eq {pop 0.5 } { div } ifelse 
taperfcn /frac xdf
bc4 0 c2 c1 sub frac mul c1 add put
bc4 1 m2 m1 sub frac mul m1 add put
bc4 2 y2 y1 sub frac mul y1 add put
bc4 3 k2 k1 sub frac mul k1 add put
bc4 vc
1 index setflat 
{ 
mark {newpath left bottom right delta rectfill}stopped
{cleartomark exch 1.3 mul dup setflat exch 2 copy gt{stop}if}
{cleartomark exit}ifelse
}loop
/bottom bottom delta sub def
}for
}
gsave stopped grestore
{exch pop 2 index exch 1.3 mul dup 100 gt{cleartomark setflat stop}if}
{exit}ifelse
 }loop
 cleartomark setflat
 iosv setinkoverprint
 cosv setoverprint
}bdf
/bcs [0 0] def 
/_lfs4 {
 /iosv inkoverprint def
 /cosv currentoverprint def
 /yt xdf       
 /xt xdf       
 /ang xdf      
 storerect
 /taperfcn xdf
 /tint2 xdf      
 /tint1 xdf      
 bcs exch 1 exch put    
 tint1 tint2 sub abs    
 bcs 1 get maxspot    
 calcgraysteps mul abs round  
 height abs adjnumsteps   
 dup 2 lt {pop 2} if    
 1 sub /numsteps1 xdf
 currentflat mark    
 currentflat clipflatness  
 /delta top bottom sub numsteps1 1 add div def 
 /right right left sub def  
 /botsv top delta sub def  
 {
{
W
xt yt translate 
ang rotate
xt neg yt neg translate 
dup setflat 
/bottom botsv def
0 1 numsteps1 
{
numsteps1 div taperfcn /frac xdf
bcs 0
1.0 tint2 tint1 sub frac mul tint1 add sub
put bcs vc
1 index setflat 
{ 
mark {newpath left bottom right delta rectfill}stopped
{cleartomark exch 1.3 mul dup setflat exch 2 copy gt{stop}if}
{cleartomark exit}ifelse
}loop
/bottom bottom delta sub def
}for
}
gsave stopped grestore
{exch pop 2 index exch 1.3 mul dup 100 gt{cleartomark setflat stop}if}
{exit}ifelse
 }loop
 cleartomark setflat
 iosv setinkoverprint
 cosv setoverprint
}bdf
/_rfs4 {
 /iosv inkoverprint def
 /cosv currentoverprint def
 /tint2 xdf      
 /tint1 xdf      
 bcs exch 1 exch put    
 /radius xdf      
 /yt xdf       
 /xt xdf       
 tint1 tint2 sub abs    
 bcs 1 get maxspot    
 calcgraysteps mul abs round  
 radius abs adjnumsteps   
 dup 2 lt {pop 2} if    
 1 sub /numsteps1 xdf
 radius numsteps1 div 2 div /halfstep xdf 
 currentflat mark    
 currentflat clipflatness  
 {
{
dup setflat 
W 
0 1 numsteps1 
{
dup /radindex xdf
numsteps1 div /frac xdf
bcs 0
tint2 tint1 sub frac mul tint1 add
put bcs vc
1 index setflat 
{ 
newpath mark xt yt radius 1 frac sub mul halfstep add 0 360
{ arc
radindex numsteps1 ne 
{
xt yt 
radindex 1 add numsteps1 
div 1 exch sub 
radius mul halfstep add
dup xt add yt moveto 
360 0 arcn 
} if
fill
}stopped
{cleartomark exch 1.3 mul dup setflat exch 2 copy gt{stop}if}
{cleartomark exit}ifelse
}loop
}for
}
gsave stopped grestore
{exch pop 2 index exch 1.3 mul dup 100 gt{cleartomark setflat stop}if}
{exit}ifelse
 }loop
 cleartomark setflat
 iosv setinkoverprint
 cosv setoverprint
}bdf
/_rfp4 {
 /iosv inkoverprint def
 /cosv currentoverprint def
 /k2 xdf /y2 xdf /m2 xdf /c2 xdf
 /k1 xdf /y1 xdf /m1 xdf /c1 xdf
 /radius xdf      
 /yt xdf       
 /xt xdf       
 c1 c2 sub abs
 m1 m2 sub abs
 y1 y2 sub abs
 k1 k2 sub abs
 maxcolor      
 calcgraysteps mul abs round  
 radius abs adjnumsteps   
 dup 2 lt {pop 1} if    
 1 sub /numsteps1 xdf
 radius numsteps1 dup 0 eq {pop} {div} ifelse 
 2 div /halfstep xdf 
 currentflat mark    
 currentflat clipflatness  
 {
{
dup setflat 
W 
0 1 numsteps1 
{
dup /radindex xdf
numsteps1 dup 0 eq {pop 0.5 } { div } ifelse 
/frac xdf
bc4 0 c2 c1 sub frac mul c1 add put
bc4 1 m2 m1 sub frac mul m1 add put
bc4 2 y2 y1 sub frac mul y1 add put
bc4 3 k2 k1 sub frac mul k1 add put
bc4 vc
1 index setflat 
{ 
newpath mark xt yt radius 1 frac sub mul halfstep add 0 360
{ arc
radindex numsteps1 ne 
{
xt yt 
radindex 1 add 
numsteps1 dup 0 eq {pop} {div} ifelse 
1 exch sub 
radius mul halfstep add
dup xt add yt moveto 
360 0 arcn 
} if
fill
}stopped
{cleartomark exch 1.3 mul dup setflat exch 2 copy gt{stop}if}
{cleartomark exit}ifelse
}loop
}for
}
gsave stopped grestore
{exch pop 2 index exch 1.3 mul dup 100 gt{cleartomark setflat stop}if}
{exit}ifelse
 }loop
 cleartomark setflat
 iosv setinkoverprint
 cosv setoverprint
}bdf
/lfp4{_lfp4}ndf
/lfs4{_lfs4}ndf
/rfs4{_rfs4}ndf
/rfp4{_rfp4}ndf
/cvc [0 0 0 1] def 
/vc{
 AltsysDict /cvc 2 index put 
 aload length 4 eq
 {setcmykcolor}
 {setspotcolor}
 ifelse
}bdf 
/origmtx matrix currentmatrix def
/ImMatrix matrix currentmatrix def
0 setseparationgray
/imgr {1692 1584 2304 2376 } def 
/bleed 0 def 
/clpr {1692 1584 2304 2376 } def 
/xs 1 def 
/ys 1 def 
/botx 0 def 
/overlap 0 def 
/wdist 18 def 
0 2 mul fhsetspreadsize 
0 0 ne {/df 0 def /clipflatness 0 def} if 
/maxsteps 256 def 
/forcemaxsteps false def 
vms
-1753 -1686 translate
/currentpacking defed{false setpacking}if 
/spots[
1 0 0 0 (Process Cyan) false newcmykcustomcolor
0 1 0 0 (Process Magenta) false newcmykcustomcolor
0 0 1 0 (Process Yellow) false newcmykcustomcolor
0 0 0 1 (Process Black) false newcmykcustomcolor
]def
/textopf false def
/curtextmtx{}def
/otw .25 def
/msf{dup/curtextmtx xdf makefont setfont}bdf
/makesetfont/msf load def
/curtextheight{.707104 .707104 curtextmtx dtransform
 dup mul exch dup mul add sqrt}bdf
/ta2{ 
tempstr 2 index gsave exec grestore 
cwidth cheight rmoveto 
4 index eq{5 index 5 index rmoveto}if 
2 index 2 index rmoveto 
}bdf
/ta{exch systemdict/cshow known
{{/cheight xdf/cwidth xdf tempstr 0 2 index put ta2}exch cshow} 
{{tempstr 0 2 index put tempstr stringwidth/cheight xdf/cwidth xdf ta2}forall} 
ifelse 6{pop}repeat}bdf
/sts{/textopf currentoverprint def vc setoverprint
/ts{awidthshow}def exec textopf setoverprint}bdf
/stol{/xt currentlinewidth def 
 setlinewidth vc newpath 
 /ts{{false charpath stroke}ta}def exec 
 xt setlinewidth}bdf 
 
/strk{/textopf currentoverprint def vc setoverprint
 /ts{{false charpath stroke}ta}def exec 
 textopf setoverprint
 }bdf 
n
[] 0 d
3.863708 M
1 w
0 j
0 J
false setoverprint
0 i
false eomode
[0 0 0 1] vc
vms
1819.5001 2156.025 m
2194.5001 2156.025 L
2194.5001 1781.025 L
1819.5001 1781.025 L
1819.5001 2156.025 L
0.75 w
s 
n
1812.0401 2165.1747 m
1803.0001 2165.1747 L
1803.0001 2146.175 L
1812.0401 2146.175 L
1812.0401 2165.1747 L
n
q
_bfh
_efh
{
f1 [18 0 0 18 0 0] makesetfont
1803.000122 2150.774963 m
0 0 32 0 0 (1) ts
}
true
[0 0 0 1]sts
Q
1812.7901 1797.6747 m
1803.7501 1797.6747 L
1803.7501 1778.675 L
1812.7901 1778.675 L
1812.7901 1797.6747 L
n
q
_bfh
_efh
{
f1 [18 0 0 18 0 0] makesetfont
1803.750122 1783.274963 m
0 0 32 0 0 (0) ts
}
true
[0 0 0 1]sts
Q
1826.2901 1777.4247 m
1817.2501 1777.4247 L
1817.2501 1758.425 L
1826.2901 1758.425 L
1826.2901 1777.4247 L
n
q
_bfh
_efh
{
f1 [18 0 0 18 0 0] makesetfont
1817.250122 1763.024963 m
0 0 32 0 0 (0) ts
}
true
[0 0 0 1]sts
Q
2052.0401 1776.4247 m
2034.0001 1776.4247 L
2034.0001 1757.425 L
2052.0401 1757.425 L
2052.0401 1776.4247 L
n
q
_bfh
_efh
{
f1 [18 0 0 18 0 0] makesetfont
2034.000122 1762.024963 m
0 0 32 0 0 (x*) ts
}
true
[0 0 0 1]sts
Q
2200.5401 1776.6747 m
2191.5001 1776.6747 L
2191.5001 1757.675 L
2200.5401 1757.675 L
2200.5401 1776.6747 L
n
q
_bfh
_efh
{
f1 [18 0 0 18 0 0] makesetfont
2191.500122 1762.274963 m
0 0 32 0 0 (1) ts
}
true
[0 0 0 1]sts
Q
2015.639 1713.8247 m
2004.799 1713.8247 L
2004.799 1691.145 L
2015.639 1691.145 L
2015.639 1713.8247 L
n
q
_bfh
_efh
{
f1 [21.599945 0 0 21.599945 0 0] makesetfont
2004.799042 1696.544983 m
0 -4.499985 rmoveto 0 0 32 0 0 (x) ts
0 4.499985 rmoveto }
true
[0 0 0 1]sts
Q
2290.5009 2098.3233 m
2290.5001 2098.3233 L
2290.5001 2053.9235 L
2290.5009 2053.9235 L
2290.5009 2098.3233 L
n
q
Q
1813.837 1983.85 m
1759.7588 1983.85 L
1759.7588 1959.2802 L
1813.837 1959.2802 L
1813.837 1983.85 L
n
q
_bfh
_efh
{
f1 [23.400055 0 0 23.400055 0 0] makesetfont
1759.758759 1965.130234 m
0 0 32 -0.13855 0 (R) ts
}
true
[0 0 0 1]sts
{
f1 [23.400055 0 0 23.400055 0 0] makesetfont
0 -7.800018 rmoveto 0 0 32 -0.13855 0 (1) ts
0 7.800018 rmoveto }
true
[0 0 0 1]sts
{
f1 [23.400055 0 0 23.400055 0 0] makesetfont
0 0 32 -0.13855 0 (\(x\)) ts
}
true
[0 0 0 1]sts
Q
1819.5001 2000.775 m
1929.0001 2156.025 L
3.863693 M
false setoverprint
S 
n
1929.0001 2156.025 m
2194.5001 1781.025 L
S 
n
1929.0001 1784.025 m
1929.0001 1777.275 L
S 
n
1939.5401 1777.4247 m
1921.5001 1777.4247 L
1921.5001 1758.425 L
1939.5401 1758.425 L
1939.5401 1777.4247 L
n
q
_bfh
_efh
{
f1 [18 0 0 18 0 0] makesetfont
1921.500122 1763.024963 m
0 0 32 0 0 (x*) ts
}
true
[0 0 0 1]sts
Q
1917.7501 1760.025 m
1937.2501 1760.025 L
false setoverprint
S 
n
1932.0401 1758.6747 m
1923.0001 1758.6747 L
1923.0001 1739.675 L
1932.0401 1739.675 L
1932.0401 1758.6747 L
n
q
_bfh
_efh
{
f1 [18 0 0 18 0 0] makesetfont
1923.000122 1744.274963 m
0 0 32 0 0 (2) ts
}
true
[0 0 0 1]sts
Q
1911.835 1743.8247 m
1896.1962 1743.8247 L
1896.1962 1721.145 L
1911.835 1721.145 L
1911.835 1743.8247 L
n
q
_bfh
_efh
{
f1 [21.599945 0 0 21.599945 0 0] makesetfont
1896.196243 1726.544983 m
0 0 32 0 0 (A) ts
}
true
[0 0 0 1]sts
Q
2117.7425 1745.3247 m
2103.2956 1745.3247 L
2103.2956 1722.645 L
2117.7425 1722.645 L
2117.7425 1745.3247 L
n
q
_bfh
_efh
{
f1 [21.599945 0 0 21.599945 0 0] makesetfont
2103.295563 1728.044998 m
0 0 32 0 0 (B) ts
}
true
[0 0 0 1]sts
Q
1818.5001 1782.025 m
2193.5001 2157.025 L
0.5 w
false setoverprint
S 
n
2038.5 2156 m
2038.5 1781 L
1 w
[4 4 ] 0 d
S 
[] 0 d
n
1830 1724 m
2028 1724 L
S 
n
true eomode
1830.4166 1724 m
1830.4166 1721.5 L
1824.1667 1724 L
1830.4166 1726.5 L
1830.4166 1724 L
f 
n
2027.5834 1724 m
2027.5834 1726.5 L
2033.8333 1724 L
2027.5834 1721.5 L
2027.5834 1724 L
f 
n
vmrs
2049 1724 m
2185 1724 L
3.863693 M
S 
n
true eomode
2049.4166 1724 m
2049.4166 1721.5 L
2043.1667 1724 L
2049.4166 1726.5 L
2049.4166 1724 L
f 
n
2184.5834 1724 m
2184.5834 1726.5 L
2190.8333 1724 L
2184.5834 1721.5 L
2184.5834 1724 L
f 
n
vmr
vmr
end
EndOfTheIncludedPostscriptMagicCookie

\closepsdump

\psdump{fig31.eps}
userdict begin /AltsysDict 245 dict def end
AltsysDict begin
/bdf{bind def}bind def
/xdf{exch def}bdf
/defed{where{pop true}{false}ifelse}bdf
/ndf{1 index where{pop pop pop}{dup xcheck{bind}if def}ifelse}bdf
/d{setdash}bdf
/h{closepath}bdf
/H{}bdf
/J{setlinecap}bdf
/j{setlinejoin}bdf
/M{setmiterlimit}bdf
/n{newpath}bdf
/N{newpath}bdf
/q{gsave}bdf
/Q{grestore}bdf
/w{setlinewidth}bdf
/sepdef{
 dup where not
 {
AltsysSepDict
 }
 if 
 3 1 roll exch put
}bdf
/st{settransfer}bdf
/colorimage defed /_rci xdf
/_NXLevel2 defed { 
 _NXLevel2 not {   
/colorimage where {
userdict eq {
/_rci false def 
} if
} if
 } if
} if
/md defed{ 
 md type /dicttype eq {  
/colorimage where { 
md eq { 
/_rci false def 
}if
}if
/settransfer where {
md eq {
/st systemdict /settransfer get def
}if
}if
 }if 
}if
/setstrokeadjust defed
{
 true setstrokeadjust
 /C{curveto}bdf
 /L{lineto}bdf
 /m{moveto}bdf
}
{
 /dr{transform .25 sub round .25 add 
exch .25 sub round .25 add exch itransform}bdf
 /C{dr curveto}bdf
 /L{dr lineto}bdf
 /m{dr moveto}bdf
 /setstrokeadjust{pop}bdf 
}ifelse
/rectstroke defed /xt xdf
xt {/yt save def} if
/privrectpath { 
 4 -2 roll m
 dtransform round exch round exch idtransform 
 2 copy 0 lt exch 0 lt xor
 {dup 0 exch rlineto exch 0 rlineto neg 0 exch rlineto}
 {exch dup 0 rlineto exch 0 exch rlineto neg 0 rlineto}
 ifelse
 closepath
}bdf
/rectclip{newpath privrectpath clip newpath}def
/rectfill{gsave newpath privrectpath fill grestore}def
/rectstroke{gsave newpath privrectpath stroke grestore}def
xt {yt restore} if
/_fonthacksave false def
/currentpacking defed 
{
 /_bfh {/_fonthacksave currentpacking def false setpacking} bdf
 /_efh {_fonthacksave setpacking} bdf
}
{
 /_bfh {} bdf
 /_efh {} bdf
}ifelse
/packedarray{array astore readonly}ndf
/` 
{ 
 false setoverprint

 /-save0- save def
 5 index concat
 pop
 storerect left bottom width height rectclip
 pop
 
 /dict_count countdictstack def
 /op_count count 1 sub def
 userdict begin
 
 /showpage {} def
 
 0 setgray 0 setlinecap 1 setlinewidth
 0 setlinejoin 10 setmiterlimit [] 0 setdash newpath
 
} bdf
/currentpacking defed{true setpacking}if
/min{2 copy gt{exch}if pop}bdf
/max{2 copy lt{exch}if pop}bdf
/xformfont { currentfont exch makefont setfont } bdf
/fhnumcolors 1 
 statusdict begin
/processcolors defed 
{
pop processcolors
}
{
/deviceinfo defed {
deviceinfo /Colors known {
pop deviceinfo /Colors get
} if
} if
} ifelse
 end 
def
/printerRes 
 gsave
 matrix defaultmatrix setmatrix
 72 72 dtransform
 abs exch abs
 max
 grestore
 def
/graycalcs
[
 {Angle Frequency}   
 {GrayAngle GrayFrequency} 
 {0 Width Height matrix defaultmatrix idtransform 
dup mul exch dup mul add sqrt 72 exch div} 
 {0 GrayWidth GrayHeight matrix defaultmatrix idtransform 
dup mul exch dup mul add sqrt 72 exch div} 
] def
/calcgraysteps {
 forcemaxsteps
 {
maxsteps
 }
 {
/currenthalftone defed
{currenthalftone /dicttype eq}{false}ifelse
{
currenthalftone begin
HalftoneType 4 le
{graycalcs HalftoneType 1 sub get exec}
{
HalftoneType 5 eq
{
Default begin
{graycalcs HalftoneType 1 sub get exec}
end
}
{0 60} 
ifelse
}
ifelse
end
}
{
currentscreen pop exch 
}
ifelse
 
printerRes 300 max exch div exch 
2 copy 
sin mul round dup mul 
3 1 roll 
cos mul round dup mul 
add 1 add 
dup maxsteps gt {pop maxsteps} if 
 }
 ifelse
} bdf
/nextrelease defed { 
 /languagelevel defed not {    
/framebuffer defed { 
0 40 string framebuffer 9 1 roll 8 {pop} repeat
dup 516 eq exch 520 eq or
{
/fhnumcolors 3 def
/currentscreen {60 0 {pop pop 1}}bdf
/calcgraysteps {maxsteps} bdf
}if
}if
 }if
}if
fhnumcolors 1 ne {
 /calcgraysteps {maxsteps} bdf
} if
/currentpagedevice defed {

 currentpagedevice /PreRenderingEnhance known
 {
currentpagedevice /PreRenderingEnhance get
{
/calcgraysteps 
{
forcemaxsteps 
{maxsteps}
{256 maxsteps min}
ifelse
} def
} if
 } if
} if
/gradfrequency 144 def
printerRes 1000 lt {
 /gradfrequency 72 def
} if
/adjnumsteps {
 
 dup dtransform abs exch abs max  
 
 printerRes div       
 
 gradfrequency mul      
 round        
 5 max       
 min        
}bdf
/goodsep {
 spots exch get 4 get dup sepname eq exch (_vc_Registration) eq or
}bdf
/BeginGradation defed
{/bb{BeginGradation}bdf}
{/bb{}bdf}
ifelse
/EndGradation defed
{/eb{EndGradation}bdf}
{/eb{}bdf}
ifelse
/bottom -0 def 
/delta -0 def 
/frac -0 def 
/height -0 def 
/left -0 def 
/numsteps1 -0 def 
/radius -0 def 
/right -0 def 
/top -0 def 
/width -0 def 
/xt -0 def 
/yt -0 def 
/df currentflat def 
/tempstr 1 string def 
/clipflatness currentflat def 
/inverted? 
 0 currenttransfer exec .5 ge def
/tc1 [0 0 0 1] def 
/tc2 [0 0 0 1] def 
/storerect{/top xdf /right xdf /bottom xdf /left xdf 
/width right left sub def /height top bottom sub def}bdf
/concatprocs{
 systemdict /packedarray known 
 {dup type /packedarraytype eq 2 index type /packedarraytype eq or}{false}ifelse
 { 
/proc2 exch cvlit def /proc1 exch cvlit def
proc1 aload pop proc2 aload pop
proc1 length proc2 length add packedarray cvx
 }
 { 
/proc2 exch cvlit def /proc1 exch cvlit def
/newproc proc1 length proc2 length add array def
newproc 0 proc1 putinterval newproc proc1 length proc2 putinterval
newproc cvx
 }ifelse
}bdf
/i{dup 0 eq
 {pop df dup} 
 {dup} ifelse 
 /clipflatness xdf setflat
}bdf
version cvr 38.0 le
{/setrgbcolor{
currenttransfer exec 3 1 roll
currenttransfer exec 3 1 roll
currenttransfer exec 3 1 roll
setrgbcolor}bdf}if
/vms {/vmsv save def} bdf
/vmr {vmsv restore} bdf
/vmrs{vmsv restore /vmsv save def}bdf
/eomode{ 
 {/filler /eofill load def /clipper /eoclip load def}
 {/filler /fill load def /clipper /clip load def}
 ifelse
}bdf
/normtaper{}bdf
/logtaper{9 mul 1 add log}bdf
/CD{
 /NF exch def 
 {    
exch dup 
/FID ne 1 index/UniqueID ne and
{exch NF 3 1 roll put}
{pop pop}
ifelse
 }forall 
 NF
}bdf
/MN{
 1 index length   
 /Len exch def 
 dup length Len add  
 string dup    
 Len     
 4 -1 roll    
 putinterval   
 dup     
 0      
 4 -1 roll   
 putinterval   
}bdf
/RC{4 -1 roll /ourvec xdf 256 string cvs(|______)anchorsearch
 {1 index MN cvn/NewN exch def cvn
 findfont dup maxlength dict CD dup/FontName NewN put dup
 /Encoding ourvec put NewN exch definefont pop}{pop}ifelse}bdf
/RF{ 
 dup      
 FontDirectory exch   
 known     
 {pop 3 -1 roll pop}  
 {RC}
 ifelse
}bdf
/FF{dup 256 string cvs(|______)exch MN cvn dup FontDirectory exch known
 {exch pop findfont 3 -1 roll pop}
 {pop dup findfont dup maxlength dict CD dup dup
 /Encoding exch /Encoding get 256 array copy 7 -1 roll 
 {3 -1 roll dup 4 -2 roll put}forall put definefont}
 ifelse}bdf
/RFJ{ 
 dup      
 FontDirectory exch   
 known     
 {pop 3 -1 roll pop  
 FontDirectory /Ryumin-Light-83pv-RKSJ-H known 
 {pop pop /Ryumin-Light-83pv-RKSJ-H dup}if  
 }      
 {RC}
 ifelse
}bdf
/FFJ{dup 256 string cvs(|______)exch MN cvn dup FontDirectory exch known
 {exch pop findfont 3 -1 roll pop}
 {pop
dup FontDirectory exch known not 
 {FontDirectory /Ryumin-Light-83pv-RKSJ-H known 
{pop /Ryumin-Light-83pv-RKSJ-H}if 
 }if            
 dup findfont dup maxlength dict CD dup dup
 /Encoding exch /Encoding get 256 array copy 7 -1 roll 
 {3 -1 roll dup 4 -2 roll put}forall put definefont}
 ifelse}bdf
/fps{
 currentflat   
 exch     
 dup 0 le{pop 1}if 
 {
dup setflat 3 index stopped
{1.3 mul dup 3 index gt{pop setflat pop pop stop}if} 
{exit} 
ifelse
 }loop 
 pop setflat pop pop
}bdf
/fp{100 currentflat fps}bdf
/clipper{clip}bdf 
/W{/clipper load 100 clipflatness dup setflat fps}bdf
userdict begin /BDFontDict 29 dict def end
BDFontDict begin
/bu{}def
/bn{}def
/setTxMode{av 70 ge{pop}if pop}def
/gm{m}def
/show{pop}def
/gr{pop}def
/fnt{pop pop pop}def
/fs{pop}def
/fz{pop}def
/lin{pop pop}def
/:M {pop pop} def
/sf {pop} def
/S {pop} def
/@b {pop pop pop pop pop pop pop pop} def
/_bdsave /save load def
/_bdrestore /restore load def
/save { dup /fontsave eq {null} {_bdsave} ifelse } def
/restore { dup null eq { pop } { _bdrestore } ifelse } def
/fontsave null def
end
/MacVec 256 array def 
MacVec 0 /Helvetica findfont
/Encoding get 0 128 getinterval putinterval
MacVec 127 /DEL put MacVec 16#27 /quotesingle put MacVec 16#60 /grave put
/NUL/SOH/STX/ETX/EOT/ENQ/ACK/BEL/BS/HT/LF/VT/FF/CR/SO/SI
/DLE/DC1/DC2/DC3/DC4/NAK/SYN/ETB/CAN/EM/SUB/ESC/FS/GS/RS/US
MacVec 0 32 getinterval astore pop
/Adieresis/Aring/Ccedilla/Eacute/Ntilde/Odieresis/Udieresis/aacute
/agrave/acircumflex/adieresis/atilde/aring/ccedilla/eacute/egrave
/ecircumflex/edieresis/iacute/igrave/icircumflex/idieresis/ntilde/oacute
/ograve/ocircumflex/odieresis/otilde/uacute/ugrave/ucircumflex/udieresis
/dagger/degree/cent/sterling/section/bullet/paragraph/germandbls
/registered/copyright/trademark/acute/dieresis/notequal/AE/Oslash
/infinity/plusminus/lessequal/greaterequal/yen/mu/partialdiff/summation
/product/pi/integral/ordfeminine/ordmasculine/Omega/ae/oslash 
/questiondown/exclamdown/logicalnot/radical/florin/approxequal/Delta/guillemotleft
/guillemotright/ellipsis/nbspace/Agrave/Atilde/Otilde/OE/oe
/endash/emdash/quotedblleft/quotedblright/quoteleft/quoteright/divide/lozenge
/ydieresis/Ydieresis/fraction/currency/guilsinglleft/guilsinglright/fi/fl
/daggerdbl/periodcentered/quotesinglbase/quotedblbase
/perthousand/Acircumflex/Ecircumflex/Aacute
/Edieresis/Egrave/Iacute/Icircumflex/Idieresis/Igrave/Oacute/Ocircumflex
/apple/Ograve/Uacute/Ucircumflex/Ugrave/dotlessi/circumflex/tilde
/macron/breve/dotaccent/ring/cedilla/hungarumlaut/ogonek/caron
MacVec 128 128 getinterval astore pop
end 
AltsysDict begin
_bfh
_efh
0 dict dup begin
end 
/f0 /Symbol FF def
_bfh
_efh
MacVec 256 array copy
/f1 /|______Times-Roman dup RF findfont def
end 
AltsysDict begin 
/onlyk4{false}ndf
/ccmyk{dup 5 -1 roll sub 0 max exch}ndf
/cmyk2gray{
 4 -1 roll 0.3 mul 4 -1 roll 0.59 mul 4 -1 roll 0.11 mul
 add add add 1 min neg 1 add
}bdf
/setcmykcolor{1 exch sub ccmyk ccmyk ccmyk pop setrgbcolor}ndf
/maxcolor { 
 max max max  
} ndf
/maxspot {
 pop
} ndf
/setcmykcoloroverprint{4{dup -1 eq{pop 0}if 4 1 roll}repeat setcmykcolor}ndf
/findcmykcustomcolor{5 packedarray}ndf
/setcustomcolor{exch aload pop pop 4{4 index mul 4 1 roll}repeat setcmykcolor pop}ndf
/setseparationgray{setgray}ndf
/setoverprint{pop}ndf 
/currentoverprint false ndf
/cmykbufs2gray{
 0 1 2 index length 1 sub
 { 
4 index 1 index get 0.3 mul 
4 index 2 index get 0.59 mul 
4 index 3 index get 0.11 mul 
4 index 4 index get 
add add add cvi 255 min
255 exch sub
2 index 3 1 roll put
 }for
 4 1 roll pop pop pop
}bdf
/colorimage{
 pop pop
 [
5 -1 roll/exec cvx 
6 -1 roll/exec cvx 
7 -1 roll/exec cvx 
8 -1 roll/exec cvx
/cmykbufs2gray cvx
 ]cvx 
 image
}
version cvr 47.1 le 
statusdict /product get (Lino) anchorsearch{pop pop true}{pop false}ifelse
and{userdict begin bdf end}{ndf}ifelse
fhnumcolors 1 ne {/yt save def} if
/customcolorimage{
 aload pop
 (_vc_Registration) eq 
 {
pop pop pop pop separationimage
 }
 {
/ik xdf /iy xdf /im xdf /ic xdf
ic im iy ik cmyk2gray /xt xdf
currenttransfer
{dup 1.0 exch sub xt mul add}concatprocs
st 
image
 }
 ifelse
}ndf
fhnumcolors 1 ne {yt restore} if
fhnumcolors 3 ne {/yt save def} if
/customcolorimage{
 aload pop 
 (_vc_Registration) eq 
 {
pop pop pop pop separationimage
 }
 {
/ik xdf /iy xdf /im xdf /ic xdf
1.0 dup ic ik add min sub 
1.0 dup im ik add min sub 
1.0 dup iy ik add min sub 
/ic xdf /iy xdf /im xdf
currentcolortransfer
4 1 roll 
{dup 1.0 exch sub ic mul add}concatprocs 4 1 roll 
{dup 1.0 exch sub iy mul add}concatprocs 4 1 roll 
{dup 1.0 exch sub im mul add}concatprocs 4 1 roll 
setcolortransfer
{/dummy xdf dummy}concatprocs{dummy}{dummy}true 3 colorimage
 }
 ifelse
}ndf
fhnumcolors 3 ne {yt restore} if
fhnumcolors 4 ne {/yt save def} if
/customcolorimage{
 aload pop
 (_vc_Registration) eq 
 {
pop pop pop pop separationimage
 }
 {
/ik xdf /iy xdf /im xdf /ic xdf
currentcolortransfer
{1.0 exch sub ik mul ik sub 1 add}concatprocs 4 1 roll
{1.0 exch sub iy mul iy sub 1 add}concatprocs 4 1 roll
{1.0 exch sub im mul im sub 1 add}concatprocs 4 1 roll
{1.0 exch sub ic mul ic sub 1 add}concatprocs 4 1 roll
setcolortransfer
{/dummy xdf dummy}concatprocs{dummy}{dummy}{dummy}
true 4 colorimage
 }
 ifelse
}ndf
fhnumcolors 4 ne {yt restore} if
/separationimage{image}ndf
/newcmykcustomcolor{6 packedarray}ndf
/inkoverprint false ndf
/setinkoverprint{pop}ndf 
/setspotcolor { 
 spots exch get
 dup 4 get (_vc_Registration) eq
 {pop 1 exch sub setseparationgray}
 {0 5 getinterval exch setcustomcolor}
 ifelse
}ndf
/currentcolortransfer{currenttransfer dup dup dup}ndf
/setcolortransfer{st pop pop pop}ndf
/fas{}ndf
/sas{}ndf
/fhsetspreadsize{pop}ndf
/filler{fill}bdf 
/F{gsave {filler}fp grestore}bdf
/f{closepath F}bdf
/S{gsave {stroke}fp grestore}bdf
/s{closepath S}bdf
/bc4 [0 0 0 0] def 
/_lfp4 {
 /iosv inkoverprint def
 /cosv currentoverprint def
 /yt xdf       
 /xt xdf       
 /ang xdf      
 storerect
 /taperfcn xdf
 /k2 xdf /y2 xdf /m2 xdf /c2 xdf
 /k1 xdf /y1 xdf /m1 xdf /c1 xdf
 c1 c2 sub abs
 m1 m2 sub abs
 y1 y2 sub abs
 k1 k2 sub abs
 maxcolor      
 calcgraysteps mul abs round  
 height abs adjnumsteps   
 dup 2 lt {pop 1} if    
 1 sub /numsteps1 xdf
 currentflat mark    
 currentflat clipflatness  
 /delta top bottom sub numsteps1 1 add div def 
 /right right left sub def  
 /botsv top delta sub def  
 {
{
W
xt yt translate 
ang rotate
xt neg yt neg translate 
dup setflat 
/bottom botsv def
0 1 numsteps1 
{
numsteps1 dup 0 eq {pop 0.5 } { div } ifelse 
taperfcn /frac xdf
bc4 0 c2 c1 sub frac mul c1 add put
bc4 1 m2 m1 sub frac mul m1 add put
bc4 2 y2 y1 sub frac mul y1 add put
bc4 3 k2 k1 sub frac mul k1 add put
bc4 vc
1 index setflat 
{ 
mark {newpath left bottom right delta rectfill}stopped
{cleartomark exch 1.3 mul dup setflat exch 2 copy gt{stop}if}
{cleartomark exit}ifelse
}loop
/bottom bottom delta sub def
}for
}
gsave stopped grestore
{exch pop 2 index exch 1.3 mul dup 100 gt{cleartomark setflat stop}if}
{exit}ifelse
 }loop
 cleartomark setflat
 iosv setinkoverprint
 cosv setoverprint
}bdf
/bcs [0 0] def 
/_lfs4 {
 /iosv inkoverprint def
 /cosv currentoverprint def
 /yt xdf       
 /xt xdf       
 /ang xdf      
 storerect
 /taperfcn xdf
 /tint2 xdf      
 /tint1 xdf      
 bcs exch 1 exch put    
 tint1 tint2 sub abs    
 bcs 1 get maxspot    
 calcgraysteps mul abs round  
 height abs adjnumsteps   
 dup 2 lt {pop 2} if    
 1 sub /numsteps1 xdf
 currentflat mark    
 currentflat clipflatness  
 /delta top bottom sub numsteps1 1 add div def 
 /right right left sub def  
 /botsv top delta sub def  
 {
{
W
xt yt translate 
ang rotate
xt neg yt neg translate 
dup setflat 
/bottom botsv def
0 1 numsteps1 
{
numsteps1 div taperfcn /frac xdf
bcs 0
1.0 tint2 tint1 sub frac mul tint1 add sub
put bcs vc
1 index setflat 
{ 
mark {newpath left bottom right delta rectfill}stopped
{cleartomark exch 1.3 mul dup setflat exch 2 copy gt{stop}if}
{cleartomark exit}ifelse
}loop
/bottom bottom delta sub def
}for
}
gsave stopped grestore
{exch pop 2 index exch 1.3 mul dup 100 gt{cleartomark setflat stop}if}
{exit}ifelse
 }loop
 cleartomark setflat
 iosv setinkoverprint
 cosv setoverprint
}bdf
/_rfs4 {
 /iosv inkoverprint def
 /cosv currentoverprint def
 /tint2 xdf      
 /tint1 xdf      
 bcs exch 1 exch put    
 /radius xdf      
 /yt xdf       
 /xt xdf       
 tint1 tint2 sub abs    
 bcs 1 get maxspot    
 calcgraysteps mul abs round  
 radius abs adjnumsteps   
 dup 2 lt {pop 2} if    
 1 sub /numsteps1 xdf
 radius numsteps1 div 2 div /halfstep xdf 
 currentflat mark    
 currentflat clipflatness  
 {
{
dup setflat 
W 
0 1 numsteps1 
{
dup /radindex xdf
numsteps1 div /frac xdf
bcs 0
tint2 tint1 sub frac mul tint1 add
put bcs vc
1 index setflat 
{ 
newpath mark xt yt radius 1 frac sub mul halfstep add 0 360
{ arc
radindex numsteps1 ne 
{
xt yt 
radindex 1 add numsteps1 
div 1 exch sub 
radius mul halfstep add
dup xt add yt moveto 
360 0 arcn 
} if
fill
}stopped
{cleartomark exch 1.3 mul dup setflat exch 2 copy gt{stop}if}
{cleartomark exit}ifelse
}loop
}for
}
gsave stopped grestore
{exch pop 2 index exch 1.3 mul dup 100 gt{cleartomark setflat stop}if}
{exit}ifelse
 }loop
 cleartomark setflat
 iosv setinkoverprint
 cosv setoverprint
}bdf
/_rfp4 {
 /iosv inkoverprint def
 /cosv currentoverprint def
 /k2 xdf /y2 xdf /m2 xdf /c2 xdf
 /k1 xdf /y1 xdf /m1 xdf /c1 xdf
 /radius xdf      
 /yt xdf       
 /xt xdf       
 c1 c2 sub abs
 m1 m2 sub abs
 y1 y2 sub abs
 k1 k2 sub abs
 maxcolor      
 calcgraysteps mul abs round  
 radius abs adjnumsteps   
 dup 2 lt {pop 1} if    
 1 sub /numsteps1 xdf
 radius numsteps1 dup 0 eq {pop} {div} ifelse 
 2 div /halfstep xdf 
 currentflat mark    
 currentflat clipflatness  
 {
{
dup setflat 
W 
0 1 numsteps1 
{
dup /radindex xdf
numsteps1 dup 0 eq {pop 0.5 } { div } ifelse 
/frac xdf
bc4 0 c2 c1 sub frac mul c1 add put
bc4 1 m2 m1 sub frac mul m1 add put
bc4 2 y2 y1 sub frac mul y1 add put
bc4 3 k2 k1 sub frac mul k1 add put
bc4 vc
1 index setflat 
{ 
newpath mark xt yt radius 1 frac sub mul halfstep add 0 360
{ arc
radindex numsteps1 ne 
{
xt yt 
radindex 1 add 
numsteps1 dup 0 eq {pop} {div} ifelse 
1 exch sub 
radius mul halfstep add
dup xt add yt moveto 
360 0 arcn 
} if
fill
}stopped
{cleartomark exch 1.3 mul dup setflat exch 2 copy gt{stop}if}
{cleartomark exit}ifelse
}loop
}for
}
gsave stopped grestore
{exch pop 2 index exch 1.3 mul dup 100 gt{cleartomark setflat stop}if}
{exit}ifelse
 }loop
 cleartomark setflat
 iosv setinkoverprint
 cosv setoverprint
}bdf
/lfp4{_lfp4}ndf
/lfs4{_lfs4}ndf
/rfs4{_rfs4}ndf
/rfp4{_rfp4}ndf
/cvc [0 0 0 1] def 
/vc{
 AltsysDict /cvc 2 index put 
 aload length 4 eq
 {setcmykcolor}
 {setspotcolor}
 ifelse
}bdf 
/origmtx matrix currentmatrix def
/ImMatrix matrix currentmatrix def
0 setseparationgray
/imgr {1692 1584 2304 2376 } def 
/bleed 0 def 
/clpr {1692 1584 2304 2376 } def 
/xs 1 def 
/ys 1 def 
/botx 0 def 
/overlap 0 def 
/wdist 18 def 
0 2 mul fhsetspreadsize 
0 0 ne {/df 0 def /clipflatness 0 def} if 
/maxsteps 256 def 
/forcemaxsteps false def 
vms
-1752 -1699 translate
/currentpacking defed{false setpacking}if 
/spots[
1 0 0 0 (Process Cyan) false newcmykcustomcolor
0 1 0 0 (Process Magenta) false newcmykcustomcolor
0 0 1 0 (Process Yellow) false newcmykcustomcolor
0 0 0 1 (Process Black) false newcmykcustomcolor
]def
/textopf false def
/curtextmtx{}def
/otw .25 def
/msf{dup/curtextmtx xdf makefont setfont}bdf
/makesetfont/msf load def
/curtextheight{.707104 .707104 curtextmtx dtransform
 dup mul exch dup mul add sqrt}bdf
/ta2{ 
tempstr 2 index gsave exec grestore 
cwidth cheight rmoveto 
4 index eq{5 index 5 index rmoveto}if 
2 index 2 index rmoveto 
}bdf
/ta{exch systemdict/cshow known
{{/cheight xdf/cwidth xdf tempstr 0 2 index put ta2}exch cshow} 
{{tempstr 0 2 index put tempstr stringwidth/cheight xdf/cwidth xdf ta2}forall} 
ifelse 6{pop}repeat}bdf
/sts{/textopf currentoverprint def vc setoverprint
/ts{awidthshow}def exec textopf setoverprint}bdf
/stol{/xt currentlinewidth def 
 setlinewidth vc newpath 
 /ts{{false charpath stroke}ta}def exec 
 xt setlinewidth}bdf 
 
/strk{/textopf currentoverprint def vc setoverprint
 /ts{{false charpath stroke}ta}def exec 
 textopf setoverprint
 }bdf 
n
[] 0 d
3.863708 M
1 w
0 j
0 J
false setoverprint
0 i
false eomode
[0 0 0 1] vc
vms
1805.0521 2155.1735 m
2180.255 2155.1735 L
2180.255 1780.3764 L
1805.0521 1780.3764 L
1805.0521 2155.1735 L
0.75 w
s 
n
1805.255 2000.025 m
1914.755 1780.275 L
S 
n
1914.755 1780.275 m
2102.255 2155.1735 L
S 
n
2102.255 2155.1735 m
2180.255 2000.025 L
3.863693 M
S 
n
1805.255 2000.025 m
2180.255 2000.025 L
[4 4 ] 0 d
S 
[] 0 d
n
2024.255 2155.275 m
2024.255 1779.525 L
[4 4 ] 0 d
S 
[] 0 d
n
1938.775 1724.1747 m
1934.255 1724.1747 L
1934.255 1705.175 L
1938.775 1705.175 L
1938.775 1724.1747 L
n
q
Q
2084.255 1760.775 m
2121.4363 1760.775 L
false setoverprint
S 
n
2106.045 1756.4247 m
2097.005 1756.4247 L
2097.005 1737.425 L
2106.045 1737.425 L
2106.045 1756.4247 L
n
q
_bfh
_efh
{
f1 [18 0 0 18 0 0] makesetfont
2097.00499 1742.024963 m
0 0 32 0 0 (2) ts
}
true
[0 0 0 1]sts
Q
2037.295 1775.1747 m
2019.255 1775.1747 L
2019.255 1756.175 L
2037.295 1756.175 L
2037.295 1775.1747 L
n
q
_bfh
_efh
{
f1 [18 0 0 18 0 0] makesetfont
2019.25499 1760.774963 m
0 0 32 0 0 (x*) ts
}
true
[0 0 0 1]sts
Q
1801.795 2006.1747 m
1783.755 2006.1747 L
1783.755 1987.175 L
1801.795 1987.175 L
1801.795 2006.1747 L
n
q
_bfh
_efh
{
f1 [18 0 0 18 0 0] makesetfont
1783.75499 1991.774963 m
0 0 32 0 0 (x*) ts
}
true
[0 0 0 1]sts
Q
1797.795 1791.6747 m
1788.755 1791.6747 L
1788.755 1772.675 L
1797.795 1772.675 L
1797.795 1791.6747 L
n
q
_bfh
_efh
{
f1 [18 0 0 18 0 0] makesetfont
1788.75499 1777.274963 m
0 0 32 0 0 (0) ts
}
true
[0 0 0 1]sts
Q
1797.795 2160.6747 m
1788.755 2160.6747 L
1788.755 2141.675 L
1797.795 2141.675 L
1797.795 2160.6747 L
n
q
_bfh
_efh
{
f1 [18 0 0 18 0 0] makesetfont
1788.75499 2146.274963 m
0 0 32 0 0 (1) ts
}
true
[0 0 0 1]sts
Q
2190.795 1773.6747 m
2181.755 1773.6747 L
2181.755 1754.675 L
2190.795 1754.675 L
2190.795 1773.6747 L
n
q
_bfh
_efh
{
f1 [18 0 0 18 0 0] makesetfont
2181.75499 1759.274963 m
0 0 32 0 0 (1) ts
}
true
[0 0 0 1]sts
Q
1815.795 1773.6747 m
1806.755 1773.6747 L
1806.755 1754.675 L
1815.795 1754.675 L
1815.795 1773.6747 L
n
q
_bfh
_efh
{
f1 [18 0 0 18 0 0] makesetfont
1806.75499 1759.274963 m
0 0 32 0 0 (0) ts
}
true
[0 0 0 1]sts
Q
2102.255 1786.275 m
2102.255 1774.275 L
false setoverprint
S 
n
1931.5917 1778.1747 m
1910.255 1778.1747 L
1910.255 1727.84 L
1931.5917 1727.84 L
1931.5917 1778.1747 L
n
q
q
[1 0 0 1 1910.25499 1737.907547] concat
[0.75 0 0 0.75 0 0] concat
vms
q
vms
25.04 50.335 m
1 50.335 L
1 25.1353 L
25.04 25.1353 L
25.04 50.335 L
n
q
_bfh
_efh
{
f1 [24 0 0 24 0 0] makesetfont
1 31.135269 m
0 0 32 0 0 (x*) ts
}
true
[0 0 0 1]sts
Q
-3 27 m
21 27 L
1 w
false setoverprint
S 
n
12.04 25.1997 m
0 25.1997 L
0 0 L
12.04 0 L
12.04 25.1997 L
n
q
_bfh
_efh
{
f1 [24 0 0 24 0 0] makesetfont
0 5.999985 m
0 0 32 0 0 (2) ts
}
true
[0 0 0 1]sts
Q
vmr
Q
vmr
Q
Q
2124.4463 1776.6747 m
2087.255 1776.6747 L
2087.255 1757.675 L
2124.4463 1757.675 L
2124.4463 1776.6747 L
n
q
_bfh
_efh
{
f1 [18 0 0 18 0 0] makesetfont
2087.25499 1762.274963 m
0 0 32 0 0 (1+x*) ts
}
true
[0 0 0 1]sts
Q
2002.6909 1722.3247 m
1991.851 1722.3247 L
1991.851 1699.645 L
2002.6909 1699.645 L
2002.6909 1722.3247 L
n
q
_bfh
_efh
{
f1 [21.599945 0 0 21.599945 0 0] makesetfont
1991.850983 1705.044998 m
0 0 32 0 0 (x) ts
}
true
[0 0 0 1]sts
Q
1901.8398 1749.0747 m
1886.2011 1749.0747 L
1886.2011 1726.395 L
1901.8398 1726.395 L
1901.8398 1749.0747 L
n
q
_bfh
_efh
{
f1 [21.599945 0 0 21.599945 0 0] makesetfont
1886.20108 1731.794998 m
0 0 32 0 0 (A) ts
}
true
[0 0 0 1]sts
Q
1 w
false setoverprint
S 
n
q
_bfh
_efh
{
f1 [21.599945 0 0 21.599945 0 0] makesetfont
2122.8004 1731.545029 m
0 0 32 0 0 (B) ts
}
true
[0 0 0 1]sts
Q
1802.3744 1974.4232 m
1758.1508 1974.4232 L
1758.1508 1949.8535 L
1802.3744 1949.8535 L
1802.3744 1974.4232 L
n
q
_bfh
_efh
{
f1 [23.400055 0 0 23.400055 0 0] makesetfont
1758.150848 1955.703491 m
0 0 32 0 0 (G\(x\)) ts
}
true
[0 0 0 1]sts
Q
1815 1729 m
2015 1729 L
false setoverprint
S 
n
true eomode
1815.4166 1729 m
1815.4166 1726.5 L
1809.1667 1729 L
1815.4166 1731.5 L
1815.4166 1729 L
f 
n
2014.5834 1729 m
2014.5834 1731.5 L
2020.8333 1729 L
2014.5834 1726.5 L
2014.5834 1729 L
f 
n
2036 1729 m
2171 1729 L
S 
n
2036.4166 1729 m
2036.4166 1726.5 L
2030.1667 1729 L
2036.4166 1731.5 L
2036.4166 1729 L
f 
n
2170.5834 1729 m
2170.5834 1731.5 L
2176.8333 1729 L
2170.5834 1726.5 L
2170.5834 1729 L
f 
n
vmr
vmr
end
EndOfTheIncludedPostscriptMagicCookie

\closepsdump

\psdump{fig41.eps}
userdict begin /AltsysDict 245 dict def end
AltsysDict begin
/bdf{bind def}bind def
/xdf{exch def}bdf
/defed{where{pop true}{false}ifelse}bdf
/ndf{1 index where{pop pop pop}{dup xcheck{bind}if def}ifelse}bdf
/d{setdash}bdf
/h{closepath}bdf
/H{}bdf
/J{setlinecap}bdf
/j{setlinejoin}bdf
/M{setmiterlimit}bdf
/n{newpath}bdf
/N{newpath}bdf
/q{gsave}bdf
/Q{grestore}bdf
/w{setlinewidth}bdf
/sepdef{
 dup where not
 {
AltsysSepDict
 }
 if 
 3 1 roll exch put
}bdf
/st{settransfer}bdf
/colorimage defed /_rci xdf
/_NXLevel2 defed { 
 _NXLevel2 not {   
/colorimage where {
userdict eq {
/_rci false def 
} if
} if
 } if
} if
/md defed{ 
 md type /dicttype eq {  
/colorimage where { 
md eq { 
/_rci false def 
}if
}if
/settransfer where {
md eq {
/st systemdict /settransfer get def
}if
}if
 }if 
}if
/setstrokeadjust defed
{
 true setstrokeadjust
 /C{curveto}bdf
 /L{lineto}bdf
 /m{moveto}bdf
}
{
 /dr{transform .25 sub round .25 add 
exch .25 sub round .25 add exch itransform}bdf
 /C{dr curveto}bdf
 /L{dr lineto}bdf
 /m{dr moveto}bdf
 /setstrokeadjust{pop}bdf 
}ifelse
/rectstroke defed /xt xdf
xt {/yt save def} if
/privrectpath { 
 4 -2 roll m
 dtransform round exch round exch idtransform 
 2 copy 0 lt exch 0 lt xor
 {dup 0 exch rlineto exch 0 rlineto neg 0 exch rlineto}
 {exch dup 0 rlineto exch 0 exch rlineto neg 0 rlineto}
 ifelse
 closepath
}bdf
/rectclip{newpath privrectpath clip newpath}def
/rectfill{gsave newpath privrectpath fill grestore}def
/rectstroke{gsave newpath privrectpath stroke grestore}def
xt {yt restore} if
/_fonthacksave false def
/currentpacking defed 
{
 /_bfh {/_fonthacksave currentpacking def false setpacking} bdf
 /_efh {_fonthacksave setpacking} bdf
}
{
 /_bfh {} bdf
 /_efh {} bdf
}ifelse
/packedarray{array astore readonly}ndf
/` 
{ 
 false setoverprint

 /-save0- save def
 5 index concat
 pop
 storerect left bottom width height rectclip
 pop
 
 /dict_count countdictstack def
 /op_count count 1 sub def
 userdict begin
 
 /showpage {} def
 
 0 setgray 0 setlinecap 1 setlinewidth
 0 setlinejoin 10 setmiterlimit [] 0 setdash newpath
 
} bdf
/currentpacking defed{true setpacking}if
/min{2 copy gt{exch}if pop}bdf
/max{2 copy lt{exch}if pop}bdf
/xformfont { currentfont exch makefont setfont } bdf
/fhnumcolors 1 
 statusdict begin
/processcolors defed 
{
pop processcolors
}
{
/deviceinfo defed {
deviceinfo /Colors known {
pop deviceinfo /Colors get
} if
} if
} ifelse
 end 
def
/printerRes 
 gsave
 matrix defaultmatrix setmatrix
 72 72 dtransform
 abs exch abs
 max
 grestore
 def
/graycalcs
[
 {Angle Frequency}   
 {GrayAngle GrayFrequency} 
 {0 Width Height matrix defaultmatrix idtransform 
dup mul exch dup mul add sqrt 72 exch div} 
 {0 GrayWidth GrayHeight matrix defaultmatrix idtransform 
dup mul exch dup mul add sqrt 72 exch div} 
] def
/calcgraysteps {
 forcemaxsteps
 {
maxsteps
 }
 {
/currenthalftone defed
{currenthalftone /dicttype eq}{false}ifelse
{
currenthalftone begin
HalftoneType 4 le
{graycalcs HalftoneType 1 sub get exec}
{
HalftoneType 5 eq
{
Default begin
{graycalcs HalftoneType 1 sub get exec}
end
}
{0 60} 
ifelse
}
ifelse
end
}
{
currentscreen pop exch 
}
ifelse
 
printerRes 300 max exch div exch 
2 copy 
sin mul round dup mul 
3 1 roll 
cos mul round dup mul 
add 1 add 
dup maxsteps gt {pop maxsteps} if 
 }
 ifelse
} bdf
/nextrelease defed { 
 /languagelevel defed not {    
/framebuffer defed { 
0 40 string framebuffer 9 1 roll 8 {pop} repeat
dup 516 eq exch 520 eq or
{
/fhnumcolors 3 def
/currentscreen {60 0 {pop pop 1}}bdf
/calcgraysteps {maxsteps} bdf
}if
}if
 }if
}if
fhnumcolors 1 ne {
 /calcgraysteps {maxsteps} bdf
} if
/currentpagedevice defed {

 currentpagedevice /PreRenderingEnhance known
 {
currentpagedevice /PreRenderingEnhance get
{
/calcgraysteps 
{
forcemaxsteps 
{maxsteps}
{256 maxsteps min}
ifelse
} def
} if
 } if
} if
/gradfrequency 144 def
printerRes 1000 lt {
 /gradfrequency 72 def
} if
/adjnumsteps {
 
 dup dtransform abs exch abs max  
 
 printerRes div       
 
 gradfrequency mul      
 round        
 5 max       
 min        
}bdf
/goodsep {
 spots exch get 4 get dup sepname eq exch (_vc_Registration) eq or
}bdf
/BeginGradation defed
{/bb{BeginGradation}bdf}
{/bb{}bdf}
ifelse
/EndGradation defed
{/eb{EndGradation}bdf}
{/eb{}bdf}
ifelse
/bottom -0 def 
/delta -0 def 
/frac -0 def 
/height -0 def 
/left -0 def 
/numsteps1 -0 def 
/radius -0 def 
/right -0 def 
/top -0 def 
/width -0 def 
/xt -0 def 
/yt -0 def 
/df currentflat def 
/tempstr 1 string def 
/clipflatness currentflat def 
/inverted? 
 0 currenttransfer exec .5 ge def
/tc1 [0 0 0 1] def 
/tc2 [0 0 0 1] def 
/storerect{/top xdf /right xdf /bottom xdf /left xdf 
/width right left sub def /height top bottom sub def}bdf
/concatprocs{
 systemdict /packedarray known 
 {dup type /packedarraytype eq 2 index type /packedarraytype eq or}{false}ifelse
 { 
/proc2 exch cvlit def /proc1 exch cvlit def
proc1 aload pop proc2 aload pop
proc1 length proc2 length add packedarray cvx
 }
 { 
/proc2 exch cvlit def /proc1 exch cvlit def
/newproc proc1 length proc2 length add array def
newproc 0 proc1 putinterval newproc proc1 length proc2 putinterval
newproc cvx
 }ifelse
}bdf
/i{dup 0 eq
 {pop df dup} 
 {dup} ifelse 
 /clipflatness xdf setflat
}bdf
version cvr 38.0 le
{/setrgbcolor{
currenttransfer exec 3 1 roll
currenttransfer exec 3 1 roll
currenttransfer exec 3 1 roll
setrgbcolor}bdf}if
/vms {/vmsv save def} bdf
/vmr {vmsv restore} bdf
/vmrs{vmsv restore /vmsv save def}bdf
/eomode{ 
 {/filler /eofill load def /clipper /eoclip load def}
 {/filler /fill load def /clipper /clip load def}
 ifelse
}bdf
/normtaper{}bdf
/logtaper{9 mul 1 add log}bdf
/CD{
 /NF exch def 
 {    
exch dup 
/FID ne 1 index/UniqueID ne and
{exch NF 3 1 roll put}
{pop pop}
ifelse
 }forall 
 NF
}bdf
/MN{
 1 index length   
 /Len exch def 
 dup length Len add  
 string dup    
 Len     
 4 -1 roll    
 putinterval   
 dup     
 0      
 4 -1 roll   
 putinterval   
}bdf
/RC{4 -1 roll /ourvec xdf 256 string cvs(|______)anchorsearch
 {1 index MN cvn/NewN exch def cvn
 findfont dup maxlength dict CD dup/FontName NewN put dup
 /Encoding ourvec put NewN exch definefont pop}{pop}ifelse}bdf
/RF{ 
 dup      
 FontDirectory exch   
 known     
 {pop 3 -1 roll pop}  
 {RC}
 ifelse
}bdf
/FF{dup 256 string cvs(|______)exch MN cvn dup FontDirectory exch known
 {exch pop findfont 3 -1 roll pop}
 {pop dup findfont dup maxlength dict CD dup dup
 /Encoding exch /Encoding get 256 array copy 7 -1 roll 
 {3 -1 roll dup 4 -2 roll put}forall put definefont}
 ifelse}bdf
/RFJ{ 
 dup      
 FontDirectory exch   
 known     
 {pop 3 -1 roll pop  
 FontDirectory /Ryumin-Light-83pv-RKSJ-H known 
 {pop pop /Ryumin-Light-83pv-RKSJ-H dup}if  
 }      
 {RC}
 ifelse
}bdf
/FFJ{dup 256 string cvs(|______)exch MN cvn dup FontDirectory exch known
 {exch pop findfont 3 -1 roll pop}
 {pop
dup FontDirectory exch known not 
 {FontDirectory /Ryumin-Light-83pv-RKSJ-H known 
{pop /Ryumin-Light-83pv-RKSJ-H}if 
 }if            
 dup findfont dup maxlength dict CD dup dup
 /Encoding exch /Encoding get 256 array copy 7 -1 roll 
 {3 -1 roll dup 4 -2 roll put}forall put definefont}
 ifelse}bdf
/fps{
 currentflat   
 exch     
 dup 0 le{pop 1}if 
 {
dup setflat 3 index stopped
{1.3 mul dup 3 index gt{pop setflat pop pop stop}if} 
{exit} 
ifelse
 }loop 
 pop setflat pop pop
}bdf
/fp{100 currentflat fps}bdf
/clipper{clip}bdf 
/W{/clipper load 100 clipflatness dup setflat fps}bdf
userdict begin /BDFontDict 29 dict def end
BDFontDict begin
/bu{}def
/bn{}def
/setTxMode{av 70 ge{pop}if pop}def
/gm{m}def
/show{pop}def
/gr{pop}def
/fnt{pop pop pop}def
/fs{pop}def
/fz{pop}def
/lin{pop pop}def
/:M {pop pop} def
/sf {pop} def
/S {pop} def
/@b {pop pop pop pop pop pop pop pop} def
/_bdsave /save load def
/_bdrestore /restore load def
/save { dup /fontsave eq {null} {_bdsave} ifelse } def
/restore { dup null eq { pop } { _bdrestore } ifelse } def
/fontsave null def
end
/MacVec 256 array def 
MacVec 0 /Helvetica findfont
/Encoding get 0 128 getinterval putinterval
MacVec 127 /DEL put MacVec 16#27 /quotesingle put MacVec 16#60 /grave put
/NUL/SOH/STX/ETX/EOT/ENQ/ACK/BEL/BS/HT/LF/VT/FF/CR/SO/SI
/DLE/DC1/DC2/DC3/DC4/NAK/SYN/ETB/CAN/EM/SUB/ESC/FS/GS/RS/US
MacVec 0 32 getinterval astore pop
/Adieresis/Aring/Ccedilla/Eacute/Ntilde/Odieresis/Udieresis/aacute
/agrave/acircumflex/adieresis/atilde/aring/ccedilla/eacute/egrave
/ecircumflex/edieresis/iacute/igrave/icircumflex/idieresis/ntilde/oacute
/ograve/ocircumflex/odieresis/otilde/uacute/ugrave/ucircumflex/udieresis
/dagger/degree/cent/sterling/section/bullet/paragraph/germandbls
/registered/copyright/trademark/acute/dieresis/notequal/AE/Oslash
/infinity/plusminus/lessequal/greaterequal/yen/mu/partialdiff/summation
/product/pi/integral/ordfeminine/ordmasculine/Omega/ae/oslash 
/questiondown/exclamdown/logicalnot/radical/florin/approxequal/Delta/guillemotleft
/guillemotright/ellipsis/nbspace/Agrave/Atilde/Otilde/OE/oe
/endash/emdash/quotedblleft/quotedblright/quoteleft/quoteright/divide/lozenge
/ydieresis/Ydieresis/fraction/currency/guilsinglleft/guilsinglright/fi/fl
/daggerdbl/periodcentered/quotesinglbase/quotedblbase
/perthousand/Acircumflex/Ecircumflex/Aacute
/Edieresis/Egrave/Iacute/Icircumflex/Idieresis/Igrave/Oacute/Ocircumflex
/apple/Ograve/Uacute/Ucircumflex/Ugrave/dotlessi/circumflex/tilde
/macron/breve/dotaccent/ring/cedilla/hungarumlaut/ogonek/caron
MacVec 128 128 getinterval astore pop
end 
AltsysDict begin
_bfh
_efh
0 dict dup begin
end 
/f0 /Symbol FF def
_bfh
_efh
MacVec 256 array copy
/f1 /|______Times-Roman dup RF findfont def
end 
AltsysDict begin 
/onlyk4{false}ndf
/ccmyk{dup 5 -1 roll sub 0 max exch}ndf
/cmyk2gray{
 4 -1 roll 0.3 mul 4 -1 roll 0.59 mul 4 -1 roll 0.11 mul
 add add add 1 min neg 1 add
}bdf
/setcmykcolor{1 exch sub ccmyk ccmyk ccmyk pop setrgbcolor}ndf
/maxcolor { 
 max max max  
} ndf
/maxspot {
 pop
} ndf
/setcmykcoloroverprint{4{dup -1 eq{pop 0}if 4 1 roll}repeat setcmykcolor}ndf
/findcmykcustomcolor{5 packedarray}ndf
/setcustomcolor{exch aload pop pop 4{4 index mul 4 1 roll}repeat setcmykcolor pop}ndf
/setseparationgray{setgray}ndf
/setoverprint{pop}ndf 
/currentoverprint false ndf
/cmykbufs2gray{
 0 1 2 index length 1 sub
 { 
4 index 1 index get 0.3 mul 
4 index 2 index get 0.59 mul 
4 index 3 index get 0.11 mul 
4 index 4 index get 
add add add cvi 255 min
255 exch sub
2 index 3 1 roll put
 }for
 4 1 roll pop pop pop
}bdf
/colorimage{
 pop pop
 [
5 -1 roll/exec cvx 
6 -1 roll/exec cvx 
7 -1 roll/exec cvx 
8 -1 roll/exec cvx
/cmykbufs2gray cvx
 ]cvx 
 image
}
version cvr 47.1 le 
statusdict /product get (Lino) anchorsearch{pop pop true}{pop false}ifelse
and{userdict begin bdf end}{ndf}ifelse
fhnumcolors 1 ne {/yt save def} if
/customcolorimage{
 aload pop
 (_vc_Registration) eq 
 {
pop pop pop pop separationimage
 }
 {
/ik xdf /iy xdf /im xdf /ic xdf
ic im iy ik cmyk2gray /xt xdf
currenttransfer
{dup 1.0 exch sub xt mul add}concatprocs
st 
image
 }
 ifelse
}ndf
fhnumcolors 1 ne {yt restore} if
fhnumcolors 3 ne {/yt save def} if
/customcolorimage{
 aload pop 
 (_vc_Registration) eq 
 {
pop pop pop pop separationimage
 }
 {
/ik xdf /iy xdf /im xdf /ic xdf
1.0 dup ic ik add min sub 
1.0 dup im ik add min sub 
1.0 dup iy ik add min sub 
/ic xdf /iy xdf /im xdf
currentcolortransfer
4 1 roll 
{dup 1.0 exch sub ic mul add}concatprocs 4 1 roll 
{dup 1.0 exch sub iy mul add}concatprocs 4 1 roll 
{dup 1.0 exch sub im mul add}concatprocs 4 1 roll 
setcolortransfer
{/dummy xdf dummy}concatprocs{dummy}{dummy}true 3 colorimage
 }
 ifelse
}ndf
fhnumcolors 3 ne {yt restore} if
fhnumcolors 4 ne {/yt save def} if
/customcolorimage{
 aload pop
 (_vc_Registration) eq 
 {
pop pop pop pop separationimage
 }
 {
/ik xdf /iy xdf /im xdf /ic xdf
currentcolortransfer
{1.0 exch sub ik mul ik sub 1 add}concatprocs 4 1 roll
{1.0 exch sub iy mul iy sub 1 add}concatprocs 4 1 roll
{1.0 exch sub im mul im sub 1 add}concatprocs 4 1 roll
{1.0 exch sub ic mul ic sub 1 add}concatprocs 4 1 roll
setcolortransfer
{/dummy xdf dummy}concatprocs{dummy}{dummy}{dummy}
true 4 colorimage
 }
 ifelse
}ndf
fhnumcolors 4 ne {yt restore} if
/separationimage{image}ndf
/newcmykcustomcolor{6 packedarray}ndf
/inkoverprint false ndf
/setinkoverprint{pop}ndf 
/setspotcolor { 
 spots exch get
 dup 4 get (_vc_Registration) eq
 {pop 1 exch sub setseparationgray}
 {0 5 getinterval exch setcustomcolor}
 ifelse
}ndf
/currentcolortransfer{currenttransfer dup dup dup}ndf
/setcolortransfer{st pop pop pop}ndf
/fas{}ndf
/sas{}ndf
/fhsetspreadsize{pop}ndf
/filler{fill}bdf 
/F{gsave {filler}fp grestore}bdf
/f{closepath F}bdf
/S{gsave {stroke}fp grestore}bdf
/s{closepath S}bdf
/bc4 [0 0 0 0] def 
/_lfp4 {
 /iosv inkoverprint def
 /cosv currentoverprint def
 /yt xdf       
 /xt xdf       
 /ang xdf      
 storerect
 /taperfcn xdf
 /k2 xdf /y2 xdf /m2 xdf /c2 xdf
 /k1 xdf /y1 xdf /m1 xdf /c1 xdf
 c1 c2 sub abs
 m1 m2 sub abs
 y1 y2 sub abs
 k1 k2 sub abs
 maxcolor      
 calcgraysteps mul abs round  
 height abs adjnumsteps   
 dup 2 lt {pop 1} if    
 1 sub /numsteps1 xdf
 currentflat mark    
 currentflat clipflatness  
 /delta top bottom sub numsteps1 1 add div def 
 /right right left sub def  
 /botsv top delta sub def  
 {
{
W
xt yt translate 
ang rotate
xt neg yt neg translate 
dup setflat 
/bottom botsv def
0 1 numsteps1 
{
numsteps1 dup 0 eq {pop 0.5 } { div } ifelse 
taperfcn /frac xdf
bc4 0 c2 c1 sub frac mul c1 add put
bc4 1 m2 m1 sub frac mul m1 add put
bc4 2 y2 y1 sub frac mul y1 add put
bc4 3 k2 k1 sub frac mul k1 add put
bc4 vc
1 index setflat 
{ 
mark {newpath left bottom right delta rectfill}stopped
{cleartomark exch 1.3 mul dup setflat exch 2 copy gt{stop}if}
{cleartomark exit}ifelse
}loop
/bottom bottom delta sub def
}for
}
gsave stopped grestore
{exch pop 2 index exch 1.3 mul dup 100 gt{cleartomark setflat stop}if}
{exit}ifelse
 }loop
 cleartomark setflat
 iosv setinkoverprint
 cosv setoverprint
}bdf
/bcs [0 0] def 
/_lfs4 {
 /iosv inkoverprint def
 /cosv currentoverprint def
 /yt xdf       
 /xt xdf       
 /ang xdf      
 storerect
 /taperfcn xdf
 /tint2 xdf      
 /tint1 xdf      
 bcs exch 1 exch put    
 tint1 tint2 sub abs    
 bcs 1 get maxspot    
 calcgraysteps mul abs round  
 height abs adjnumsteps   
 dup 2 lt {pop 2} if    
 1 sub /numsteps1 xdf
 currentflat mark    
 currentflat clipflatness  
 /delta top bottom sub numsteps1 1 add div def 
 /right right left sub def  
 /botsv top delta sub def  
 {
{
W
xt yt translate 
ang rotate
xt neg yt neg translate 
dup setflat 
/bottom botsv def
0 1 numsteps1 
{
numsteps1 div taperfcn /frac xdf
bcs 0
1.0 tint2 tint1 sub frac mul tint1 add sub
put bcs vc
1 index setflat 
{ 
mark {newpath left bottom right delta rectfill}stopped
{cleartomark exch 1.3 mul dup setflat exch 2 copy gt{stop}if}
{cleartomark exit}ifelse
}loop
/bottom bottom delta sub def
}for
}
gsave stopped grestore
{exch pop 2 index exch 1.3 mul dup 100 gt{cleartomark setflat stop}if}
{exit}ifelse
 }loop
 cleartomark setflat
 iosv setinkoverprint
 cosv setoverprint
}bdf
/_rfs4 {
 /iosv inkoverprint def
 /cosv currentoverprint def
 /tint2 xdf      
 /tint1 xdf      
 bcs exch 1 exch put    
 /radius xdf      
 /yt xdf       
 /xt xdf       
 tint1 tint2 sub abs    
 bcs 1 get maxspot    
 calcgraysteps mul abs round  
 radius abs adjnumsteps   
 dup 2 lt {pop 2} if    
 1 sub /numsteps1 xdf
 radius numsteps1 div 2 div /halfstep xdf 
 currentflat mark    
 currentflat clipflatness  
 {
{
dup setflat 
W 
0 1 numsteps1 
{
dup /radindex xdf
numsteps1 div /frac xdf
bcs 0
tint2 tint1 sub frac mul tint1 add
put bcs vc
1 index setflat 
{ 
newpath mark xt yt radius 1 frac sub mul halfstep add 0 360
{ arc
radindex numsteps1 ne 
{
xt yt 
radindex 1 add numsteps1 
div 1 exch sub 
radius mul halfstep add
dup xt add yt moveto 
360 0 arcn 
} if
fill
}stopped
{cleartomark exch 1.3 mul dup setflat exch 2 copy gt{stop}if}
{cleartomark exit}ifelse
}loop
}for
}
gsave stopped grestore
{exch pop 2 index exch 1.3 mul dup 100 gt{cleartomark setflat stop}if}
{exit}ifelse
 }loop
 cleartomark setflat
 iosv setinkoverprint
 cosv setoverprint
}bdf
/_rfp4 {
 /iosv inkoverprint def
 /cosv currentoverprint def
 /k2 xdf /y2 xdf /m2 xdf /c2 xdf
 /k1 xdf /y1 xdf /m1 xdf /c1 xdf
 /radius xdf      
 /yt xdf       
 /xt xdf       
 c1 c2 sub abs
 m1 m2 sub abs
 y1 y2 sub abs
 k1 k2 sub abs
 maxcolor      
 calcgraysteps mul abs round  
 radius abs adjnumsteps   
 dup 2 lt {pop 1} if    
 1 sub /numsteps1 xdf
 radius numsteps1 dup 0 eq {pop} {div} ifelse 
 2 div /halfstep xdf 
 currentflat mark    
 currentflat clipflatness  
 {
{
dup setflat 
W 
0 1 numsteps1 
{
dup /radindex xdf
numsteps1 dup 0 eq {pop 0.5 } { div } ifelse 
/frac xdf
bc4 0 c2 c1 sub frac mul c1 add put
bc4 1 m2 m1 sub frac mul m1 add put
bc4 2 y2 y1 sub frac mul y1 add put
bc4 3 k2 k1 sub frac mul k1 add put
bc4 vc
1 index setflat 
{ 
newpath mark xt yt radius 1 frac sub mul halfstep add 0 360
{ arc
radindex numsteps1 ne 
{
xt yt 
radindex 1 add 
numsteps1 dup 0 eq {pop} {div} ifelse 
1 exch sub 
radius mul halfstep add
dup xt add yt moveto 
360 0 arcn 
} if
fill
}stopped
{cleartomark exch 1.3 mul dup setflat exch 2 copy gt{stop}if}
{cleartomark exit}ifelse
}loop
}for
}
gsave stopped grestore
{exch pop 2 index exch 1.3 mul dup 100 gt{cleartomark setflat stop}if}
{exit}ifelse
 }loop
 cleartomark setflat
 iosv setinkoverprint
 cosv setoverprint
}bdf
/lfp4{_lfp4}ndf
/lfs4{_lfs4}ndf
/rfs4{_rfs4}ndf
/rfp4{_rfp4}ndf
/cvc [0 0 0 1] def 
/vc{
 AltsysDict /cvc 2 index put 
 aload length 4 eq
 {setcmykcolor}
 {setspotcolor}
 ifelse
}bdf 
/origmtx matrix currentmatrix def
/ImMatrix matrix currentmatrix def
0 setseparationgray
/imgr {1692 1584 2304 2376 } def 
/bleed 0 def 
/clpr {1692 1584 2304 2376 } def 
/xs 1 def 
/ys 1 def 
/botx 0 def 
/overlap 0 def 
/wdist 18 def 
0 2 mul fhsetspreadsize 
0 0 ne {/df 0 def /clipflatness 0 def} if 
/maxsteps 256 def 
/forcemaxsteps false def 
vms
-1755 -1595 translate
/currentpacking defed{false setpacking}if 
/spots[
1 0 0 0 (Process Cyan) false newcmykcustomcolor
0 1 0 0 (Process Magenta) false newcmykcustomcolor
0 0 1 0 (Process Yellow) false newcmykcustomcolor
0 0 0 1 (Process Black) false newcmykcustomcolor
]def
/textopf false def
/curtextmtx{}def
/otw .25 def
/msf{dup/curtextmtx xdf makefont setfont}bdf
/makesetfont/msf load def
/curtextheight{.707104 .707104 curtextmtx dtransform
 dup mul exch dup mul add sqrt}bdf
/ta2{ 
tempstr 2 index gsave exec grestore 
cwidth cheight rmoveto 
4 index eq{5 index 5 index rmoveto}if 
2 index 2 index rmoveto 
}bdf
/ta{exch systemdict/cshow known
{{/cheight xdf/cwidth xdf tempstr 0 2 index put ta2}exch cshow} 
{{tempstr 0 2 index put tempstr stringwidth/cheight xdf/cwidth xdf ta2}forall} 
ifelse 6{pop}repeat}bdf
/sts{/textopf currentoverprint def vc setoverprint
/ts{awidthshow}def exec textopf setoverprint}bdf
/stol{/xt currentlinewidth def 
 setlinewidth vc newpath 
 /ts{{false charpath stroke}ta}def exec 
 xt setlinewidth}bdf 
 
/strk{/textopf currentoverprint def vc setoverprint
 /ts{{false charpath stroke}ta}def exec 
 textopf setoverprint
 }bdf 
n
[] 0 d
3.863708 M
1 w
0 j
0 J
false setoverprint
0 i
false eomode
[0 0 0 1] vc
vms
1819.5001 2165.883 m
2194.5001 2165.883 L
2194.5001 1790.883 L
1819.5001 1790.883 L
1819.5001 2165.883 L
0.75 w
s 
n
1819.5001 1874.883 m
1879.5001 1790.883 L
S 
n
1879.5001 1790.883 m
2143.5001 2165.883 L
S 
n
2143.5001 2165.883 m
2194.5001 2095.383 L
S 
n
1904.2501 2165.883 m
1904.2501 1790.883 L
3.863693 M
[4 4 ] 0 d
S 
[] 0 d
n
2123.2501 2165.883 m
2123.2501 1790.883 L
[4 4 ] 0 d
S 
[] 0 d
n
1812.0401 2175.0328 m
1803.0001 2175.0328 L
1803.0001 2156.0331 L
1812.0401 2156.0331 L
1812.0401 2175.0328 L
n
q
_bfh
_efh
{
f1 [18 0 0 18 0 0] makesetfont
1803.000122 2160.633041 m
0 0 32 0 0 (1) ts
}
true
[0 0 0 1]sts
Q
1812.7901 1807.5328 m
1803.7501 1807.5328 L
1803.7501 1788.5331 L
1812.7901 1788.5331 L
1812.7901 1807.5328 L
n
q
_bfh
_efh
{
f1 [18 0 0 18 0 0] makesetfont
1803.750122 1793.133041 m
0 0 32 0 0 (0) ts
}
true
[0 0 0 1]sts
Q
1826.2901 1787.2828 m
1817.2501 1787.2828 L
1817.2501 1768.2831 L
1826.2901 1768.2831 L
1826.2901 1787.2828 L
n
q
_bfh
_efh
{
f1 [18 0 0 18 0 0] makesetfont
1817.250122 1772.883041 m
0 0 32 0 0 (0) ts
}
true
[0 0 0 1]sts
Q
2200.5401 1786.5328 m
2191.5001 1786.5328 L
2191.5001 1767.5331 L
2200.5401 1767.5331 L
2200.5401 1786.5328 L
n
q
_bfh
_efh
{
f1 [18 0 0 18 0 0] makesetfont
2191.500122 1772.133041 m
0 0 32 0 0 (1) ts
}
true
[0 0 0 1]sts
Q
2145.5824 1781.5728 m
2104.0374 1781.5728 L
2104.0374 1763.293 L
2145.5824 1763.293 L
2145.5824 1781.5728 L
n
q
_bfh
_efh
{
f1 [17.280029 0 0 17.280029 0 0] makesetfont
2104.037415 1767.748962 m
0 -1.800003 rmoveto 0 0 32 0 0 (R) ts
0 1.800003 rmoveto }
true
[0 0 0 1]sts
{
f1 [10.799973 0 0 10.799973 0 0] makesetfont
0 10.799973 rmoveto 0 0 32 0 0 (\(3\)) ts
0 -10.799973 rmoveto }
true
[0 0 0 1]sts
{
f1 [10.799973 0 0 10.799973 0 0] makesetfont
0 0 32 0 0 (\(x) ts
}
true
[0 0 0 1]sts
{
f1 [10.799973 0 0 10.799973 0 0] makesetfont
0 -3.149994 rmoveto 0 0 32 0 0 (c) ts
0 3.149994 rmoveto }
true
[0 0 0 1]sts
{
f1 [10.799973 0 0 10.799973 0 0] makesetfont
0 0 32 0 0 (\)) ts
}
true
[0 0 0 1]sts
Q
1927.3324 1780.8228 m
1885.7874 1780.8228 L
1885.7874 1762.543 L
1927.3324 1762.543 L
1927.3324 1780.8228 L
n
q
_bfh
_efh
{
f1 [17.280029 0 0 17.280029 0 0] makesetfont
1885.787399 1766.998947 m
0 -1.800003 rmoveto 0 0 32 0 0 (R) ts
0 1.800003 rmoveto }
true
[0 0 0 1]sts
{
f1 [10.799973 0 0 10.799973 0 0] makesetfont
0 10.799973 rmoveto 0 0 32 0 0 (\(4\)) ts
0 -10.799973 rmoveto }
true
[0 0 0 1]sts
{
f1 [10.799973 0 0 10.799973 0 0] makesetfont
0 0 32 0 0 (\(x) ts
}
true
[0 0 0 1]sts
{
f1 [10.799973 0 0 10.799973 0 0] makesetfont
0 -3.149994 rmoveto 0 0 32 0 0 (c) ts
0 3.149994 rmoveto }
true
[0 0 0 1]sts
{
f1 [10.799973 0 0 10.799973 0 0] makesetfont
0 0 32 0 0 (\)) ts
}
true
[0 0 0 1]sts
Q
1862.835 1762.1828 m
1847.1962 1762.1828 L
1847.1962 1739.5031 L
1862.835 1739.5031 L
1862.835 1762.1828 L
n
q
_bfh
_efh
{
f1 [21.599945 0 0 21.599945 0 0] makesetfont
1847.196243 1744.903076 m
0 -4.499985 rmoveto 0 0 32 0 0 (X) ts
0 4.499985 rmoveto }
true
[0 0 0 1]sts
Q
1885.6902 1751.8928 m
1863.3912 1751.8928 L
1863.3912 1735.7731 L
1885.6902 1735.7731 L
1885.6902 1751.8928 L
n
q
_bfh
_efh
{
f1 [15.119888 0 0 15.119888 0 0] makesetfont
1863.39122 1739.797043 m
0 -3.149979 rmoveto 0 0 32 0 0 (2,A) ts
0 3.149979 rmoveto }
true
[0 0 0 1]sts
Q
2163.585 1762.4328 m
2147.9462 1762.4328 L
2147.9462 1739.7531 L
2163.585 1739.7531 L
2163.585 1762.4328 L
n
q
_bfh
_efh
{
f1 [21.599945 0 0 21.599945 0 0] makesetfont
2147.946228 1745.153076 m
0 -4.499985 rmoveto 0 0 32 0 0 (X) ts
0 4.499985 rmoveto }
true
[0 0 0 1]sts
Q
2186.1755 1752.8928 m
2164.7107 1752.8928 L
2164.7107 1736.7731 L
2186.1755 1736.7731 L
2186.1755 1752.8928 L
n
q
_bfh
_efh
{
f1 [15.119888 0 0 15.119888 0 0] makesetfont
2164.710739 1740.797043 m
0 -3.149979 rmoveto 0 0 32 0 0 (2,B) ts
0 3.149979 rmoveto }
true
[0 0 0 1]sts
Q
2015.639 1730.6828 m
2004.799 1730.6828 L
2004.799 1708.0031 L
2015.639 1708.0031 L
2015.639 1730.6828 L
n
q
_bfh
_efh
{
f1 [21.599945 0 0 21.599945 0 0] makesetfont
2004.799042 1713.403091 m
0 -4.499985 rmoveto 0 0 32 0 0 (x) ts
0 4.499985 rmoveto }
true
[0 0 0 1]sts
Q
1827.9159 2006.1814 m
1784.8987 2006.1814 L
1784.8987 1961.7816 L
1827.9159 1961.7816 L
1827.9159 2006.1814 L
n
q
Q
2290.5009 2108.1814 m
2290.5001 2108.1814 L
2290.5001 2063.7816 L
2290.5009 2063.7816 L
2290.5009 2108.1814 L
n
q
Q
1816.9695 1993.8828 m
1761.0459 1993.8828 L
1761.0459 1969.313 L
1816.9695 1969.313 L
1816.9695 1993.8828 L
n
q
_bfh
_efh
{
f1 [23.400055 0 0 23.400055 0 0] makesetfont
1761.045929 1975.163055 m
0 0 32 0 0 (G) ts
}
true
[0 0 0 1]sts
{
f1 [23.400055 0 0 23.400055 0 0] makesetfont
0 -7.800018 rmoveto 0 0 32 0 0 (2) ts
0 7.800018 rmoveto }
true
[0 0 0 1]sts
{
f1 [23.400055 0 0 23.400055 0 0] makesetfont
0 0 32 0 0 (\(x\)) ts
}
true
[0 0 0 1]sts
Q
vmrs
2063.001 1770.5798 m
2063 1770.5798 L
2063 1756.98 L
2063.001 1756.98 L
2063.001 1770.5798 L
n
q
Q
1830 1729 m
1895 1729 L
3.863693 M
false setoverprint
S 
n
true eomode
1830.4166 1729 m
1830.4166 1726.5 L
1824.1667 1729 L
1830.4166 1731.5 L
1830.4166 1729 L
f 
n
1894.5834 1729 m
1894.5834 1731.5 L
1900.8333 1729 L
1894.5834 1726.5 L
1894.5834 1729 L
f 
n
1915 1729 m
2114 1729 L
S 
n
1915.4166 1729 m
1915.4166 1726.5 L
1909.1667 1729 L
1915.4166 1731.5 L
1915.4166 1729 L
f 
n
2113.5834 1729 m
2113.5834 1731.5 L
2119.8333 1729 L
2113.5834 1726.5 L
2113.5834 1729 L
f 
n
2134 1729 m
2185 1729 L
S 
n
2134.4166 1729 m
2134.4166 1726.5 L
2128.1667 1729 L
2134.4166 1731.5 L
2134.4166 1729 L
f 
n
2184.5834 1729 m
2184.5834 1731.5 L
2190.8333 1729 L
2184.5834 1726.5 L
2184.5834 1729 L
f 
n
1819.5001 1790.883 m
2194.5001 2165.883 L
0.5 w
S 
n
1819.5001 1874.883 m
1904.5 1875 L
0.75 w
[4 4 ] 0 d
S 
[] 0 d
n
2194.5001 2095.383 m
2123 2095.5 L
[4 4 ] 0 d
S 
[] 0 d
n
2036.0635 1764.1997 m
1999 1764.1997 L
1999 1741.5201 L
2036.0635 1741.5201 L
2036.0635 1764.1997 L
n
q
_bfh
_efh
{
f1 [21.599945 0 0 21.599945 0 0] makesetfont
1999 1746.920029 m
0 -6 rmoveto 0 0 32 0 0 (X) ts
0 6 rmoveto }
true
[0 0 0 1]sts
{
f1 [15.119888 0 0 15.119888 0 0] makesetfont
0 -10 rmoveto 0 0 32 0 0 (2,C) ts
0 10 rmoveto }
true
[0 0 0 1]sts
Q
2064.001 1906.2797 m
2064 1906.2797 L
2064 1883.6 L
2064.001 1883.6 L
2064.001 1906.2797 L
n
q
Q
2036.0635 1739.5201 m
1992 1739.5201 L
1992 1595.5201 L
2036.0635 1595.5201 L
2036.0635 1739.5201 L
n
q
Q
vmr
vmr
end
EndOfTheIncludedPostscriptMagicCookie

\closepsdump

\psdump{fig51.eps}
userdict begin /AltsysDict 245 dict def end
AltsysDict begin
/bdf{bind def}bind def
/xdf{exch def}bdf
/defed{where{pop true}{false}ifelse}bdf
/ndf{1 index where{pop pop pop}{dup xcheck{bind}if def}ifelse}bdf
/d{setdash}bdf
/h{closepath}bdf
/H{}bdf
/J{setlinecap}bdf
/j{setlinejoin}bdf
/M{setmiterlimit}bdf
/n{newpath}bdf
/N{newpath}bdf
/q{gsave}bdf
/Q{grestore}bdf
/w{setlinewidth}bdf
/sepdef{
 dup where not
 {
AltsysSepDict
 }
 if 
 3 1 roll exch put
}bdf
/st{settransfer}bdf
/colorimage defed /_rci xdf
/_NXLevel2 defed { 
 _NXLevel2 not {   
/colorimage where {
userdict eq {
/_rci false def 
} if
} if
 } if
} if
/md defed{ 
 md type /dicttype eq {  
/colorimage where { 
md eq { 
/_rci false def 
}if
}if
/settransfer where {
md eq {
/st systemdict /settransfer get def
}if
}if
 }if 
}if
/setstrokeadjust defed
{
 true setstrokeadjust
 /C{curveto}bdf
 /L{lineto}bdf
 /m{moveto}bdf
}
{
 /dr{transform .25 sub round .25 add 
exch .25 sub round .25 add exch itransform}bdf
 /C{dr curveto}bdf
 /L{dr lineto}bdf
 /m{dr moveto}bdf
 /setstrokeadjust{pop}bdf 
}ifelse
/rectstroke defed /xt xdf
xt {/yt save def} if
/privrectpath { 
 4 -2 roll m
 dtransform round exch round exch idtransform 
 2 copy 0 lt exch 0 lt xor
 {dup 0 exch rlineto exch 0 rlineto neg 0 exch rlineto}
 {exch dup 0 rlineto exch 0 exch rlineto neg 0 rlineto}
 ifelse
 closepath
}bdf
/rectclip{newpath privrectpath clip newpath}def
/rectfill{gsave newpath privrectpath fill grestore}def
/rectstroke{gsave newpath privrectpath stroke grestore}def
xt {yt restore} if
/_fonthacksave false def
/currentpacking defed 
{
 /_bfh {/_fonthacksave currentpacking def false setpacking} bdf
 /_efh {_fonthacksave setpacking} bdf
}
{
 /_bfh {} bdf
 /_efh {} bdf
}ifelse
/packedarray{array astore readonly}ndf
/` 
{ 
 false setoverprint

 /-save0- save def
 5 index concat
 pop
 storerect left bottom width height rectclip
 pop
 
 /dict_count countdictstack def
 /op_count count 1 sub def
 userdict begin
 
 /showpage {} def
 
 0 setgray 0 setlinecap 1 setlinewidth
 0 setlinejoin 10 setmiterlimit [] 0 setdash newpath
 
} bdf
/currentpacking defed{true setpacking}if
/min{2 copy gt{exch}if pop}bdf
/max{2 copy lt{exch}if pop}bdf
/xformfont { currentfont exch makefont setfont } bdf
/fhnumcolors 1 
 statusdict begin
/processcolors defed 
{
pop processcolors
}
{
/deviceinfo defed {
deviceinfo /Colors known {
pop deviceinfo /Colors get
} if
} if
} ifelse
 end 
def
/printerRes 
 gsave
 matrix defaultmatrix setmatrix
 72 72 dtransform
 abs exch abs
 max
 grestore
 def
/graycalcs
[
 {Angle Frequency}   
 {GrayAngle GrayFrequency} 
 {0 Width Height matrix defaultmatrix idtransform 
dup mul exch dup mul add sqrt 72 exch div} 
 {0 GrayWidth GrayHeight matrix defaultmatrix idtransform 
dup mul exch dup mul add sqrt 72 exch div} 
] def
/calcgraysteps {
 forcemaxsteps
 {
maxsteps
 }
 {
/currenthalftone defed
{currenthalftone /dicttype eq}{false}ifelse
{
currenthalftone begin
HalftoneType 4 le
{graycalcs HalftoneType 1 sub get exec}
{
HalftoneType 5 eq
{
Default begin
{graycalcs HalftoneType 1 sub get exec}
end
}
{0 60} 
ifelse
}
ifelse
end
}
{
currentscreen pop exch 
}
ifelse
 
printerRes 300 max exch div exch 
2 copy 
sin mul round dup mul 
3 1 roll 
cos mul round dup mul 
add 1 add 
dup maxsteps gt {pop maxsteps} if 
 }
 ifelse
} bdf
/nextrelease defed { 
 /languagelevel defed not {    
/framebuffer defed { 
0 40 string framebuffer 9 1 roll 8 {pop} repeat
dup 516 eq exch 520 eq or
{
/fhnumcolors 3 def
/currentscreen {60 0 {pop pop 1}}bdf
/calcgraysteps {maxsteps} bdf
}if
}if
 }if
}if
fhnumcolors 1 ne {
 /calcgraysteps {maxsteps} bdf
} if
/currentpagedevice defed {

 currentpagedevice /PreRenderingEnhance known
 {
currentpagedevice /PreRenderingEnhance get
{
/calcgraysteps 
{
forcemaxsteps 
{maxsteps}
{256 maxsteps min}
ifelse
} def
} if
 } if
} if
/gradfrequency 144 def
printerRes 1000 lt {
 /gradfrequency 72 def
} if
/adjnumsteps {
 
 dup dtransform abs exch abs max  
 
 printerRes div       
 
 gradfrequency mul      
 round        
 5 max       
 min        
}bdf
/goodsep {
 spots exch get 4 get dup sepname eq exch (_vc_Registration) eq or
}bdf
/BeginGradation defed
{/bb{BeginGradation}bdf}
{/bb{}bdf}
ifelse
/EndGradation defed
{/eb{EndGradation}bdf}
{/eb{}bdf}
ifelse
/bottom -0 def 
/delta -0 def 
/frac -0 def 
/height -0 def 
/left -0 def 
/numsteps1 -0 def 
/radius -0 def 
/right -0 def 
/top -0 def 
/width -0 def 
/xt -0 def 
/yt -0 def 
/df currentflat def 
/tempstr 1 string def 
/clipflatness currentflat def 
/inverted? 
 0 currenttransfer exec .5 ge def
/tc1 [0 0 0 1] def 
/tc2 [0 0 0 1] def 
/storerect{/top xdf /right xdf /bottom xdf /left xdf 
/width right left sub def /height top bottom sub def}bdf
/concatprocs{
 systemdict /packedarray known 
 {dup type /packedarraytype eq 2 index type /packedarraytype eq or}{false}ifelse
 { 
/proc2 exch cvlit def /proc1 exch cvlit def
proc1 aload pop proc2 aload pop
proc1 length proc2 length add packedarray cvx
 }
 { 
/proc2 exch cvlit def /proc1 exch cvlit def
/newproc proc1 length proc2 length add array def
newproc 0 proc1 putinterval newproc proc1 length proc2 putinterval
newproc cvx
 }ifelse
}bdf
/i{dup 0 eq
 {pop df dup} 
 {dup} ifelse 
 /clipflatness xdf setflat
}bdf
version cvr 38.0 le
{/setrgbcolor{
currenttransfer exec 3 1 roll
currenttransfer exec 3 1 roll
currenttransfer exec 3 1 roll
setrgbcolor}bdf}if
/vms {/vmsv save def} bdf
/vmr {vmsv restore} bdf
/vmrs{vmsv restore /vmsv save def}bdf
/eomode{ 
 {/filler /eofill load def /clipper /eoclip load def}
 {/filler /fill load def /clipper /clip load def}
 ifelse
}bdf
/normtaper{}bdf
/logtaper{9 mul 1 add log}bdf
/CD{
 /NF exch def 
 {    
exch dup 
/FID ne 1 index/UniqueID ne and
{exch NF 3 1 roll put}
{pop pop}
ifelse
 }forall 
 NF
}bdf
/MN{
 1 index length   
 /Len exch def 
 dup length Len add  
 string dup    
 Len     
 4 -1 roll    
 putinterval   
 dup     
 0      
 4 -1 roll   
 putinterval   
}bdf
/RC{4 -1 roll /ourvec xdf 256 string cvs(|______)anchorsearch
 {1 index MN cvn/NewN exch def cvn
 findfont dup maxlength dict CD dup/FontName NewN put dup
 /Encoding ourvec put NewN exch definefont pop}{pop}ifelse}bdf
/RF{ 
 dup      
 FontDirectory exch   
 known     
 {pop 3 -1 roll pop}  
 {RC}
 ifelse
}bdf
/FF{dup 256 string cvs(|______)exch MN cvn dup FontDirectory exch known
 {exch pop findfont 3 -1 roll pop}
 {pop dup findfont dup maxlength dict CD dup dup
 /Encoding exch /Encoding get 256 array copy 7 -1 roll 
 {3 -1 roll dup 4 -2 roll put}forall put definefont}
 ifelse}bdf
/RFJ{ 
 dup      
 FontDirectory exch   
 known     
 {pop 3 -1 roll pop  
 FontDirectory /Ryumin-Light-83pv-RKSJ-H known 
 {pop pop /Ryumin-Light-83pv-RKSJ-H dup}if  
 }      
 {RC}
 ifelse
}bdf
/FFJ{dup 256 string cvs(|______)exch MN cvn dup FontDirectory exch known
 {exch pop findfont 3 -1 roll pop}
 {pop
dup FontDirectory exch known not 
 {FontDirectory /Ryumin-Light-83pv-RKSJ-H known 
{pop /Ryumin-Light-83pv-RKSJ-H}if 
 }if            
 dup findfont dup maxlength dict CD dup dup
 /Encoding exch /Encoding get 256 array copy 7 -1 roll 
 {3 -1 roll dup 4 -2 roll put}forall put definefont}
 ifelse}bdf
/fps{
 currentflat   
 exch     
 dup 0 le{pop 1}if 
 {
dup setflat 3 index stopped
{1.3 mul dup 3 index gt{pop setflat pop pop stop}if} 
{exit} 
ifelse
 }loop 
 pop setflat pop pop
}bdf
/fp{100 currentflat fps}bdf
/clipper{clip}bdf 
/W{/clipper load 100 clipflatness dup setflat fps}bdf
userdict begin /BDFontDict 29 dict def end
BDFontDict begin
/bu{}def
/bn{}def
/setTxMode{av 70 ge{pop}if pop}def
/gm{m}def
/show{pop}def
/gr{pop}def
/fnt{pop pop pop}def
/fs{pop}def
/fz{pop}def
/lin{pop pop}def
/:M {pop pop} def
/sf {pop} def
/S {pop} def
/@b {pop pop pop pop pop pop pop pop} def
/_bdsave /save load def
/_bdrestore /restore load def
/save { dup /fontsave eq {null} {_bdsave} ifelse } def
/restore { dup null eq { pop } { _bdrestore } ifelse } def
/fontsave null def
end
/MacVec 256 array def 
MacVec 0 /Helvetica findfont
/Encoding get 0 128 getinterval putinterval
MacVec 127 /DEL put MacVec 16#27 /quotesingle put MacVec 16#60 /grave put
/NUL/SOH/STX/ETX/EOT/ENQ/ACK/BEL/BS/HT/LF/VT/FF/CR/SO/SI
/DLE/DC1/DC2/DC3/DC4/NAK/SYN/ETB/CAN/EM/SUB/ESC/FS/GS/RS/US
MacVec 0 32 getinterval astore pop
/Adieresis/Aring/Ccedilla/Eacute/Ntilde/Odieresis/Udieresis/aacute
/agrave/acircumflex/adieresis/atilde/aring/ccedilla/eacute/egrave
/ecircumflex/edieresis/iacute/igrave/icircumflex/idieresis/ntilde/oacute
/ograve/ocircumflex/odieresis/otilde/uacute/ugrave/ucircumflex/udieresis
/dagger/degree/cent/sterling/section/bullet/paragraph/germandbls
/registered/copyright/trademark/acute/dieresis/notequal/AE/Oslash
/infinity/plusminus/lessequal/greaterequal/yen/mu/partialdiff/summation
/product/pi/integral/ordfeminine/ordmasculine/Omega/ae/oslash 
/questiondown/exclamdown/logicalnot/radical/florin/approxequal/Delta/guillemotleft
/guillemotright/ellipsis/nbspace/Agrave/Atilde/Otilde/OE/oe
/endash/emdash/quotedblleft/quotedblright/quoteleft/quoteright/divide/lozenge
/ydieresis/Ydieresis/fraction/currency/guilsinglleft/guilsinglright/fi/fl
/daggerdbl/periodcentered/quotesinglbase/quotedblbase
/perthousand/Acircumflex/Ecircumflex/Aacute
/Edieresis/Egrave/Iacute/Icircumflex/Idieresis/Igrave/Oacute/Ocircumflex
/apple/Ograve/Uacute/Ucircumflex/Ugrave/dotlessi/circumflex/tilde
/macron/breve/dotaccent/ring/cedilla/hungarumlaut/ogonek/caron
MacVec 128 128 getinterval astore pop
end 
AltsysDict begin
_bfh
_efh
0 dict dup begin
end 
/f0 /Symbol FF def
_bfh
_efh
MacVec 256 array copy
/f1 /|______Times-Roman dup RF findfont def
end 
AltsysDict begin 
/onlyk4{false}ndf
/ccmyk{dup 5 -1 roll sub 0 max exch}ndf
/cmyk2gray{
 4 -1 roll 0.3 mul 4 -1 roll 0.59 mul 4 -1 roll 0.11 mul
 add add add 1 min neg 1 add
}bdf
/setcmykcolor{1 exch sub ccmyk ccmyk ccmyk pop setrgbcolor}ndf
/maxcolor { 
 max max max  
} ndf
/maxspot {
 pop
} ndf
/setcmykcoloroverprint{4{dup -1 eq{pop 0}if 4 1 roll}repeat setcmykcolor}ndf
/findcmykcustomcolor{5 packedarray}ndf
/setcustomcolor{exch aload pop pop 4{4 index mul 4 1 roll}repeat setcmykcolor pop}ndf
/setseparationgray{setgray}ndf
/setoverprint{pop}ndf 
/currentoverprint false ndf
/cmykbufs2gray{
 0 1 2 index length 1 sub
 { 
4 index 1 index get 0.3 mul 
4 index 2 index get 0.59 mul 
4 index 3 index get 0.11 mul 
4 index 4 index get 
add add add cvi 255 min
255 exch sub
2 index 3 1 roll put
 }for
 4 1 roll pop pop pop
}bdf
/colorimage{
 pop pop
 [
5 -1 roll/exec cvx 
6 -1 roll/exec cvx 
7 -1 roll/exec cvx 
8 -1 roll/exec cvx
/cmykbufs2gray cvx
 ]cvx 
 image
}
version cvr 47.1 le 
statusdict /product get (Lino) anchorsearch{pop pop true}{pop false}ifelse
and{userdict begin bdf end}{ndf}ifelse
fhnumcolors 1 ne {/yt save def} if
/customcolorimage{
 aload pop
 (_vc_Registration) eq 
 {
pop pop pop pop separationimage
 }
 {
/ik xdf /iy xdf /im xdf /ic xdf
ic im iy ik cmyk2gray /xt xdf
currenttransfer
{dup 1.0 exch sub xt mul add}concatprocs
st 
image
 }
 ifelse
}ndf
fhnumcolors 1 ne {yt restore} if
fhnumcolors 3 ne {/yt save def} if
/customcolorimage{
 aload pop 
 (_vc_Registration) eq 
 {
pop pop pop pop separationimage
 }
 {
/ik xdf /iy xdf /im xdf /ic xdf
1.0 dup ic ik add min sub 
1.0 dup im ik add min sub 
1.0 dup iy ik add min sub 
/ic xdf /iy xdf /im xdf
currentcolortransfer
4 1 roll 
{dup 1.0 exch sub ic mul add}concatprocs 4 1 roll 
{dup 1.0 exch sub iy mul add}concatprocs 4 1 roll 
{dup 1.0 exch sub im mul add}concatprocs 4 1 roll 
setcolortransfer
{/dummy xdf dummy}concatprocs{dummy}{dummy}true 3 colorimage
 }
 ifelse
}ndf
fhnumcolors 3 ne {yt restore} if
fhnumcolors 4 ne {/yt save def} if
/customcolorimage{
 aload pop
 (_vc_Registration) eq 
 {
pop pop pop pop separationimage
 }
 {
/ik xdf /iy xdf /im xdf /ic xdf
currentcolortransfer
{1.0 exch sub ik mul ik sub 1 add}concatprocs 4 1 roll
{1.0 exch sub iy mul iy sub 1 add}concatprocs 4 1 roll
{1.0 exch sub im mul im sub 1 add}concatprocs 4 1 roll
{1.0 exch sub ic mul ic sub 1 add}concatprocs 4 1 roll
setcolortransfer
{/dummy xdf dummy}concatprocs{dummy}{dummy}{dummy}
true 4 colorimage
 }
 ifelse
}ndf
fhnumcolors 4 ne {yt restore} if
/separationimage{image}ndf
/newcmykcustomcolor{6 packedarray}ndf
/inkoverprint false ndf
/setinkoverprint{pop}ndf 
/setspotcolor { 
 spots exch get
 dup 4 get (_vc_Registration) eq
 {pop 1 exch sub setseparationgray}
 {0 5 getinterval exch setcustomcolor}
 ifelse
}ndf
/currentcolortransfer{currenttransfer dup dup dup}ndf
/setcolortransfer{st pop pop pop}ndf
/fas{}ndf
/sas{}ndf
/fhsetspreadsize{pop}ndf
/filler{fill}bdf 
/F{gsave {filler}fp grestore}bdf
/f{closepath F}bdf
/S{gsave {stroke}fp grestore}bdf
/s{closepath S}bdf
/bc4 [0 0 0 0] def 
/_lfp4 {
 /iosv inkoverprint def
 /cosv currentoverprint def
 /yt xdf       
 /xt xdf       
 /ang xdf      
 storerect
 /taperfcn xdf
 /k2 xdf /y2 xdf /m2 xdf /c2 xdf
 /k1 xdf /y1 xdf /m1 xdf /c1 xdf
 c1 c2 sub abs
 m1 m2 sub abs
 y1 y2 sub abs
 k1 k2 sub abs
 maxcolor      
 calcgraysteps mul abs round  
 height abs adjnumsteps   
 dup 2 lt {pop 1} if    
 1 sub /numsteps1 xdf
 currentflat mark    
 currentflat clipflatness  
 /delta top bottom sub numsteps1 1 add div def 
 /right right left sub def  
 /botsv top delta sub def  
 {
{
W
xt yt translate 
ang rotate
xt neg yt neg translate 
dup setflat 
/bottom botsv def
0 1 numsteps1 
{
numsteps1 dup 0 eq {pop 0.5 } { div } ifelse 
taperfcn /frac xdf
bc4 0 c2 c1 sub frac mul c1 add put
bc4 1 m2 m1 sub frac mul m1 add put
bc4 2 y2 y1 sub frac mul y1 add put
bc4 3 k2 k1 sub frac mul k1 add put
bc4 vc
1 index setflat 
{ 
mark {newpath left bottom right delta rectfill}stopped
{cleartomark exch 1.3 mul dup setflat exch 2 copy gt{stop}if}
{cleartomark exit}ifelse
}loop
/bottom bottom delta sub def
}for
}
gsave stopped grestore
{exch pop 2 index exch 1.3 mul dup 100 gt{cleartomark setflat stop}if}
{exit}ifelse
 }loop
 cleartomark setflat
 iosv setinkoverprint
 cosv setoverprint
}bdf
/bcs [0 0] def 
/_lfs4 {
 /iosv inkoverprint def
 /cosv currentoverprint def
 /yt xdf       
 /xt xdf       
 /ang xdf      
 storerect
 /taperfcn xdf
 /tint2 xdf      
 /tint1 xdf      
 bcs exch 1 exch put    
 tint1 tint2 sub abs    
 bcs 1 get maxspot    
 calcgraysteps mul abs round  
 height abs adjnumsteps   
 dup 2 lt {pop 2} if    
 1 sub /numsteps1 xdf
 currentflat mark    
 currentflat clipflatness  
 /delta top bottom sub numsteps1 1 add div def 
 /right right left sub def  
 /botsv top delta sub def  
 {
{
W
xt yt translate 
ang rotate
xt neg yt neg translate 
dup setflat 
/bottom botsv def
0 1 numsteps1 
{
numsteps1 div taperfcn /frac xdf
bcs 0
1.0 tint2 tint1 sub frac mul tint1 add sub
put bcs vc
1 index setflat 
{ 
mark {newpath left bottom right delta rectfill}stopped
{cleartomark exch 1.3 mul dup setflat exch 2 copy gt{stop}if}
{cleartomark exit}ifelse
}loop
/bottom bottom delta sub def
}for
}
gsave stopped grestore
{exch pop 2 index exch 1.3 mul dup 100 gt{cleartomark setflat stop}if}
{exit}ifelse
 }loop
 cleartomark setflat
 iosv setinkoverprint
 cosv setoverprint
}bdf
/_rfs4 {
 /iosv inkoverprint def
 /cosv currentoverprint def
 /tint2 xdf      
 /tint1 xdf      
 bcs exch 1 exch put    
 /radius xdf      
 /yt xdf       
 /xt xdf       
 tint1 tint2 sub abs    
 bcs 1 get maxspot    
 calcgraysteps mul abs round  
 radius abs adjnumsteps   
 dup 2 lt {pop 2} if    
 1 sub /numsteps1 xdf
 radius numsteps1 div 2 div /halfstep xdf 
 currentflat mark    
 currentflat clipflatness  
 {
{
dup setflat 
W 
0 1 numsteps1 
{
dup /radindex xdf
numsteps1 div /frac xdf
bcs 0
tint2 tint1 sub frac mul tint1 add
put bcs vc
1 index setflat 
{ 
newpath mark xt yt radius 1 frac sub mul halfstep add 0 360
{ arc
radindex numsteps1 ne 
{
xt yt 
radindex 1 add numsteps1 
div 1 exch sub 
radius mul halfstep add
dup xt add yt moveto 
360 0 arcn 
} if
fill
}stopped
{cleartomark exch 1.3 mul dup setflat exch 2 copy gt{stop}if}
{cleartomark exit}ifelse
}loop
}for
}
gsave stopped grestore
{exch pop 2 index exch 1.3 mul dup 100 gt{cleartomark setflat stop}if}
{exit}ifelse
 }loop
 cleartomark setflat
 iosv setinkoverprint
 cosv setoverprint
}bdf
/_rfp4 {
 /iosv inkoverprint def
 /cosv currentoverprint def
 /k2 xdf /y2 xdf /m2 xdf /c2 xdf
 /k1 xdf /y1 xdf /m1 xdf /c1 xdf
 /radius xdf      
 /yt xdf       
 /xt xdf       
 c1 c2 sub abs
 m1 m2 sub abs
 y1 y2 sub abs
 k1 k2 sub abs
 maxcolor      
 calcgraysteps mul abs round  
 radius abs adjnumsteps   
 dup 2 lt {pop 1} if    
 1 sub /numsteps1 xdf
 radius numsteps1 dup 0 eq {pop} {div} ifelse 
 2 div /halfstep xdf 
 currentflat mark    
 currentflat clipflatness  
 {
{
dup setflat 
W 
0 1 numsteps1 
{
dup /radindex xdf
numsteps1 dup 0 eq {pop 0.5 } { div } ifelse 
/frac xdf
bc4 0 c2 c1 sub frac mul c1 add put
bc4 1 m2 m1 sub frac mul m1 add put
bc4 2 y2 y1 sub frac mul y1 add put
bc4 3 k2 k1 sub frac mul k1 add put
bc4 vc
1 index setflat 
{ 
newpath mark xt yt radius 1 frac sub mul halfstep add 0 360
{ arc
radindex numsteps1 ne 
{
xt yt 
radindex 1 add 
numsteps1 dup 0 eq {pop} {div} ifelse 
1 exch sub 
radius mul halfstep add
dup xt add yt moveto 
360 0 arcn 
} if
fill
}stopped
{cleartomark exch 1.3 mul dup setflat exch 2 copy gt{stop}if}
{cleartomark exit}ifelse
}loop
}for
}
gsave stopped grestore
{exch pop 2 index exch 1.3 mul dup 100 gt{cleartomark setflat stop}if}
{exit}ifelse
 }loop
 cleartomark setflat
 iosv setinkoverprint
 cosv setoverprint
}bdf
/lfp4{_lfp4}ndf
/lfs4{_lfs4}ndf
/rfs4{_rfs4}ndf
/rfp4{_rfp4}ndf
/cvc [0 0 0 1] def 
/vc{
 AltsysDict /cvc 2 index put 
 aload length 4 eq
 {setcmykcolor}
 {setspotcolor}
 ifelse
}bdf 
/origmtx matrix currentmatrix def
/ImMatrix matrix currentmatrix def
0 setseparationgray
/imgr {1692 1584 2304 2376 } def 
/bleed 0 def 
/clpr {1692 1584 2304 2376 } def 
/xs 1 def 
/ys 1 def 
/botx 0 def 
/overlap 0 def 
/wdist 18 def 
0 2 mul fhsetspreadsize 
0 0 ne {/df 0 def /clipflatness 0 def} if 
/maxsteps 256 def 
/forcemaxsteps false def 
vms
-1795 -1770 translate
/currentpacking defed{false setpacking}if 
/spots[
1 0 0 0 (Process Cyan) false newcmykcustomcolor
0 1 0 0 (Process Magenta) false newcmykcustomcolor
0 0 1 0 (Process Yellow) false newcmykcustomcolor
0 0 0 1 (Process Black) false newcmykcustomcolor
]def
/textopf false def
/curtextmtx{}def
/otw .25 def
/msf{dup/curtextmtx xdf makefont setfont}bdf
/makesetfont/msf load def
/curtextheight{.707104 .707104 curtextmtx dtransform
 dup mul exch dup mul add sqrt}bdf
/ta2{ 
tempstr 2 index gsave exec grestore 
cwidth cheight rmoveto 
4 index eq{5 index 5 index rmoveto}if 
2 index 2 index rmoveto 
}bdf
/ta{exch systemdict/cshow known
{{/cheight xdf/cwidth xdf tempstr 0 2 index put ta2}exch cshow} 
{{tempstr 0 2 index put tempstr stringwidth/cheight xdf/cwidth xdf ta2}forall} 
ifelse 6{pop}repeat}bdf
/sts{/textopf currentoverprint def vc setoverprint
/ts{awidthshow}def exec textopf setoverprint}bdf
/stol{/xt currentlinewidth def 
 setlinewidth vc newpath 
 /ts{{false charpath stroke}ta}def exec 
 xt setlinewidth}bdf 
 
/strk{/textopf currentoverprint def vc setoverprint
 /ts{{false charpath stroke}ta}def exec 
 textopf setoverprint
 }bdf 
/makepattern defed /xt xdf
xt not {/yt save def} if
 
 /PATmp{
exch 
dup length 3 add dict copy 
begin
currentdict /Multi known not {/Multi 1 def} if
Multi 1 ne {
/UserProc /PaintProc load def
/PaintProc {
begin
0 1 Multi 1 sub {
PaintColors 1 index get PATsc
PaintData exch get
gsave currentdict UserProc grestore
}for 
end
} bdf
} if
currentdict
end
exch makepattern
 }bdf

 /PATsp{
dup /PaintType get 2 eq
{
exch aload length 4 eq
{5 -1 roll}
{spots exch get 0 4 getinterval aload pop
4 {4 index mul 4 1 roll} repeat 6 -2 roll pop}ifelse 
[/Pattern /DeviceCMYK] setcolorspace
}if
setpattern
 }bdf
 /PATfill{{{eofill}fp}{{fill}fp}ifelse}bdf
 /PATprestroke{}bdf
 /PATstroke{stroke}bdf
xt not {yt restore} if
xt {/yt save def} if
 
 /PATtcalc{
gsave
exch concat 
matrix currentmatrix exch 
2 ne {
dup 4 get exch dup 5 get exch
XStep 0 dtransform round exch round exch
XStep div exch XStep div exch
0 YStep dtransform round exch round exch
YStep div exch YStep div exch
7 -3 roll astore
} if
grestore
 }bdf
 
 /PATmp{
exch 
dup 
length 8 add 
dict copy 
begin 
TilingType PATtcalc 
/PATcurrentMtx xdf 
currentdict /Multi known not {/Multi 1 def} if
/FontType 3 def
/Encoding 256 array def
3 string 0 1 255 
{Encoding exch dup 3 index cvs cvn put} for pop
/FontMatrix matrix def
/FontBBox BBox def
/BuildChar {
mark 3 1 roll
exch begin
Multi 1 ne {PaintData exch get}{pop}ifelse
PaintType 2 eq Multi 1 ne or
{XStep 0 FontBBox aload pop setcachedevice}
{XStep 0 setcharwidth}ifelse
currentdict 
/PaintProc load 
end
gsave exec grestore
cleartomark
}bdf
currentdict 
end
/foo exch
definefont
 }bdf

 /PATsp{
/PATcurrent xdf
PATcurrent /PaintType get 2 eq
{/PATcolor xdf}if
 }bdf
 
 /PATpcalc{
PATcurrent begin
gsave
PATcurrentMtx setmatrix
BBox aload pop pop pop translate
pathbbox
grestore
YStep div ceiling 4 1 roll
XStep div ceiling 4 1 roll
YStep div floor 4 1 roll
XStep div floor 4 1 roll
2 index sub cvi abs
exch 3 index sub cvi abs exch
4 2 roll
YStep mul exch XStep mul exch
end
 }bdf

 /PATfill{
{{eoclip}fp}{{clip}fp}ifelse
PATpcalc 
newpath
PATcurrent dup begin
setfont
PATcurrentMtx setmatrix
PaintType 2 eq {PATcolor vc} if
3 index string 
0 1 Multi 1 sub { 
3 index 3 index moveto
Multi 1 ne {dup PaintColors exch get vc} if
0 1 7 index 1 sub { 
2 index 
exch 
2 index put
} for
pop 
3 index 
{
currentpoint 
2 index show
YStep add moveto
}repeat
}for
5 {pop} repeat 
end
 }bdf
 /PATprestroke{{strokepath}fp}bdf
 /PATstroke{false PATfill}bdf
xt {yt restore} if
/makepattern defed /xt xdf
xt not {/yt save def} if
userdict begin /fhpatdict 12 dict def end
fhpatdict begin
 /PatternType 1 def
 /PaintType 1 def 
 /TilingType 1 def
 /BBox [0 0 8 8] def
 /XStep 8 def
 /YStep 8 def
 /PatMtx [1 0 0 -1 0 8] def
 /PatIMtx [1 0 0 1 0 0] def
/PaintProc
{begin FHPatColor vc 8 8 true PatMtx PatData imagemask end} bdf
 /FHPatColor [0 0 0 0] def

end
/macpatorient{1 0 dtransform 0 eq exch 0 ne and}bdf
/veccalc { dtransform round exch round exch idtransform dup mul exch dup mul exch add sqrt } bdf
gsave 
macpatorient 
initmatrix
0 0 transform 
2 copy 
round exch round exch
2 index sub exch 3 index sub exch
idtransform translate
pop pop
1 0 veccalc 0 1 veccalc 
scale
{0}{-90}ifelse
rotate
matrix currentmatrix
/PATmtx xdf
grestore
/pF{
 gsave 1 setgray filler grestore
 fhpatdict begin  
 /PatData xdf  
 /FHPatColor xdf  
 end
 save    
 PATmtx setmatrix 
 fhpatdict dup   
 /PatIMtx get   
 PATmp     
 PATsp     
 /clipper load /eoclip load eq PATfill
 restore
}bdf
/pS{
 fhpatdict begin  
 /PatData xdf  
 /FHPatColor xdf  
 end
 save
 PATprestroke    
 PATmtx setmatrix
 gsave 1 setgray stroke grestore
 fhpatdict dup 
 /PatIMtx get 
 PATmp 
 PATsp 
 PATstroke
 restore
}bdf
xt not {yt restore} if
xt {/yt save def} if
/macpatstring 8 string def
/macpattint 0 def
/macpatcol [] def
/macpatangle{1 0 matrix defaultmatrix dtransform exch atan}bdf
/macpatorient{1 0 dtransform 0 eq exch 0 ne and}bdf
/macpatcountbits 
{
 0 exch     
 {
cvi 
0 1 8 
{ 
pop
dup 1 and 
0 ne 
{
exch 
1 add 
exch 
} if
-1 bitshift 
} for
pop 
 }
 forall
}bdf
/macpatset
{
 macpatstring copy pop 
 9.375 
 macpatangle 
 macpatorient not{-90 add}if
 {
1 add 4 mul cvi macpatstring exch get exch
1 add 4 mul cvi 7 sub bitshift 1 and 
inverted? {1 exch sub} if
 }
 setscreen
 
 64 macpatstring macpatcountbits sub 64 div
 inverted? {.9921875 exch sub} if
 /macpattint xdf

 {} st
 fhnumcolors 1 ne
 {
cvc dup length array copy /macpatcol xdf 
/macpattint 1 macpattint sub def 
macpatcol length 4 eq
{
0 1 3
{
macpatcol exch 2 copy
get .25 lt{0}{macpattint}ifelse put
}for
}
{
macpatcol dup 0 get .25 lt{0}{macpattint}ifelse 0 exch put
}ifelse 
macpatcol vc
 }
 {
currentgray 1 ne {macpattint setseparationgray} if
 } 
 ifelse
}bdf
/pF{
 gsave 
 exch vc
 macpatset 
 {filler}fp 
 grestore
}bdf
/pS{
 gsave 
 exch vc
 macpatset 
 {stroke}fp 
 grestore
}bdf
xt {yt restore} if
/pf{closepath pF}bdf
/ps{closepath pS}bdf
n
[] 0 d
3.863708 M
1 w
0 j
0 J
false setoverprint
0 i
false eomode
[0 0 0 1] vc
vms
1835.0098 2156.4872 m
2185.0083 2156.4872 L
2185.0083 1806.4887 L
1835.0098 1806.4887 L
1835.0098 2156.4872 L
0.7 w
s 
n
1828.0498 2165.027 m
1819.6099 2165.027 L
1819.6099 2147.2272 L
1828.0498 2147.2272 L
1828.0498 2165.027 L
n
q
_bfh
_efh
{
f1 [16.799927 0 0 16.799927 0 0] makesetfont
1819.609879 2151.587234 m
0 0 32 0 0 (1) ts
}
true
[0 0 0 1]sts
Q
1828.7498 1822.0285 m
1820.3099 1822.0285 L
1820.3099 1804.2287 L
1828.7498 1804.2287 L
1828.7498 1822.0285 L
n
q
_bfh
_efh
{
f1 [16.799927 0 0 16.799927 0 0] makesetfont
1820.309875 1808.58873 m
0 0 32 0 0 (0) ts
}
true
[0 0 0 1]sts
Q
1812.411 2018.1989 m
1801.451 2018.1989 L
1801.451 1973.4273 L
1812.411 1973.4273 L
1812.411 2018.1989 L
n
q
_bfh
_efh
{
f1 [21.839951 0 0 21.839951 0 0] makesetfont
1801.451035 1978.887283 m
0 -4.549988 rmoveto 0 0 32 0 0 (x) ts
0 4.549988 rmoveto }
true
[0 0 0 1]sts
Q
2274.6086 2102.6325 m
2274.6079 2102.6325 L
2274.6079 2059.1928 L
2274.6086 2059.1928 L
2274.6086 2102.6325 L
n
q
Q
2047.1089 1806.4887 m
2047.1089 2157.1872 L
3.863693 M
false setoverprint
[2 2 ] 0 d
S 
[] 0 d
n
1972.9092 2156.4872 m
1972.9092 1806.4887 L
[2 2 ] 0 d
S 
[] 0 d
n
1930.9094 1806.4887 m
1930.9094 2156.4872 L
[2 2 ] 0 d
S 
[] 0 d
n
1856.0097 2156.4872 m
1856.0097 1806.4887 L
[2 2 ] 0 d
S 
[] 0 d
n
1972.063 1839.3886 m
2047.0043 1839.3886 L
[4 4 ] 0 d
S 
[] 0 d
n
1972.9092 1885.5884 m
2047.8089 1885.5884 L
S 
n
2089.1087 2010.8878 m
2164.0084 2010.8878 L
[4 4 ] 0 d
S 
[] 0 d
n
1972.9092 2089.9875 m
2047.1089 2089.9875 L
S 
n
1972.9092 2128.4873 m
2047.1089 2128.4873 L
[4 4 ] 0 d
S 
[] 0 d
n
1856.0097 1809.2887 m
1930.9094 1809.2887 L
[4 4 ] 0 d
S 
[] 0 d
n
1856.0097 1813.4887 m
1930.9094 1813.4887 L
S 
n
1856.0097 1833.0886 m
1930.9094 1833.0886 L
S 
n
1856.0097 1840.7886 m
1930.9094 1840.7886 L
S 
n
1856.0097 1837.9886 m
1930.9094 1837.9886 L
[4 4 ] 0 d
S 
[] 0 d
n
1856.0097 2123.5874 m
1930.9094 2123.5874 L
S 
n
vmrs
1856.0097 2131.2873 m
1930.9094 2131.2873 L
0.7 w
3.863693 M
S 
n
1856.0097 2127.0873 m
1931.6094 2127.0873 L
[4 4 ] 0 d
S 
[] 0 d
n
1856.0097 2152.9872 m
1930.9094 2152.9872 L
[4 4 ] 0 d
S 
[] 0 d
n
1856.0097 2150.1872 m
1930.9094 2150.1872 L
S 
n
1856.0097 1808.5887 m
1930.9094 1808.5887 L
1930.9094 1806.4887 L
1856.0097 1806.4887 L
1856.0097 1808.5887 L
[0 0 0 1] <8822882288228822> pf 
0.35 w
S 
n
vmrs
1856.0097 1813.4887 m
1930.9094 1813.4887 L
1930.9094 1810.1637 L
1856.0097 1810.1637 L
1856.0097 1813.4887 L
[0 0 0 1] <8822882288228822> pf 
0.35 w
3.863693 M
S 
n
vmrs
1856.0097 1840.7886 m
1930.9094 1840.7886 L
1930.9094 1838.8636 L
1856.0097 1838.8636 L
1856.0097 1840.7886 L
[0 0 0 1] <8822882288228822> pf 
0.35 w
3.863693 M
S 
n
vmrs
1856.0097 1837.2886 m
1930.9094 1837.2886 L
1930.9094 1833.0886 L
1856.0097 1833.0886 L
1856.0097 1837.2886 L
[0 0 0 1] <8822882288228822> pf 
0.35 w
3.863693 M
S 
n
vmrs
1856.0097 2131.2873 m
1930.9094 2131.2873 L
1930.9094 2127.9623 L
1856.0097 2127.9623 L
1856.0097 2131.2873 L
[0 0 0 1] <8822882288228822> pf 
0.35 w
3.863693 M
S 
n
vmrs
1856.0097 2126.3873 m
1930.9094 2126.3873 L
1930.9094 2123.5874 L
1856.0097 2123.5874 L
1856.0097 2126.3873 L
[0 0 0 1] <8822882288228822> pf 
0.35 w
3.863693 M
S 
n
vmrs
1856.0097 2152.2872 m
1930.9094 2152.2872 L
1930.9094 2150.1872 L
1856.0097 2150.1872 L
1856.0097 2152.2872 L
[0 0 0 1] <8822882288228822> pf 
0.35 w
3.863693 M
S 
n
vmrs
1856.0097 2156.4872 m
1930.9094 2156.4872 L
1930.9094 2153.6872 L
1856.0097 2153.6872 L
1856.0097 2156.4872 L
[0 0 0 1] <8822882288228822> pf 
0.35 w
3.863693 M
S 
n
vmrs
1974.6592 2127.2623 m
1974.6619 2127.2623 L
1974.6619 2127.0873 L
1974.6592 2127.0873 L
1974.6592 2127.2623 L
[0 0 0 1] <AA55AA55AA55AA55> pf 
0.35 w
3.863693 M
S 
n
vmrs
1972.9092 2156.4872 m
2047.1089 2156.4872 L
2047.1089 2129.1873 L
1972.9092 2129.1873 L
1972.9092 2156.4872 L
[0 0 0 1] <8000000008000000> pf 
0.35 w
3.863693 M
S 
n
vmrs
1972.9092 2127.7873 m
2046.9339 2127.7873 L
2046.9339 2089.9875 L
1972.9092 2089.9875 L
1972.9092 2127.7873 L
[0 0 0 1] <8000000008000000> pf 
0.35 w
3.863693 M
S 
n
vmrs
1972.9092 1885.5884 m
2046.9339 1885.5884 L
2046.9339 1840.0886 L
1972.9092 1840.0886 L
1972.9092 1885.5884 L
[0 0 0 1] <8000000008000000> pf 
0.35 w
3.863693 M
S 
n
vmrs
1972.9092 1838.6886 m
2046.9339 1838.6886 L
2046.9339 1806.4887 L
1972.9092 1806.4887 L
1972.9092 1838.6886 L
[0 0 0 1] <8000000008000000> pf 
0.35 w
3.863693 M
S 
n
vmrs
2088.4087 2010.1879 m
2164.0084 2010.1879 L
2164.0084 1806.6637 L
2088.4087 1806.6637 L
2088.4087 2010.1879 L
[0 0 0 1] <8000000008000000> pf 
0.35 w
3.863693 M
S 
n
vmrs
2088.4087 2156.8372 m
2164.0084 2156.8372 L
2164.0084 2011.7628 L
2088.4087 2011.7628 L
2088.4087 2156.8372 L
[0 0 0 1] <8000000008000000> pf 
0.35 w
3.863693 M
S 
n
vmrs
2002.109 1981.488 m
1883.8096 2142.8373 L
0.7 w
3.863693 M
S 
n
true eomode
1884.0104 2142.5609 m
1882.3519 2141.3559 L
1880.9981 2146.7069 L
1885.6688 2143.7658 L
1884.0104 2142.5609 L
f 
n
2002.109 1981.488 m
1883.4595 1823.2887 L
S 
n
1883.6651 1823.5615 m
1885.3023 1822.3278 L
1880.5809 1819.4687 L
1882.028 1824.7952 L
1883.6651 1823.5615 L
f 
n
2141.0601 1793.1746 m
2106.8632 1793.1746 L
2106.8632 1770.2429 L
2141.0601 1770.2429 L
2141.0601 1793.1746 L
n
q
_bfh
_efh
{
f1 [21.839951 0 0 21.839951 0 0] makesetfont
2106.86319 1775.702881 m
0 0 32 0 0 (n=1) ts
}
true
[0 0 0 1]sts
Q
2027.6606 1793.8746 m
1993.4637 1793.8746 L
1993.4637 1770.9429 L
2027.6606 1770.9429 L
2027.6606 1793.8746 L
n
q
_bfh
_efh
{
f1 [21.839951 0 0 21.839951 0 0] makesetfont
1993.463684 1776.402878 m
0 0 32 0 0 (n=2) ts
}
true
[0 0 0 1]sts
Q
1908.6611 1793.1746 m
1874.4642 1793.1746 L
1874.4642 1770.2429 L
1908.6611 1770.2429 L
1908.6611 1793.1746 L
n
q
_bfh
_efh
{
f1 [21.839951 0 0 21.839951 0 0] makesetfont
1874.464203 1775.702881 m
0 0 32 0 0 (n=3) ts
}
true
[0 0 0 1]sts
Q
1922.1963 1993.6998 m
1885.1914 1993.6998 L
1885.1914 1975.4199 L
1922.1963 1975.4199 L
1922.1963 1993.6998 L
n
q
_bfh
_efh
{
f1 [17.280136 0 0 17.280136 0 0] makesetfont
1885.191391 1979.8759 m
0 0 32 0 0 (X) ts
}
true
[0 0 0 1]sts
{
f1 [17.280136 0 0 17.280136 0 0] makesetfont
0 -7 rmoveto 0 0 32 0 0 (3,C) ts
0 7 rmoveto }
true
[0 0 0 1]sts
Q
2043.0049 1991.8239 m
2006 1991.8239 L
2006 1973.544 L
2043.0049 1973.544 L
2043.0049 1991.8239 L
n
q
_bfh
_efh
{
f1 [17.280136 0 0 17.280136 0 0] makesetfont
2006 1978 m
0 0 32 0 0 (X) ts
}
true
[0 0 0 1]sts
{
f1 [17.280136 0 0 17.280136 0 0] makesetfont
0 -7 rmoveto 0 0 32 0 0 (2,C) ts
0 7 rmoveto }
true
[0 0 0 1]sts
Q
vmr
vmr
end
EndOfTheIncludedPostscriptMagicCookie

\closepsdump

\usepackage{graphicx}

\newcounter{saveeqn}%
\newcommand{\alpheqn}{\setcounter{saveeqn}{\value{equation}}%
\stepcounter{saveeqn}\setcounter{equation}{0}%
\renewcommand{\theequation}
      {\mbox{\arabic{saveeqn}\alph{equation}}}}%
\newcommand{\reseteqn}{\setcounter{equation}{\value{saveeqn}}%
\renewcommand{\theequation}{\arabic{equation}}}%

\begin{document}

\setlength{\baselineskip}{19pt}

\title{Spectral Decomposition of the Tent Map \\ with Varying Height}
\author{Suresh Subbiah$^1$ and Dean J. Driebe$^{1,2}$ \\ 
{\em $^1$Center for Studies in Statistical Mechanics and Complex Systems} \\
{\em The University of Texas at Austin} \\
{\em Austin, Texas 78712 USA} \\
{\em $^2$International Solvay Institutes for Physics and Chemistry} \\
{\em 1050 Brussels, Belgium}  }

\date{}

\maketitle

\bibliography{tent}
\bibliographystyle{unsrt}

\begin{abstract}
The generalized spectral decomposition of the Frobenius--Perron operator of
the tent map with varying height is determined at the band-splitting points.  The
decomposition includes both decay onto the attracting set and the approach to the
asymptotically periodic state on the attractor.  Explicit compact
expressions for the polynomial eigenstates are obtained using algebraic techniques.  
\end{abstract}
\section{Introduction}

A natural description of the dynamics of chaotic systems is in 
terms of evolving probability densities~\cite{Lasota}. 
On this level the  time evolution in maps is governed by the linear Frobenius--Perron
operator and the dynamical problem is solved by the determination of the spectral decomposition of this
operator. In recent years, several authors have constructed complete and explicit spectral
decompositions of the Frobenius--Perron operator of a variety of model
systems~\cite{firstgenspec,scndgenspec}.  The most useful decompositions contain in
their spectrum the decay rates characterizing the approach to equilibrium of the system.  
For one-dimensional piecewise-linear Markov maps such decompositions
are constructed in function spaces spanned by polynomials.  The dual space of these polynomials
are generalized function spaces and so the decompositions are known as generalized
spectral decompositions~\cite{Deanbook}.  

Chaotic systems often contain a control parameter that characterizes
the strength of the chaos.  A simple model system with such a control
parameter is the well-known tent map with varying height~\cite{Schuster}.  
The tent map with height $h$ on the unit interval is given by
\begin{equation} \label{generaltent}
{\rm T}(x) = \left\{ \begin{array}{lc} 
\alpha x & 0 \leq x < \frac{1}{2} \\ \noalign{\vskip4pt} 
\alpha(1-x) & \frac{1}{2} \leq x < 1,
\end{array} \right.
\end{equation}
where the parameter $\alpha \equiv 2 h$. In Figure 1 the map (\ref{generaltent}) with 
height $\sqrt{2}/2$ is shown.  Note that the map acts on the unit interval $[0,1)$ but
has images on $[0,h]$. 
\begin{figure}[htb]
\begin{center}
\scalebox{.5}[.5]{\includegraphics{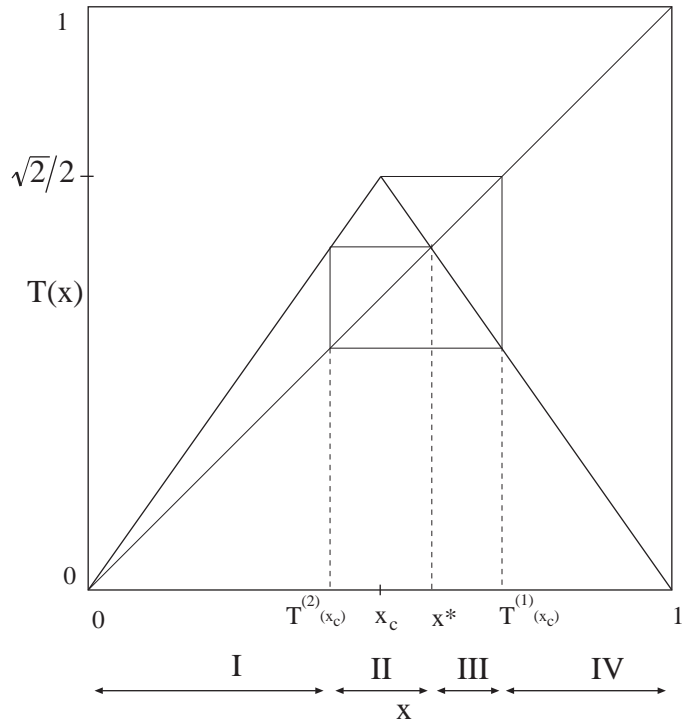}}
\parbox{5in}{\caption{\small The tent map at $\alpha = \protect\sqrt{2}$.  For this value of
$\alpha$ iterates of the critical point define four intervals around which the dynamics is
organized, as discussed in Section 2.}}
\end{center}
\end{figure}
As characterized by its Lyapunov exponent, $\log\alpha$,
the map (\ref{generaltent}) switches abruptly chaotic as the
height is raised past $1/2$.   We are interested in the chaotic regime where $1/2 < h \leq 1$, i.e.,
$1 < \alpha \leq 2$.  Similar to the well-known universality of the quadratic map, it has recently 
been reported~\cite{Moon} that the tent map also governs the low-dimensional behavior of a
wide class of nonlinear phenomena.  Specifically, it was found that the dynamics  of the
Ginzburg--Landau equation, in its description of the modulational instability of a wave train, is
reducible to the tent map.  Under this reduction, varying the height in
(\ref{generaltent}) corresponds to varying the wavelength of the initial modulational instability.   

Our interest is in the
statistical properties of the iterates of the tent map and in evolving probability
densities.  The Frobenius--Perron operator, $U$, corresponding to a map, ${\rm S}(x)$, defined on the
unit interval evolves a probability density, $\rho(x,t)$ by one time step as
\begin{equation}
\rho(x,t+1) = U \rho(x,t) \equiv \int_0^1 dx'\, \delta(x - {\rm S}(x')) \, \rho(x',t).
\end{equation}
Evaluating the integral gives a sum of contributions from the inverse branches of ${\rm S}(x)$.  
The Frobenius--Perron operator corresponding to the map (\ref{generaltent}) acts explicitly on a
density $\rho(x)$ as
\begin{equation} \label{gententfpop}
U_{\rm T} \rho(x) = \frac{1}{\alpha}\left[ \rho\left(
\frac{x}{\alpha} \right) + \rho\left(\frac{\alpha - x}{\alpha} 
\right) \right]\Theta \left(\frac{\alpha}{2} - x \right),   
\end{equation}
where 
\begin{equation} 
\Theta(a-x) = \left\{ \begin{array}{lc} 
1 & x \leq a \\
0 & x > a.
\end{array} \right. 
\end{equation}
The step function appears here because the map has
no inverse images for $x>\alpha/2$.
For $\alpha=2$ the map has images on the
whole unit interval.  For this value of $\alpha$ the invariant density (being the stationary
solution of (\ref{gententfpop})) is
uniform on the whole unit interval.  As $\alpha$ is lowered the
invariant density is supported only on a subset of the interval
$[\alpha(1-\alpha/2),\alpha/2]$, i.e., from the
second iterate to the first iterate of the critical point $x_c \equiv 1/2$. The invariant density is
discontinuous at all values of the  trajectory of the critical point. If the critical trajectory is
periodic (or eventually periodic) there will be a finite number of discontinuities. 

For $\alpha \geq \sqrt{2}$ the invariant density has nonvanishing support on all of
$[\alpha(1-\alpha/2),\alpha/2]$.  As $\alpha$ is decreased past this critical value,
$\alpha_{1}$, the invariant density breaks up into two bands with a gap in the
middle. Decreasing $\alpha$ past $\alpha_{2} = 2^{1/4}$ causes the invariant
density to break up into 4 bands. In general the
invariant density has
$2^n$ bands as $\alpha$ is decreased past $\alpha_{n} = 2^{2^{-n}}$. 
These values of $\alpha$ are called the band-splitting 
points~\cite{bsps}. 
This is illustrated in Figure 2.
\begin{figure}[htb]
\begin{center}
\scalebox{0.73}[0.73]{\includegraphics{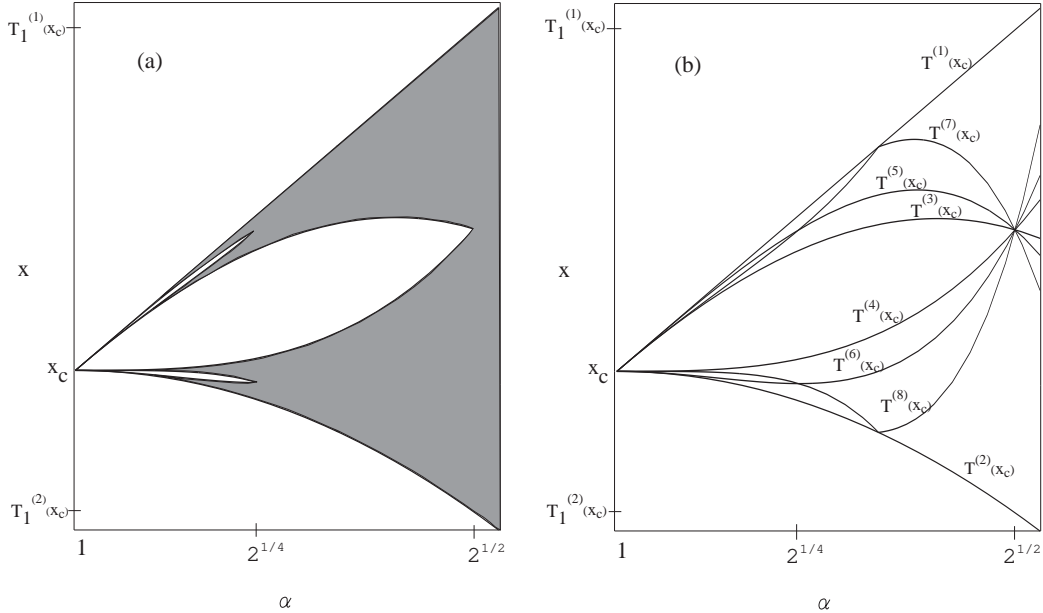}}
\parbox{5in}{\caption{\small Bifurcation plot of the tent map with varying height showing
the formation of bands as the height, parametrized by $\alpha=2h$ on the horizontal axis,
is lowered.  The vertical axis is the subset $[\alpha(1-\alpha/2),\alpha/2]$ of the
one-dimensional phase space.  In (a) the shaded regions indicate where the invariant
density has nonvanishing support.  At this resolution we see up to the formation of the
four bands at $\alpha=2^{1/4}$ but higher band-splitting points are not resolved.  In (b) iterates
of the critical trajectory as a function of $\alpha$ are plotted.  These are seen to
determine the band structure in (a).}}
\end{center}
\end{figure}
At the band-splitting points the critical trajectory is eventually 
periodic and the invariant density is constant in each band and thus piecewise-constant
over the unit interval.  But a general initial density will also have persistent oscillating
components among the bands.  This feature is known as asymptotic periodicity~\cite{Lasota}.

We want to determine the spectral decomposition of the Frobenius--Perron operator so that we may
expand a density or correlation function in terms of its eigenmodes.  In order to do this we need
the dual states or left eigenstates of $U$.  These correspond to the right eigenstates of the
adjoint of $U$, which is known as the Koopman operator~\cite{Lasota}.  The Koopman operator,
$K=U^\dagger$, acts on a phase space function $A(x)$ as
\begin{equation} \label{koopman}
K A(x) = A({\rm T}(x)),
\end{equation}
where ${\rm T}(x)$ is the rule for the map, such as~(\ref{generaltent}).
For decompositions of one-dimensional, chaotic, Markov maps in spaces spanned by polynomials the
Koopman operator has eigenstates that are generalized functions or
eigenfunctionals~\cite{firstgenspec,scndgenspec,Deanbook}.  

The decay of the $x$-autocorrelation function for the tent map at the 
band-splitting points and at values of $\alpha$
close to band-splitting points was calculated in~\cite{Mori}.   
The tent map has also been studied by
Dorfle~\cite{Dorfle} at arbitrary values of $\alpha$ in several function spaces
but he does not provide explicit complete spectral decompositions.
The asymptotic periodicity of the system has been studied by Provatas and Mackey~\cite{ProvMac}. 
Since as $t \to \infty$ the density is supported only within $[\alpha(1-\alpha/2),\alpha/2]$
all these authors have only considered the map in that region and neglected transient behavior onto
it.   That is sufficient if one only considers the behavior of time correlation functions; but for
the general evolution of densities and observables in a nonequilibrium statistical mechanics context
this transient behavior must not be neglected since, as will be seen, the slowest decay
modes originate from this part of the dynamics.

In the next section we construct the spectral decomposition of the
Frobenius--Perron operator at the first band-splitting point.   In Section 3 the decomposition at all
the band-splitting points is constructed using the self-similarity of the map at higher
band-splitting points to the map at lower band-splitting points.  \section{The first band-splitting point}

From (\ref{generaltent}) the map at the first band-splitting point corresponding to the
height $h=\sqrt{2}/2$ or $\alpha=\sqrt{2}$ is
\begin{equation} \label{mapatfbsp}
{\rm{T}}_{1}(x) = \left\{ \begin{array}{lc} 
\sqrt{2} \, x & 0 \leq x < \frac{1}{2} \\ \noalign{\vskip4pt} 
\sqrt{2} \, (1-x) & \frac{1}{2} \leq x < 1,
\end{array} \right.
\end{equation}
which is shown in Figure 1.
The dynamics is organized around four
intervals determined by the trajectory of the critical point.  At the third iteration
the critical trajectory settles onto the fixed point, $x^*=2-\sqrt{2}$.  The four
intervals: ${\rm{I}}=[0,{\rm{T}}_{1}^{(2)}(x_{\rm c}))$, 
${\rm{II}}=[{\rm{T}}_{1}^{(2)}(x_{\rm c}),{\rm{T}}_{1}^{(3)}(x_{\rm c}))$
${\rm{III}}=[{\rm{T}}_{1}^{(3)}(x_{\rm c}),{\rm{T}}_{1}^{(1)}(x_{\rm c}))$ and
${\rm{IV}}=[{\rm{T}}_{1}^{(1)}(x_{\rm c}),1)$,  define a minimal Markov partition
for the map. (These intervals are indicated in Figure 1.)  Any point in the interior of
interval ${\rm{IV}}$ is mapped onto some point in interval ${\rm{I}}$ in one
iteration.  Under successive iterations all points in the interior of ${\rm{I}}$ are
eventually mapped into ${\rm{II}}$. Any point in interval $\rm{II}$ maps into interval
$\rm{III}$ in one iteration and any point in interval $\rm{III}$ maps into
$\rm{II}$ in one iteration.  Thus the union of the intervals $\rm{II}$ and $\rm{III}$ 
form the attracting set $\Omega$. 

The Frobenius--Perron operator for the map (\ref{mapatfbsp}) acts on 
a density as 
\begin{equation} \label{op}
U_{\rm T_1} \rho(x) = \frac{1}{\sqrt{2}}\left[ \rho\bigg(
\frac{x}{\sqrt{2}} \bigg) + \rho\bigg(\frac{\sqrt{2} - x}{\sqrt{2}} 
\bigg) \right]\Theta \bigg( \frac{\sqrt{2}}{2} - x \bigg).   
\end{equation}
A general initial density continuous over the unit interval develops
discontinuities at the endpoints of the four intervals described above. 
We thus choose a function space to consider $U_{\rm T_1}$ in as the space
of piecewise-polynomial functions where the pieces are  the four
intervals described above.  The invariant density has support only in
$\Omega$. 

The fact that points oscillate between the intervals $\rm{II}$ and $\rm{III}$ means
that a general density will have a persistent oscillating component between these two
intervals under time evolution.  This property  we will sometimes refer to in the
present context as the ``flip property", since the part of the density in $\rm{II}$
will all be in $\rm{III}$ (with stretching) in the next time step and vice-versa. The parts of the
initial density with support in the intervals
${\rm{I}}$ and $\rm{IV}$ will decay onto $\Omega$.  Since we will use the eigenfunctions on
$\Omega$ to determine those on its complement, we first consider the decomposition on $\Omega$.

\subsection{Evolution on the attractor}

For convenience $\Omega$ is stretched
onto the interval $[0,1)$. At the end of the computation we will rescale all
results back to $\Omega$. The linear function that makes the stretch is
\begin{equation} \label{phi}
\phi(x) = 
(2/x^{*}) x - \sqrt{2},
\end{equation}
where ${\rm{T}}_{1}^{(2)}(x_{\rm c}) \leq x < {\rm{T}}_{1}^{(1)}(x_{\rm c})$.
The transformation (\ref{phi}) is a homeomorphism so that it begets a new
map ${\rm R_1}$ topologically conjugate to the part of ${\rm T_1}$ on $\Omega$ as
$\rm{R}_{1} = \phi \circ {\rm{T}}_{1} \circ \phi^{-1}$
given by 
\begin{equation} \label{rescaled}
{\rm{R}}_{1}(x) = \left\{ \begin{array}{lc}
\sqrt{2} \, x + x^{*} & 0 \leq x < x^{*} \! /2 \\ \noalign{\vskip4pt}
\sqrt{2} \, (1-x) & x^{*} \! /2 \leq x < 1, 
\end{array} \right.
\end{equation}
where $x^{*} = 2 - \sqrt{2}$ is the fixed point of the map ${\rm{R}}_{1}(x)$,
which is the same as
the fixed point of the map ${\rm{T}}_{1}(x)$. Under this transformation, the
intervals $\rm{II}$ and $\rm{III}$
are stretched to the intervals ${\rm{A}} \equiv [0,x^{*})$ and ${\rm{B}} \equiv
[x^{*},1)$ respectively. The map ${\rm{R}}_{1}(x)$ is shown in Figure 3.   
\begin{figure}[htb]
\begin{center}
\scalebox{.5}[.5]{\includegraphics{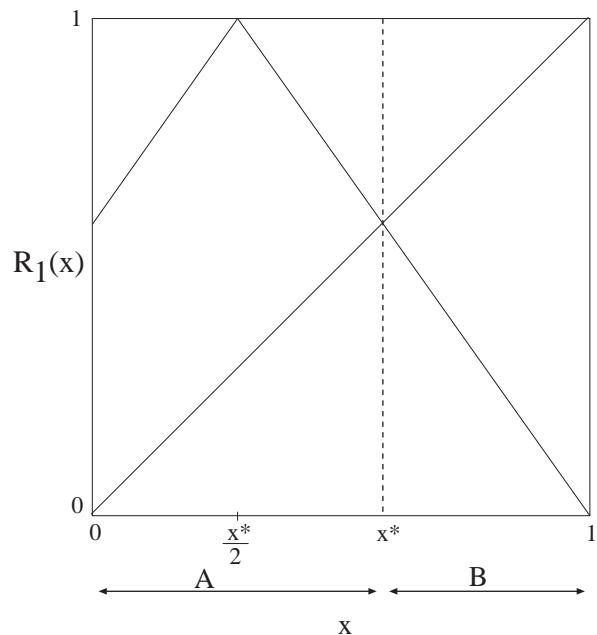}}
\parbox{5in}{\caption{\small The rescaled Tent map at the first band-splitting point.}}
\end{center}
\end{figure}

The Frobenius--Perron operator corresponding to the
rescaled map
$\rm{R}_{1}$ acts on a density as
\begin{equation} \label{ur1}
U_{\rm{R}_{1}} \rho(x) = \frac{1}{\sqrt{2}}\left[ \rho\bigg(
\frac{\sqrt{2} - x}{\sqrt{2}} \bigg) + \rho\bigg(\frac{x - x^{*}}{\sqrt{2}} 
\bigg)\Theta(x - x^{*}) \right].  
\end{equation}
The flip property of $\rm T_1$ is inherited by $\rm R_1$ in that the inverse image of
$\rm A$ is $\rm B$ and vice-versa.  This suggests that a simpler analysis will be
obtained by considering the map corresponding to two iterations of $\rm R_1$.  This map,
${\rm{G}}_{1} \equiv \rm{R}_{1} \circ \rm{R}_{1}$, is given by 
\begin{equation}
{\rm G}_{1}(x) = \left\{ 
\begin{array}{lc} 
-2x + x^{*} & 0 \leq x < x^{*} \! /2 \\ \noalign{\vskip4pt}
 2x - x^{*}  & x^{*} \! /2 \leq x < (1+x^*)/2 \\ \noalign{\vskip4pt}
-2x + (2 + x^{*}) & (1+x^*)/2 \leq x < 1.
\end{array} \right. 
\end{equation}
The flip property of $\rm{R}_{1}$ means that ${\rm{G}}_{1}$
is metrically  decomposable into two independent maps on the intervals
$\rm{A}$ and
$\rm{B}$, as is clear from Figure 4. From now on in this section we shall
drop the subscript $1$ on the maps $\rm R_1$ and $\rm G_1$; it being
understood that we are referring to these maps at the first
band-splitting point. 
\begin{figure}[htb]
\begin{center}
\scalebox{.5}[.5]{\includegraphics{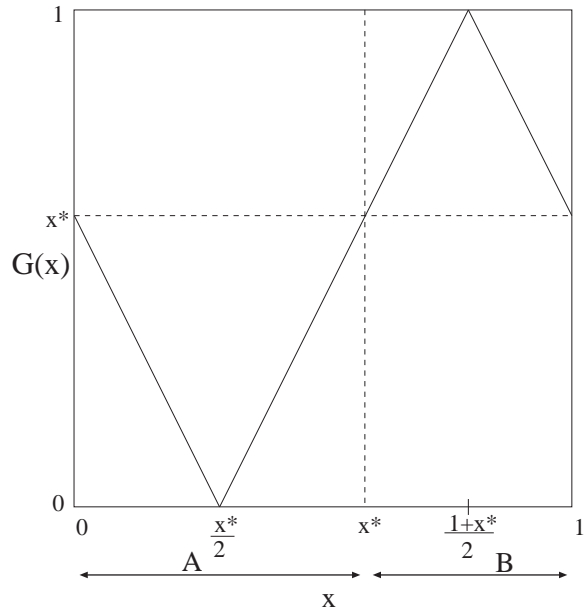}}
\end{center}
\parbox{5in}{\caption{\small The map ${\rm{G_{1}}}={\rm{R_{1} \circ R_{1}}}$ is metrically
decomposable into two parts each conjugate to the tent map with unit height.} }
\end{figure}

The map ${\rm G}$ restricted to $\rm{A}$ is just a rescaling (with a
flip) of the tent map with full height, ${\rm T}_0$,  i.e., the map
(\ref{generaltent}) with $\alpha=2$. This is expressed in terms of a
topological conjugacy as
\begin{equation}
{\rm{G}}_{\rm{A}}(x) = \phi^{-1}_{\rm{A}}(x) \circ {\rm{T_0}}(x) \circ
\phi_{\rm{A}}(x),
\end{equation}
where
the conjugating function $\phi_{\rm{A}}(x)$ is 
\begin{equation}
\phi_{\rm{A}}(x) = 1 - \frac{x}{x^*},
\end{equation}
and $x \in [0, x^{*})$.
Similarly the map on $\rm B$ is topologically conjugate to ${\rm T}_0$ as
\begin{equation}
{\rm{G}}_{\rm{B}}(x) = \phi^{-1}_{\rm{B}}(x) \circ {\rm{T}}_{0}(x) \circ
\phi_{\rm{B}}(x),
\end{equation}
where the conjugating function $\phi_{\rm B}(x)$ is 
\begin{equation}
\begin{array}{lc}
\phi_{\rm{B}}(x) = 
(\sqrt{2}/x^{*})x - \sqrt{2} ,
\end{array} 
\end{equation}
and here $x \in [x^{*}, 1)$.
These conjugacies are useful for us because the
spectral decompositions of maps that are topologically conjugate are simply
related, as is reviewed in Appendix A.
The generalized spectral decomposition of $\rm{T_{0}}$ has been
previously determined~\cite{Gonzalo,fox} and is reviewed
in Appendix B.  

Following the discussion in those appendices gives the right eigenvectors of
$\rm{G}_{\rm{A}}$ and $\rm{G}_{\rm{B}}$ as
\alpheqn
\begin{eqnarray} \label{Gvecsaa}
| 2^{-2j} \rangle_{\rm{G}_{\rm{A}}} & = & \frac{1}{x^{*}}
B_{2j}\left( \frac{x^{*} - x}{2x^{*}} \right)\chi_{\rm{A}} \\ \label{Gvecsab}
| 0_{2j+1} \rangle_{\rm{G}_{\rm{A}}} & = & \frac{1}{x^{*}}
E_{2j+1}\left( \frac{x}{x^{*}} \right)\chi_{\rm{A}}, 
\end{eqnarray}
\reseteqn
\alpheqn
\begin{eqnarray} \label{Gvecsba}
| 2^{-2j} \rangle_{\rm{G}_{\rm{B}}} & = & \frac{\sqrt{2}}{x^{*}}
B_{2j}\left( \frac{x - x^{*}}{\sqrt{2}x^{*}} \right)\chi_{\rm{B}} \\ \label{Gvecsbb}
| 0_{2j+1} \rangle_{\rm{G}_{\rm{B}}} & = & \frac{\sqrt{2}}{x^{*}}
E_{2j+1}\left( \frac{\sqrt{2}}{x^{*}}\left( x-x^{*} \right) \right)
\chi_{\rm{B}},
\end{eqnarray}
\reseteqn
where the associated eigenvalue is the argument of the
ket vector with $|0_{2j+1}\rangle$ meaning a null eigenpolynomial of degree $2j+1$
and $\chi_{\rm{A}}$ and $\chi_{\rm{B}}$ are indicator functions on the intervals
$\rm{A}$ and $\rm{B}$ respectively.  Due to the metric decomposability of $\rm G$ these
states are eigenstates of $U_{\rm G}$ as well.

Similarly, the left eigenvectors of
$\rm{G}_{\rm{A}}$ and $\rm{G}_{\rm{B}}$ are the generalized functions
\alpheqn 
\begin{eqnarray} \label{lgvecaa}
\langle 2^{-2j} |_{\rm{G}_{\rm{A}}} & = & 
\frac{(-1)^{2j-1}\left( 2x^{*} \right)^{2j}}{(2j)!}\left[
\delta^{(2j-1)}_{-}(x - x^{*}) - \delta^{(2j-1)}_{+}(x) \right] \\ \label{lgvecab}
\langle 0_{2j+1} |_{\rm{G}_{\rm{A}}} & = & -\frac{\left(x^{*}\right)^{2j+2}}
{(2j+1)!}\delta^{(2j+1)}_{+}\left( x \right),  
\end{eqnarray}
\reseteqn
\alpheqn
\begin{eqnarray} \label{lgvecba}
\langle 2^{-2j} |_{\rm G_B} 
& = & \frac{(-1)^{2j-1}\left( \sqrt{2} \, x^{*} \right)^{2j}}{(2j)!}\left[
\delta^{(2j-1)}_{-}(x - 1) - \delta^{(2j-1)}_{+}
(x - x^{*}) \right] \\ \label{lgvecbb}
\langle 0_{2j+1} |_{\rm G_B} & = & \frac{-1}{(2j+1)!}
\left( \frac{x^{*}}{\sqrt{2}} \right)^{2j+2}
\delta^{(2j+1)}_{-}\left( x - 1 \right),
\end{eqnarray}
\reseteqn
where the definitions of $\delta_{\pm}$ are given in Appendix B.

Since $U_{\rm{G}} = U^{2}_{\rm{R}}$ the spectrum of $U_{\rm{R}}$ is a subset of
$\{0,\pm 2^{-j} \}$. Consider a non-zero eigenvalue, 
${2}^{-2j}$, of $U_{\rm{G}}$. There are two
eigenvectors associated with this eigenvalue, each a polynomial of
order $2j$. Since the function space on which $U_{\rm{R}}$ acts has two basis
elements for each degree $j$, i.e., an $j^{\rm th}$ degree polynomial in ${\rm A}$ and an
$j^{\rm th}$ degree polynomial in
${\rm B}$, there should be either two eigenvectors or one eigenvector and one Jordan
vector that are polynomials of degree $2j$, associated with either one or both the
eigenvalues $\{ +2^{-j},-2^{-j} \}$. Since $U_{\rm{G}}$ does not have any Jordan vectors
it follows that $U_{\rm{R}}$ doesn't either (for non-zero eigenvalues). 
The eigenvalues of $U_{\rm R}$ cannot be twofold degenerate since that
would imply that all the eigenvectors of
$U_{\rm{G}}$ are also  eigenvectors of $U_{\rm{R}}$, which is impossible since $\rm{G}$ is
metrically decomposable and $\rm{R}$ has the flip property. Therefore the
non-zero eigenvalues of $\rm{R}$ are  $+2^{-j}$ and $-2^{-j}$.

The eigenvectors of
$U_{\rm{R}}$ with eigenvalue $\pm 2^{-j}$ are in the eigenspace 
spanned by the two eigenvectors of $U_{\rm{G}}$ corresponding to
$+ 2^{-2j}$. Thus they will be linear combinations as
\alpheqn 
\begin{equation} \label{reigenforma}
|{+2^{-j}} \rangle_{\rm{R}} = \frac{1}{2} \left( |{+2^{-2j}}
\rangle_{\rm{G}_{\rm{A}}} + c_{j}|{+2^{-2j}} \rangle_{\rm{G}_{\rm{B}}} \right)
\end{equation}
and
\begin{equation}
|{-2^{-j}} \rangle_{\rm{R}} = \frac{1}{2} \left( |{+2^{-2j}} \label{reigenformb}
\rangle_{\rm{G}_{\rm{A}}} + d_{j}|{+2^{-2j}} \rangle_{\rm{G}_{\rm{B}}} \right) , 
\end{equation}
\reseteqn
where the coefficent of $1/2$ is put for convenient normalization.  To determine
$c_j$ we use that $|{+2^{-j}} \rangle_{\rm R}$ is an eigenvector of 
$U_{\rm{R}}$ as
\begin{equation}
U_{\rm{R}}\left[ |{+2^{-2j}} \rangle_{\rm{G}_{\rm{A}}} + 
c_{j}|{+2^{-2j}} \rangle_{\rm{G}_{\rm{B}}} \right] =
2^{-j}\left[ |{+2^{-2j}}\rangle_{\rm{G}_{\rm{A}}} + 
c_{j}|{+2^{-2j}} \rangle_{\rm{G}_{\rm{B}}} \right]. 
\end{equation}
The flip property tells us that 
\begin{equation} \label{nn}
U_{\rm{R}}\left[ c_{j}|{+2^{-2j}} \rangle_{\rm{G}_{\rm{B}}} \right]
 = 2^{-j}|{+2^{-2j}}\rangle_{\rm{G}_{\rm{A}}}. 
\end{equation}
Substituting the explicit form (\ref{Gvecsaa}) of $|{+2^{-2j}} \rangle_{\rm{G}_{\rm{A}}}$
and (\ref{Gvecsba}) of $|{+2^{-2j}} \rangle_{\rm{G}_{\rm{B}}}$ in (\ref{nn})  and solving
for $c_{j}$ we find that
$c_{j} = 2^{-j}$. A similar analysis  shows that $d_{j} = -2^{-j}$. Hence 
\begin{equation} \label{fbspurstates}
|{\pm 2^{-j}} \rangle_{\rm R}  =  \frac{1}{2 x^*}\left( B_{2j}\left( \frac
{x^{*} - x}{2x^{*}} \right)\chi_{\rm{A}} \pm \frac{\sqrt{2}}{2^{j}}B_{2j}\left(
\frac {x - x^{*}}{\sqrt{2}x^{*}} \right)\chi_{\rm{B}} \right). 
\end{equation}

The invariant state, corresponding to the invariant density of $U_{\rm R}$ is
\begin{equation}
|{+1}\rangle_{\rm R} = \frac{1}{2 x^*}(\chi_{\rm A} + \sqrt{2} \, \chi_{\rm B}).
\end{equation}
This state carries all the probability under evolution of $U_{\rm R}$ and any density
will have this component.  The state
\begin{equation}
|{-1}\rangle_{\rm R} = \frac{1}{2 x^*}(\chi_{\rm A} - \sqrt{2} \, \chi_{\rm B})
\end{equation}
is the asymptotically periodic state.  Only it and the invariant density survives as $t
\to \infty$; but $|{-1}\rangle_{\rm R}$, like decaying states, doesn't carry any probability. 
It does keep the memory though of the projection of the initial density on
$|{-1}\rangle_{\rm R}$, which is a special property of asymptotically periodic
systems~\cite{Lasota}.

Now we consider the null space of $\rm{U_{R}}$.  The map $\rm{G}$ has two independent
null vectors (one in $\rm A$ and one in $\rm B$) for each odd degree.  
This implies that
$\rm{R}$ can have either a corresponding $2 \times 2$ Jordan block 
or have 2 independent eigenvectors for each odd degree. The latter case is not possible
since null vectors of $\rm{R}$ cannot have support in interval $\rm{B}$ because only one
of the terms on the rhs of (\ref{ur1}) acts on  functions in $\rm B$.
Thus there is a $2 \times 2$ Jordan block for each odd degree associated with
eigenvalue $0$.

Consider the action of $U_{\rm G}
=U^{2}_{\rm{R}}$ on a null state, $| 0_{2j+1}\rangle_{\rm{G_{A}}}$, of
$\rm{G_A}$ as  
\begin{equation} 
U_{\rm{R}}\Big[ U_{\rm{R}}|0_{2j+1} \rangle_{\rm{G}_A} \Big] = 0.
\end{equation} 
The function inside the square brackets has support only in $\rm B$, and
$U_{\rm{R}}$ acting on any non-zero function with support in $\rm B$ cannot vanish
in one iteration. Thus $|0_{2j+1} \rangle_{\rm R}=|0_{2j+1} \rangle_{\rm
G_A}$ is a null vector of $U_{\rm{R}}$ with explicit form given in
(\ref{Gvecsab})

The Jordan vector, $|0_{J_{2j+1}} \rangle_{\rm R}$, associated with this eigenvector
satisfies
\begin{equation} \label{q}
 U_{\rm{R}}| 0_{J_{2j+1}} \rangle_{\rm{R}} = | 0_{2j+1}
\rangle_{\rm{R}}.
\end{equation} 
We may choose the Jordan vector to have support only
in ${\rm B}$ as
\begin{equation} \label{qq}
| 0_{J_{2j+1}} \rangle_{\rm{R}} = \eta_{2j+1}| 0_{2j+1}
\rangle_{\rm{G}_{\rm{B}}},
\end{equation} 
where $\eta_{2j+1}$ is a constant to be determined.  
To determine $\eta_{2j+1}$ we apply $U_{\rm R}$ on (\ref{qq}) and use (\ref{q}) and the
explicit forms (\ref{Gvecsab}) and (\ref{Gvecsbb}) (remembering that
$|0_{2j+1} \rangle_{\rm R}=|0_{2j+1} \rangle_{\rm G_A}$) to obtain
$\eta_{2j+1} = -1$.

\subsubsection{Left eigenstates of $U_{\rm{R}}$}

The left eigenstates of $U_{\rm{R}}$ with non-zero eigenvalues may be determined by an
approach similar to the one used to find the right eigenstates.  
The result is
\begin{eqnarray}
\langle \pm 2^{-j}|_{\rm{R}} & = & \frac{\left( 2x^{*} \right)^{2j}}
{(2j)!} \left[ \delta^{(2j-1)}_{+}(x) - \delta^{(2j-1)}_{-}(x - x^{*})
\right. \nonumber \\ & & \left. \hskip50pt
\pm \delta^{(2j-1)}_{+}(x - x^{*}) \mp \delta^{(2j-1)}_{-}(x - 1) \right].
\end{eqnarray} 
The left states with zero eigenvalues are like the right states identical to those
of $U_{\rm G}$ (except for the factor of $-1$ for the dual state of the Jordan vector) as
\alpheqn  
\begin{eqnarray}
\left\langle 0_{2j+1} \right|_{\rm R} & = &  
\left\langle 0_{2j+1} \right|_{\rm G_A} \\
\langle 0_{J_{2j+1}} |_{\rm{R}} & = & 
- \left\langle 0_{2j+1} \right|_{\rm G_B}. 
\end{eqnarray}
\reseteqn
Note that $\langle 0_{J_{2j+1}} |_{\rm{R}}$ is an eigenstate of the Koopman operator
and $\left\langle 0_{2j+1} \right|_{\rm{R}}$ is a Jordan state.

\subsection{Decay onto the attractor}

As noted before, initial densities with support in the intervals $\rm{I}$ and/or
$\rm{IV}$ will decay into $\Omega$. Consider a density with support
only in $\rm{IV}$ at $t=0$.  At $t=1$  the density has support only in ${\rm{I}}$.  Thus
any eigenvector of $U_{\rm T_1}$ with support in $\rm{IV}$ can only have
eigenvalue
$0$.  To determine such an eigenvector we write an ansatz for it as 
\begin{equation} \label{aa}
| 0_j \rangle_{\rm{IV}} =
f_{{\rm{I}},j}(x)\chi_{{\rm{I}}} +
f_{{\rm{II}},j}(x)\chi_{\rm{II}}
+ f_{{\rm{III}},j}(x)\chi_{\rm{III}} + f_{{\rm{IV}},j}(x)\chi_{\rm{IV}},
\end{equation}
where the subscript $\rm IV$ on the ket denotes that it
describes decay out of interval $\rm IV$ and we take $f_{{\rm IV},j}(x)$ as a
polynomial of order $j$. Applying $U_{\rm T_1}$ to (\ref{aa}) and collecting terms that are
multiplied  by the same indicator function we get
\begin{eqnarray} \label{aaa}
U_{\rm T_1}|0_j \rangle_{\rm{IV}} & = & 
\left[ f_{{\rm{I}},j} \left( \frac{x}{\sqrt{2}} 
\right) + f_{{\rm{IV}},j}\left(1- \frac{x}{\sqrt{2}} 
\right) \right]\chi_{{\rm{I}}} \nonumber \\
 &  & \mbox{} + \left[ f_{{\rm{I}},j} \left(
\frac{x}{\sqrt{2}} 
\right) + f_{{\rm{III}},j}\left(1- \frac{x}{\sqrt{2}} \right) \right]
\chi_{\rm{II}} \nonumber \\ 
&  & \mbox{} + \left[ f_{{\rm{II}},j} \left( \frac{x}{\sqrt{2}} 
\right) + f_{{\rm{II}},j}\left(1- \frac{x}{\sqrt{2}} \right)
\right]\chi_{\rm{III}}.
\end{eqnarray} 
Since $|0_j \rangle_{\rm{IV}}$ is a null eigenstate the coefficients of each of the
indicator functions must vanish. Since
$f_{{\rm{IV}},j}$ is a polynomial of degree
$j$, it follows that 
$f_{{\rm{I}},j}$ and
$f_{{\rm{III}},j}$ are also polynomials of degree $j$. We choose
$f_{{\rm{II}},j}$ to be zero.  Clearly,
$f_{{\rm{III}},j}=f_{{\rm{IV}},j}$ and choosing it as
$(x-1)^j$ fixes $f_{{\rm{I}},j}$.  We thus have 
\begin{equation} \label{rt1} 
| 0_j  \rangle_{\rm{IV}} = (-1)^{j+1}x^j\chi_{\rm{\rm{I}}} +
(x-1)^j(\chi_{\rm{III}} + \chi_{\rm{IV}}). 
\end{equation} 
We note that because of the degeneracy associated
with eigenvalue $0$ in interval $\rm{IV}$, this choice of eigenvectors is not
unique. The left eigenstates (given below) associated with this choice of
right  eigenvectors is unique and therefore when we expand an arbitrary 
density in terms of the right eigenvectors, the expansion coefficents 
are uniquely defined.

The action of $U_{\rm T_1}$ on a function with support only 
on $\rm{\rm{I}}$ is given by
\begin{equation} U_{\rm T_1} [f(x)\chi_{\rm{I}}] = \frac{1}{\sqrt{2}}
f(x/\sqrt{2}) ( \chi_{\rm{I}} + \chi_{\rm{II}}).
\end{equation} 
Acting with $U_{\rm T_1}$ on a monomial in $\rm{I}$ gives
\begin{equation} \label{ind}
 U_{\rm T_1} [x^{j}\chi_{\rm I}] =
\frac{x^{j}}{(\sqrt{2})^{j+1}} (\chi_{\rm{\rm{I}}} + \chi_{\rm{II}}), 
\end{equation}
so that the eigenvectors of $U_{\rm T_1}$ with support in $\rm{I}$
are monomials in $\rm{I}$.
To determine their form in the other intervals we again write an ansatz as
we did for the eigenvectors with support in $\rm{IV}$ as
\begin{equation} \label{haha} 
| 2^{-(j+1)/2}
\rangle_{{\rm{I}}} = x^{j}\chi_{{\rm{I}}} + g_{{\rm II},j}(x) \chi_{\rm{II}} +
g_{{\rm III},j}(x)\chi_{\rm{III}},
\end{equation}
where the associated eigenvalue, appearing as the argument of the ket, is seen from (\ref{ind}). In
equation (\ref{haha}) and below the subscript $\rm I$ on a ket implies that
it describes decay out of region ${\rm I}$. Note that the
$j=0$ mode here is the slowest decay mode in this system. Since $U_{\rm T_1}$
does not raise the degree of the polynomial it acts on, 
$g_{{\rm II},j}(x)$ and $g_{{\rm III},j}(x)$ are polynomials of degree $j$. Applying
$U_{\rm T_1}$ to (\ref{haha}) and using (\ref{ind}) and the fact that the function on
the rhs of (\ref{haha}) is an eigenvector with eigenvalue
$2^{-(j+1)/2}$, we find that 
\begin{equation}  \label{form1}
U_{\rm T_1} ( g_{{\rm II},j}(x)\chi_{\rm{II}} + g_{{\rm III},j}(x)\chi_{\rm{III}}) =
2^{-(j+1)/2}  (g_{{\rm II},j}(x)\chi_{\rm{II}} + g_{{\rm III},j}(x)\chi_{\rm{III}}
-x^{j}\chi_{\rm{II}}). 
\end{equation}
The determination of $g_{{\rm II},j}(x)$ 
and $g_{{\rm III},j}(x)$ is described in Appendix C 

The explicit form of the first few of these eigenvectors is
\alpheqn
\begin{eqnarray} \label{firstfew}
| 2^{-1/2}\rangle_{\rm I} & = & \chi_{{\rm{I}}} +
\frac{1}{2(1-\sqrt{2})}|1\rangle_{\Omega} + \frac{1}{2(1+\sqrt{2})}|{-1\rangle_\Omega} \\
| 2^{-1}\rangle_{\rm I} & = & x \, \chi_{\rm{I}} -
\frac{1}{4(x^*)}|1\rangle_\Omega + \frac{1}{12}|{-1\rangle_\Omega} +
\frac{(x^*)^2}{2}|0\rangle_\Omega \\ 
|2^{-3/2}\rangle_{\rm I} & = & x^2 \,\chi_{\rm{I}} -
\frac{\sqrt{2}x^*}{12}|1\rangle_\Omega +
\frac{(9\sqrt{2}-8)x^*}{84}|-1\rangle_\Omega + 
\frac{\sqrt{2}(x^*)^3}{2}|0\rangle_{\Omega} \\ 
& & \mbox{}\hspace{1cm} - \frac{\sqrt{2}(x^*)^3}{2}|2^{-1}\rangle_\Omega +
\frac{(x^*)^4}{2(1+\sqrt{2})}|{-2^{-1}\rangle_\Omega}.
\end{eqnarray}
\reseteqn 
The subscript $\Omega$ on a ket implies that it is a right state of 
$U_{\rm R}$ which has been rescaled into the attractor, $\Omega$, of $U_{\rm T_1}$.
The explicit form of these states is given in Table~1.

\subsection{Left eigenstates of $U_{\rm T_1}$}

It is easily seen that the left states given by 
\begin{equation} \label{lt1}
\langle 0_j |_{\rm{IV}} = \frac{(-1)^j}{j!}\delta_{-}^{(j)}(x-1)
\end{equation}
form a bi-orthonormal set with the null states in~(\ref{rt1}).
These left states are also orthogonal to all the other right states of $U_{\rm T_1}$.
The left states associated with the transient decay states out of $\rm{I}$ are given by
\begin{equation} \label{l1}
\langle {2}^{-(j+1)/2}|_{\rm I} = \frac{1}{j!}
\left[ (-1)^j\delta_{+}^{(j)}(x) + \delta_{-}^{(j)}(x-1) \right]. 
\end{equation}

The left states of $U_{\rm R}$ are not the left states of $U_{\rm T_1}$, even though
the right states of $U_{\rm R}$ are also right states of $U_{\rm T_1}$. This is
because $U_{\rm T_1}$ acting on a density contained within $\Omega$ will
continue to be in $\Omega$.  But in general the
Koopman operator, $K_{\rm T_1}$, acting on a function with support only in $\Omega$
will result in the function having support outside $\Omega$ too.
The left states of $U_{\rm R}$ scaled back to $\Omega$ form a bi-orthonormal set with the
right states contained in $\Omega$, but they are not orthogonal to the transient
states that decay into $\Omega$.   To make them so we use Gram--Schmidt orthogonalization.  The
results are given in Table 1.  

\subsection{The spectral decomposition}

Using all the eigenstates and eigenvalues given in Table~1 we may write
the action of $U_{\rm T_1}^t$ in terms of its spectral decomposition as
\begin{eqnarray} \label{specdec}
U_{\rm T_1}^{t} & = & | 1 \rangle\langle 1 |_{\Omega} + 
(-1)^t|{-1 \rangle}\langle{-1 |_\Omega} + \sum_{j=0}^{\infty}
\left( 2^{-(2j+1)/2} \right)^t|2^{-(2j+1)/2}\rangle\langle
2^{-(2j+1)/2}|_{\rm{I}} \nonumber \\ 
& & + \sum_{j=1}^{\infty}(2^{-j})^t \left[ | {+2^{-j}}
\rangle\langle {+2^{-j} |_\Omega} +
|2^{-j}\rangle\langle 2^{-j}|_{\rm{I}} \right] \nonumber \\ 
& & + \sum_{j=1}^{\infty}(-2^{-j})^t | {-2^{-j}}
\rangle\langle {-2^{-j} |_\Omega} +
\sum_{j=0}^{\infty}\delta_{1,t} | 0_{2j+1}
\rangle\langle 0_{J_{2j+1}} |_\Omega \nonumber \\ 
& & + \sum_{j=0}^{\infty}\delta_{0,t} \left[ |0_{2j+1}
\rangle\langle 0_{2j+1} |_{\Omega} + |0_{J_{2j+1}}
\rangle\langle 0_{J_{2j+1}} |_{\Omega} + |0_j
\rangle\langle 0_j|_{\rm{IV}} \right], 
\end{eqnarray}
where the subscript on the bra states also identifies their dual ket states.

\clearpage

\begin{table}[p!] 
\[
\begin{array}{|c|c|l|l|} \hline 
\mbox{eigenvalue} & \mbox{degeneracy} & \mbox{symbol} & \mbox{eigenvector} \\ \hline 
1 & 1 & | 1 \rangle_{\Omega} & \chi_{\rm{II}} + \sqrt{2}\chi_{\rm{III}} \\
\vspace{-1mm}& \vspace{-1mm} & \vspace{-1mm} & \vspace{-1mm}\\ 
 &  & \langle 1 |_{\Omega} & \frac{(x^*)^2}{2}(\chi_{\rm{II}} + \chi_{\rm{III}}) \\
& & & \mbox{}-\sum_{k=0}^{\infty}\left(a_{0,k}\langle 2^{-(k+1)/2}|_{\rm{I}} +
\alpha_{0,k}'\left\langle 0_k \right|_{\rm{IV}} \right) \\
\vspace{-1mm}& \vspace{-1mm} & \vspace{-1mm} & \vspace{-1mm}\\ 
\hline 
\vspace{-1mm}& \vspace{-1mm} & \vspace{-1mm} & \vspace{-1mm}\\ 
-1 & 1 & | {-1 \rangle_\Omega} & \chi_{\rm{II}} - \sqrt{2}\chi_{\rm{III}} \\
\vspace{-1mm}& \vspace{-1mm} & \vspace{-1mm} & \vspace{-1mm}\\ 
 &  & \langle -1 |_{\Omega} & \frac{(x^*)^2}{2}(\chi_{\rm{II}} - \chi_{\rm{III}}) \\
& & &\mbox{}-
\sum_{k=0}^{\infty}\left( b_{0,k}\langle 2^{-(k+1)/2}|_{\rm{I}} -
\alpha_{0,k}'\left\langle 0_k \right|_{\rm{IV}} \right) \\
\vspace{-1mm}& \vspace{-1mm} & \vspace{-1mm} & \vspace{-1mm}\\
\hline
\vspace{-1mm}& \vspace{-1mm} & \vspace{-1mm} & \vspace{-1mm}\\ 
+\frac{1}{2^j} & 1 & |{+ 2^{-j} \rangle_\Omega} &
B_{2j}\left( \frac{x^* - \phi(x)}{2x^*} \right)\chi_{\rm{II}} +
\frac{\sqrt{2}}{2^j}B_{2j}\left( \frac{\phi(x) -
x^*}{\sqrt{2}x^*}\right)\chi_{\rm{III}} \\
\vspace{-1mm}& \vspace{-1mm} & \vspace{-1mm} & \vspace{-1mm}\\ 
 &  & \langle +2^{-j} |_{\Omega} & \frac{\left( x^{*} \right)^{4j-2}}
{(2j)!} \left[ \delta^{(2j-1)}_{+}(x-{\rm{T}}^{(2)}(x_c))
 - \delta^{(2j-1)}_{-}(x -x^{*}) \right. \\ 
& & & \mbox{} \left. 
 + \delta^{(2j-1)}_{+}(x - x^{*}) -
\delta^{(2j-1)}_{-}(x - {\rm{T}}^{(1)}(x_c)) \right] \\ & & & - 
\sum_{k=2j}^{\infty}\left( a_{j,k}\langle 2^{-(k+1)/2}|_{\rm{{I}}} +
\alpha_{j,k}'\left\langle 0_k \right|_{\rm{IV}} \right) \\
\vspace{-1mm}& \vspace{-1mm} & \vspace{-1mm} & \vspace{-1mm}\\
 \hline   
\vspace{-1mm}& \vspace{-1mm} & \vspace{-1mm} & \vspace{-1mm}\\
-\frac{1}{2^j} & 1 & |{- 2^{-j} \rangle_\Omega} &
B_{2j}\left( \frac{x^* - \phi(x)}{2x^*} \right)\chi_{\rm{II}} -
\frac{\sqrt{2}}{2^j}B_{2j}\left( \frac{\phi(x) -
x^*}{\sqrt{2}x^*}\right)\chi_{\rm{III}} \\
\vspace{-1mm}& \vspace{-1mm} & \vspace{-1mm} & \vspace{-1mm}\\ 
 &  & \langle -2^{-j} |_{\Omega} &  \frac{\left( x^{*}
\right)^{4j-2}} {(2j)!} \left[ \delta^{(2j-1)}_{+}(x-{\rm{T}}^{(2)}(x_c))
- \delta^{(2j-1)}_{-}(x - x^{*}) \right. \\
& &  & \mbox{} \left. 
 - \delta^{(2j-1)}_{+}(x - x^{*})  +
\delta^{(2j-1)}_{-}(x - {\rm{T}}^{(1)}(x_c)) \right] \\ & & & - 
\sum_{k=2j}^{\infty}\left( b_{j,k}\langle 2^{-(k+1)/2}|_{\rm{I}} - \alpha_{j,k}'\left\langle
0_k \right|_{\rm{IV}} \right) \\
\vspace{-1mm}& \vspace{-1mm} & \vspace{-1mm} & \vspace{-1mm}\\
 \hline  
\vspace{-1mm}& \vspace{-1mm} & \vspace{-1mm} & \vspace{-1mm}\\
0 & \infty & | 0_{2j+1} \rangle_{\Omega} & E_{2j+1}\left(
\frac{\phi(x)}{x^*} \right)\chi_{\rm{II}} \\
\vspace{-1mm}& \vspace{-1mm} & \vspace{-1mm} & \vspace{-1mm}\\ 
 & ( 2 \times 2 & \left\langle 0_{2j+1} \right|_{\Omega} &
-\left(\frac{x^*}{\sqrt{2}}\right)^{4j+2}\frac{1}{(2j+1)!}
\delta^{(2j+1)}_{+}(x-{\rm{T}}^{(2)}(x_c)) \\
& \mbox{Jordan} & &\mbox{} - \sum_{k=2j+1}^{\infty}c_{j,k}
\langle 2^{-(k+1)/2}|_{\rm{I}} \\
\vspace{-1mm}& \vspace{-1mm} & \vspace{-1mm} & \vspace{-1mm}\\ 
 & \mbox{blocks})  & | 0_{J_{2j+1}} \rangle_{\Omega} &
-\sqrt{2}\left(E_{2j+1}\left(
\frac{\sqrt{2}}{x^*} (\phi(x) - x^*)\right)\right)\chi_{\rm{III}} \\
\vspace{-1mm}& \vspace{-1mm} & \vspace{-1mm} & \vspace{-1mm}\\ 
 & & \left\langle 0_{J_{2j+1}} \right|_{\Omega} &
\frac{(x^*)^{4j+2}}{\sqrt{2}(2\sqrt{2})^{2j+1}}\frac{1}{(2j+1)!}
\delta^{(2j+1)}_{-}(x-{\rm{T}}^{(1)}(x_c)) \\
& & & \mbox{}-
\sum_{k=2j+1}^{\infty}\gamma^{'}_{j,k}\left\langle 0_{k}\right|_{\rm{IV}}
\\
\vspace{-1mm}& \vspace{-1mm} & \vspace{-1mm} & \vspace{-1mm}\\
 \hline 
\vspace{-1mm}& \vspace{-1mm} & \vspace{-1mm} & \vspace{-1mm}\\
\left( \frac{1}{\sqrt{2}}\right)^{j+1} & 1 & |2^{-(j+1)/2}\rangle_{\rm{I}} &
x^j\chi_{\rm{I}} +
\sum_{i=0}a_{i,j}| 2^{-i}\rangle + b_{i,j}|
-2^{-i}\rangle + c_{i,j}|0_i\rangle \\
\vspace{-1mm}& \vspace{-1mm} & \vspace{-1mm} & \vspace{-1mm}\\ 
 & & \langle 2^{-(j+1)/2}|_{\rm{I}} & \frac{1}{j!}
\left( (-1)^j\delta^{(j)}(x) + \delta^{(j)}(x-1) \right) \\ 
\vspace{-1mm}& \vspace{-1mm} & \vspace{-1mm} & \vspace{-1mm}\\
\hline 
\vspace{-1mm}& \vspace{-1mm} & \vspace{-1mm} & \vspace{-1mm}\\
0 & \infty & |
0_j \rangle_{\rm{IV}} &   (-1)^{j+1}x^j\chi_{\rm{I}} + (x-1)^{j}\left(
\chi_{\rm{III}} + \chi_{\rm{IV}} \right) \\
\vspace{-1mm}& \vspace{-1mm} & \vspace{-1mm} & \vspace{-1mm}\\ 
 & & \left\langle 0_j \right|_{\rm{IV}} & \frac{(-1)^j}{j!}
\delta^{(j)}(x-1)  \\ \hline
 \end{array} \]
\caption{\small Elements of the spectral decomposition of the tent map at the first band splitting
point. The constants $a_{i,j}\equiv\langle
{2^{-i}}|_{\Omega}(2)^{-(j+1)/2}
\rangle_{\rm{I}}$, $b_{i,j}\equiv\langle
-{2^{-i}}|_{\Omega}(2)^{-(j+1)/2}
\rangle_{\rm{I}}$ and $c_{i,j}\equiv\langle
0_i|_{\Omega}(2)^{-(j+1)/2}
\rangle_{\rm{I}}$ and are given in Appendix C.  The
constants $\alpha_{i,j}'\equiv\langle
+{2^{-i}}|_\Omega 0_j \rangle_{\rm{IV}}$, $\beta_{i,j}'\equiv\langle -{2^{-i}}|_\Omega 0_j
\rangle_{\rm{IV}}=-\alpha_{i,j}'$ and $\gamma_{i,j}'\equiv\langle 0_{J_i}|_\Omega 0_j
\rangle_{\rm{IV}}$ and
$\phi(x)$ is defined in (\ref{phi}).}
\end{table}

\clearpage

\section{Higher band-splitting points}

We can determine the spectral decomposition of the tent map at any
band-splitting point (bsp) by generalizing the approach used in the previous
section for the first bsp. For the rescaled map at
$\alpha = \sqrt{2}$ we found that by considering its square
it separated into two parts that were
directly related by a simple change of scale (including a reflection for one
part) to the tent map at full height where
$\alpha = 2$. This relationship also holds generally~\cite{ProvMac,Heidel}
between the map at the
$n^{\rm{th}}$ bsp,
$(\alpha_{n} = 2^{2^{-n}})$, and the $(n-1)^{\rm{th}}$ bsp, 
$(\alpha_{n-1} = 2^{2^{-(n-1)}} = \alpha_{n}^2)$.

The tent map at $\alpha_n$ on the interval
$[ {\rm{T}}^{(2)}_n (x_c), {\rm{T}}^{(1)}_n (x_c))$,
which contains the attractor, is first stretched to the 
unit interval $[0,1)$ to make the rescaled map ${\rm R}_n$. 
The linear function
that makes this  stretch is 
\begin{equation} \label{genphi}
\phi_n (x) = \frac{2x - \alpha_n (2 - \alpha_n)}{\alpha_n(\alpha_n - 1)}.
\end{equation} 
Under (\ref{genphi}) the map ${\rm{T}}_{n}$ transforms to 
${\rm{R}}_n = \phi_n \circ {\rm{T}}_n \circ \phi^{-1}_n$ and
is  given by 
\begin{equation} \label{genr}
{\rm{R}}_n (x) = \left\{ \begin{array}{lc}
2-\alpha_n (1-x) & 0 \leq x < (\alpha_n -1)/ \alpha_n \\
\noalign{\vskip4pt}
\alpha_n (1- x) & (\alpha_n -1)/ \alpha_n \leq x < 1. 
\end{array} \right.
\end{equation}

We then compose ${\rm{R}}_n$ with itself to obtain
${\rm{G}}_n \equiv {\rm{R}}_n \circ {\rm{R}}_n$. 
As is illustrated in Figure 5 for the $2^{\rm{nd}}$ bsp, 1t can be shown easily that in general: 
\newline
(a) ${\rm{G}}_n(x)$ in the interval 
$X_{n, {\rm A}} \equiv [0,{\rm{R}}_n^{(4)}(x_c))$ is
topologically conjugate to
${\rm{R}}_{n-1}$ (the rescaled map at higher height) in the interval
$[0,1)$ as 
\begin{equation} \label{imp1}
{\rm{R}}_{n-1}  =  \phi_{n,\rm{A}}
\circ {\rm{G}}_{n,\rm{A}} \circ \phi_{n,\rm{A}}^{-1} , 
\end{equation}
where
\begin{equation} \label{impc1}
\phi_{n,\rm{A}}  =  1 - \frac{x}{\alpha_n (\alpha_n -1)},
 \;\;\;\;\; x \in X_{n, {\rm A}}.
\end{equation}
(b) ${\rm{G}}_n(x)$ in the interval 
$X_{n, {\rm B}} \equiv [{\rm{R}}_n^{(3)}(x_c),1)$ is
topologically conjugate to
${\rm{R}}_{n-1}$ in the interval $[0,1)$ as
\begin{equation} \label{imp2}
{\rm{R}}_{n-1}  = \phi_{n,\rm{B}} \circ
{\rm{G}}_{n,\rm{B}} \circ \phi_{n,\rm{B}}^{-1},  
\end{equation}
where
\begin{equation} \label{impc2}
\phi_{n,\rm{B}}  =  \frac{x-(2-\alpha_n)}{\alpha_n -1},
 \;\;\;\;\; x \in X_{n, {\rm B}}. 
\end{equation}
\begin{figure}[htb]
\begin{center}
\scalebox{.5}[.5]{\includegraphics{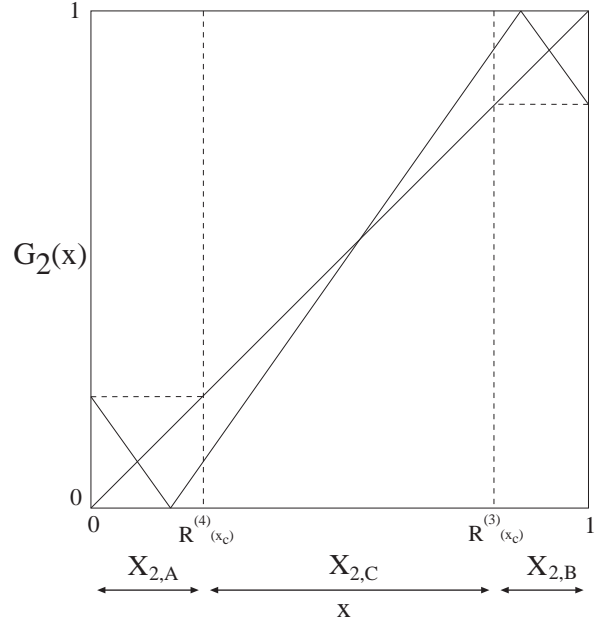}}
\parbox{5in}{\caption{\small At the second band-splitting point, ${\rm{G_{2}}}$ 
is conjugate in $X_{2,{\rm A}}$ and $X_{2,{\rm B}}$ to
$\rm{R_{1}}$ shown in  Figure 3. The central transient region 
is shown as $X_{2,{\rm C}}$.}}
\end{center}
\end{figure}

At the $n^{\rm{th}}$  bsp the critical trajectory is eventually periodic with period
$2^{n-1}$ ($n \geq 1$), and therefore the number of discontinuities that appear
under time evolution of an initially smooth density is finite. Equations (\ref{imp1}) --
(\ref{impc2}) imply that the band structure at the $n^{\rm{th}}$ bsp consists of 2 scaled
copies of the band structure at the $(n-1)^{\rm{th}}$ bsp separated by the
interval $[ {\rm{R}}_{n}^{(4)}(x_c),{\rm{R}}_{n}^{(3)}(x_c))  \equiv
X_{n, {\rm C}}$, which makes one band. Therefore we have the following
 recursion relation for $S_n$, the number of bands at the $n^{{\rm th}}$ bsp
for the map ${\rm R}_{n}$,
\begin{equation}
S_{n} = 2 S_{n-1} + 1 ,  
\end{equation}
for $n \geq 2$,
where $ S_{1} = 2$. Solving this recursion relation gives 
\begin{equation}\label{bandno}
S_{n} = 2^n + 2^{n-1} - 1, 
\end{equation}
for $n \geq 1$.
Of the $S_n$ bands, the invariant density has support only on $2^n$
bands. The density is transient on the other
$2^{n-1} -1$ bands. The function space we consider at the $n^{\rm{th}}$ bsp is
piecewise polynomial, with each piece extending over one band. 

\subsection{Decomposition on the attractor}

Associated with the $2^n$ bands that make up 
$\Omega_{n}$, the attractor at the $n^{\rm{th}}$ bsp, are  $2^n$
 eigen/Jordan vectors of each polynomial  degree
$j$. We denote an element in the spectrum of
$U_{{\rm R}_n}$ associated with states on $\Omega_{n}$ as $\lambda^{n}_{k,j}$ and the right
eigen/Jordan vector associated with it as $ |\lambda^{n}_{k,j} \rangle$. The
superscript $n$ stands for the order of the band-splitting point. The index
$k$ is an integer from $1$ to $2^n$ and distinguishes between the $2^n$
independent eigen/Jordan vectors of degree $j$, at the $n^{\rm{th}}$ bsp. 

Since ${\rm G}_{n,{\rm A}}$ and ${\rm G}_{n,{\rm B}}$  are topologically conjugate 
to ${\rm R}_{n-1}$  all three share the same spectrum. The map ${\rm G}_n$ is
the union of ${\rm G}_{n,{\rm A}}$, ${\rm G}_{n,{\rm B}}$ and the 
part of ${\rm{G}}_{n}$ on the central transient interval $X_{n,{\rm C}}$.
The spectrum of ${\rm{R}}_{n}$ on $\Omega_{n}$ is determined from the
spectrum of ${\rm{G}}_{n}$ on $\Omega_{n}$. Consider a non-zero eigenvalue
$\lambda^{n-1}_{k,2j}$ of ${\rm R}_{n-1}$ (we will show later that as at the
first bsp the right states corresponding to non-zero eigenvalues
on the attractor are even-order polynomials), which is
also an eigenvalue of ${\rm{G}}_{n}$ with degeneracy 2 
(one for ${\rm G}_{n,{\rm A}}$ and one for ${\rm G}_{n,{\rm B}}$). 
Using arguments parallel to those that we used at the first bsp
we deduce that  ${\rm{R}}_{n}$ has two distinct eigenvalues 
$+\sqrt{\lambda^{n-1}_{k,2j}}$ and $-\sqrt{\lambda^{n-1}_{k,2j}}$. By induction 
${\rm{R}}_{n+1}$ on $\Omega_{n+1}$ has in its spectrum the eigenvalues
$+(\lambda^{n-1}_{k,2j})^{1/4}$, $-(\lambda^{n-1}_{k,2j})^{1/4}$,
$+i(\lambda^{n-1}_{k,2j})^{1/4}$ and  $-i(\lambda^{n-1}_{k,2j})^{1/4}$.
Thus from
the non-zero eigenvalues at the $0^{\rm{th}}$ bsp we can determine 
the non-zero eigenvalues at the
$n^{\rm {th}}$ bsp by taking square roots of $2^{-2j}$
recursively, $n$ times. Thus 
\begin{equation} \label{eig1}
\lambda^{n}_{k,2j} =
\left(\frac{1}{2}\right)^{{2j}/{2^n}}\mbox{exp}\left(\frac{2\pi
ik}{2^n}
\right),
\end{equation}
where $k=1,2,\ldots,2^n$ and $j=0,1,2,\ldots$ .
Note that for $k \leq 2^{n-1}$ 
\begin{equation} \label{minus}
\lambda^{n}_{k,2j} = -\lambda^{n}_{k+2^{n-1},2j}.
\end{equation}

At the $0^{\rm{th}}$ bsp the eigenvalue $0$ is infinitely degenerate with
an associated eigenpolynomial of odd degree for each occurence of the
eigenvalue.  In the previous section we saw that at the $1^{\rm st}$ bsp 
there are an infinite number of  $2 \times 2$ Jordan blocks associated with
eigenvalue zero, one each for each odd degree.  We will show by explicit
construction below that this trend continues and at the $n^{\rm{th}}$ bsp, ${\rm R}_{n}$
has a $2^n \times 2^n$ Jordan block for every odd degree.  

\subsubsection{Right states on the attractor}

To determine the right states on $\Omega_{n}$ with non-zero eigenvalues we begin similar to
$(20)$ by writing
the right states at the
$n^{\rm{th}}$ bsp in terms of those at the $(n-1)^{\rm th}$ bsp as 
\alpheqn 
\begin{eqnarray}
 |\lambda^{n}_{k,2j} \rangle & = & |\lambda^{n-1}_{k,2j}
\rangle_{{\rm G}_{n,{\rm A}}} +c^{n}_{k,2j} |\lambda^{n-1}_{k,2j}
\rangle_{{\rm G}_{n,{\rm B}}} \\ 
 |\lambda^{n}_{k+2^{n-1},2j} \rangle & = & |\lambda^{n-1}_{k,2j}
\rangle_{{\rm G}_{n,{\rm A}}} +d^{n}_{k,2j} |\lambda^{n-1}_{k,2j}
\rangle_{{\rm G}_{n,{\rm B}}},
\end{eqnarray}
\reseteqn
where  $k = 1,2,\ldots,2^{n-1}$. In equation (52) and
in the remaining part of this subsection kets without
any subscript denote right states of  ${\rm{R}}$ at the appropriate bsp. 
Conjugacies (\ref{imp1}) and (\ref{imp2}) imply that the eigenvectors 
of ${\rm G}_{n,{\rm A}}$ and ${\rm G}_{n,{\rm B}}$ are related to those 
of ${\rm R}_{n-1}$ as
\alpheqn 
\begin{eqnarray}
|\lambda^{n-1}_{k,2j}\rangle_{{\rm G}_{n,{\rm A}}} & = & 
U_{\phi^{-1}_{n,\rm{A}}} |\lambda^{n-1}_{k,2j} \rangle \\ 
|\lambda^{n-1}_{k,2j}\rangle_{{\rm G}_{n,{\rm B}}} & = & 
U_{\phi^{-1}_{n,\rm{B}}} |\lambda^{n-1}_{k,2j} \rangle. 
\end{eqnarray}
\reseteqn
Using this in (52) gives
\alpheqn 
\begin{eqnarray} \label{onea} 
 |\lambda^{n}_{k,2j} \rangle & =
& U_{\phi^{-1}_{n,\rm{A}}} |
\lambda^{n-1}_{k,2j} \rangle +c^{n}_{k,2j}U_{\phi^{-1}_{n,\rm{B}}} 
 |\lambda^{n-1}_{k,2j} \rangle \\  \label{oneb}
 |\lambda^{n}_{k+2^{n-1},2j} \rangle & =
& U_{\phi^{-1}_{n,\rm{A}}} |
\lambda^{n-1}_{k,2j} \rangle +d^{n}_{k,2j}U_{\phi^{-1}_{n,\rm{B}}} 
|\lambda^{n-1}_{k,2j} \rangle.  
\end{eqnarray}
\reseteqn
The state $U_{\phi^{-1}_{n,\rm{A}}}|\lambda^{n-1}_{k,2j} \rangle$
has support only in $X_{n,{\rm{A}}}$ and 
$U_{\phi^{-1}_{n,\rm{B}}} |\lambda^{n-1}_{k,2j}
\rangle$ has support only in $X_{n,{\rm{B}}}$. 

We now use the relation
\begin{equation} \label{pmyst}
 {\rm{R}}_{n}\circ \phi_{n,\rm{B}}^{-1} = \phi_{n,\rm{A}}^{-1}, 
\end{equation}
which implies that 
\begin{equation} \label{myst}
U_{{\rm{R}}_{n}}U_{\phi_{n,\rm{B}}^{-1}} =
U_{\phi_{n,\rm{A}}^{-1}}.
\end{equation} 
(Even though (\ref{pmyst}) and (\ref{myst}) are written 
specifically at the band-splitting points, they are valid for all
$\alpha \in (1,2]$.)
Acting on (54) by $U_{{\rm{R}}_n}$ and using (\ref{myst}) we get
\alpheqn
\begin{eqnarray} \label{xiaoa}
\lambda^{n}_{k,2j} | \lambda^{n}_{k,2j} \rangle & =
& U_{{\rm{R}}_n}U_{\phi^{-1}_{n,\rm{A}}}|
\lambda^{n-1}_{k,2j} \rangle +c^{n}_{k,2j}U_{\phi^{-1}_{n,\rm{A}}} 
 |\lambda^{n-1}_{k,2j} \rangle \\ \label{xiaob} 
-\lambda^{n}_{k,2j}| \lambda^{n}_{k+2^{n-1},2j} \rangle & =
& U_{{\rm{R}}_n}U_{\phi^{-1}_{n,\rm{A}}} |
\lambda^{n-1}_{k,2j} \rangle +d^{n}_{k,2j}U_{\phi^{-1}_{n,\rm{A}}} 
 |\lambda^{n-1}_{k,2j} \rangle,  
\end{eqnarray}
\reseteqn
where (\ref{minus}) has been used on the lhs of (\ref{xiaob}).  Since
$U_{{\rm{R}}_n}$ has the flip property that any function with support 
only in $X_{n,\rm{B}}$ will go entirely to
$X_{n,\rm{A}}$ in one iteration and vice-versa we know that
$U_{{\rm{R}}_n}U_{\phi^{-1}_{n,\rm{A}}}|\lambda^{n-1}_{k,2j} \rangle$
 has support only in $X_{n,{\rm{B}}}$ and $U_{\phi^{-1}_{n,\rm{A}}} 
|\lambda^{n-1}_{k,2j} \rangle$ has support only in $X_{n,\rm{A}}$.
Multiplying (\ref{onea}) by $\lambda^{n}_{k,2j}$ and (\ref{oneb}) by
$-\lambda^{n}_{k,2j}$ and identifying the components with support only
in $X_{n,{\rm A}}$ with the corresponding components in (57)
gives
\alpheqn
\begin{eqnarray}
c^{n}_{k,2j}U_{\phi^{-1}_{n,\rm{A}}} 
 |\lambda^{n-1}_{k,2j} \rangle & = & \lambda^{n}_{k,2j}
U_{\phi^{-1}_{n,\rm{A}}} |\lambda^{n-1}_{k,2j} \rangle \\ 
d^{n}_{k,2j}U_{\phi^{-1}_{n,\rm{A}}} 
 |\lambda^{n-1}_{k,2j} \rangle & = & -\lambda^{n}_{k,2j}
U_{\phi^{-1}_{n,\rm{A}}}|\lambda^{n-1}_{k,2j} \rangle,
\end{eqnarray}
\reseteqn
showing that $c^{n}_{k,2j} = \lambda^{n}_{k,2j}$ and $d^{n}_{k,2j} = -\lambda^{n}_{k,2j}$. 
Using this result gives the pair of recursion relations
\alpheqn
\begin{eqnarray} \label{three}
| \lambda^{n}_{k,2j} \rangle & =
& \left( U_{\phi^{-1}_{n,\rm{A}}}
+\lambda^{n}_{k,2j}U_{\phi^{-1}_{n,\rm{B}}} \right) 
 |\lambda^{n-1}_{k,2j} \rangle \\ 
| \lambda^{n}_{k+2^{n-1},2j} \rangle & =
& \left( U_{\phi^{-1}_{n,\rm{A}}} 
-\lambda^{n}_{k,2j}U_{\phi^{-1}_{n,\rm{B}}} \right)
 |\lambda^{n-1}_{k,2j} \rangle ,
\end{eqnarray}
\reseteqn 
which express the right states at the $n^{\rm{th}}$ bsp in terms 
of those at the $(n-1)^{\rm{th}}$ bsp. 

These recursion relations can be solved to write the right
states at the
$n^{\rm{th}}$ bsp in terms of the right states at the $0^{\rm{th}}$ ($\alpha =
2$) bsp. For notational convenience we define
\begin{equation} \label{four}
\begin{array}{lcl}
\widehat{{\rm{A}}}_{i} & \equiv & U_{\phi^{-1}_{i,{\rm{A}}}} \\
\widehat{{\rm B}}_{i} & \equiv & U_{\phi^{-1}_{i,{\rm{B}}}} \\
\widehat{{\rm B}}^{n}_{k,2j,i} & \equiv &
\left(\lambda^{n}_{k,2j}\right)^{i}\widehat{{\rm{B}}}_{i}
\end{array}\end{equation}
Let $\sigma_i$ ($i=1,2,\dots,n$) be either $0$ or $1$ and we define $\widehat
{\Pi}_{\sigma_n\sigma_{n-1}\dots\sigma_1}$ to be an ordered 
$n$-product of $\widehat{{\rm{A}}}_{i}$'s and $\widehat{{\rm
B}}^{n}_{k,2j,i}$'s, where if $\sigma_i = 1$ then the $i^{\rm {th}}$
location in the $n$-product (counting from the right) will be taken by
$\widehat{{\rm B}}^{n}_{k,2j,i}$  and if $\sigma_i = 0$ then the $i^{\rm {th}}$
location will be taken by  $\widehat{{\rm{A}}}_{i}$. Solving (59)
gives  
\begin{equation} \label{five}
| \lambda^{n}_{k,2j} \rangle = 
\sum_{\{\sigma \}}\widehat{\Pi}_{\sigma_n\sigma_{n-1}...\sigma_1} |
\lambda^{0}_{1,2j} \rangle.
\end{equation}
The sum in (\ref{five}) is over all possible $\sigma$-strings 
of $0$'s and $1$'s of length $n$ and so consists of
$2^n$ terms ($n$-products). The order of the operators in each $n$-product 
must be strictly observed 
since the operators involved do not commute. 

To illustrate (\ref{five}) we write it out explicitly for $n = 1$ and $2$.
For $n=1$ (\ref{five}) gives
\begin{equation} \label{example1}
| \lambda^{1}_{k,2j} \rangle = \widehat{\rm{A}}_{1} |
\lambda^{0}_{1,2j} \rangle + \widehat{\rm B}^{1}_{k,2j,1} |
\lambda^{0}_{1,2j} \rangle.
\end{equation}
This agrees with the expression (\ref{fbspurstates})
(corresponding to $k=1$ and $k=2$) we had for the
right eigenstates at the first bsp. 
For  $n=2$ (\ref{five}) gives
\begin{eqnarray}
| \lambda^{2}_{k,2j}\rangle & = &
\widehat{\rm{A}}_{2}\widehat{\rm{A}}_{1} |
\lambda^{0}_{1,2j} \rangle + \widehat{\rm{A}}_{2}\widehat{\rm B}^{2}_{k,2j,1} |
\lambda^{0}_{1,2j} \rangle  \nonumber \\
& & \mbox{} + \widehat{\rm B}^{2}_{k,2j,2} \widehat{\rm A}_{1} |
\lambda^{0}_{1,2j} \rangle + \widehat{\rm B}^{2}_{k,2j,2}\widehat{\rm B}^{2}_{k,2j,1}
 | \lambda^{0}_{1,2j} \rangle.
\end{eqnarray}

Now we prove by induction that there is a $2^n \times
2^n$ Jordan block associated with the eigenvalue $0$ at the $n^{\rm{th}}$ bsp for
each odd order $2j+1$. In section 2 it was shown that this statement is
true for the first bsp.  Assume that this statement is true at the $(n-1)^{\rm{th}}$ bsp.
We denote the Jordan vectors as $| 0^{n-1}_{k,2j+1} \rangle$ where 
$k=2,3,\dots,2^{n-1}$ ($| 0^{n-1}_{1,2j+1} \rangle$ is the eigenvector of the block).  They satisfy
\alpheqn
\begin{eqnarray} \label{jorddef}
U_{{\rm R}_n}| 0^{n-1}_{k,2j+1} \rangle & = & | 0^{n-1}_{k-1,2j+1}
\rangle, \;\;\;\;\;\;\;\; k \neq 1  \\ 
U_{{\rm R}_n} | 0^{n-1}_{1,2j+1} \rangle & = & 0.
\end{eqnarray}
\reseteqn 
Since we are assuming that $U_{{\rm{R}}_{n-1}}$ has $2^{n-1} \times 2^{n-1}$ 
Jordan blocks, the conjugacies 
(\ref{imp1}) and (\ref{imp2}) imply that ${\rm{G}}_{n,{\rm A}}$ and ${\rm{G}}_{n,{\rm B}}$
both have $2^{n-1} \times 2^{n-1}$ Jordan blocks with states given by
\alpheqn 
\begin{eqnarray} 
| 0^{n}_{k,2j+1} \rangle_{{\rm{G}}_{n,A}} & = & 
\widehat{{\rm{A}}}_{n}| 0^{n-1}_{k,2j+1} \rangle \\ 
| 0^{n}_{k,2j+1} \rangle_{{\rm{G}}_{n,B}} & = & \label{jzerob}
\widehat{{\rm{B}}}_{n}| 0^{n-1}_{k,2j+1} \rangle,
\end{eqnarray}
\reseteqn
where $k=1,2,\dots,2^{n-1}$. Since 
${{\rm{G}}_{n}}$ on $\Omega_{n}$ has two $2^{n-1} \times 2^{n-1}$ Jordan blocks
for each $j$, ${{\rm{R}}_{n}}$ on $\Omega_{n}$ can either have two 
$2^{n-1} \times 2^{n-1}$ Jordan blocks for each $j$ or have one $2^{n}
\times 2^{n}$ Jordan block for each $j$. The first case implies that 
$U_{{\rm R}_n}$ have two null vectors $| 0^{n}_{1,2j+1}
\rangle_{{\rm{G}}_{n,{\rm A}}}$ and $| 0^{n}_{1,2j+1}
\rangle_{{\rm{G}}_{n,{\rm B}}}$for each $j$. But $U_{{\rm R}_n}
| 0^{n}_{1,2j+1} \rangle_{{\rm{G}}_{n,{\rm B}}} \neq 0$, since no function with
support in $X_{n,\rm{B}}$ can vanish in one iteration under
$U_{{\rm R}_n}$.  Therefore 
$U_{{\rm R}_n}$ has a $2^n \times 2^n$ Jordan block for 
each odd degree $2j+1$. This
completes the proof by induction. 

A null state of $U_{{\rm R}_n}$ has to be a null state of
$U_{{\rm G}_n}$ too. Therefore the null state of
$U_{{\rm R}_n}$ for each $j$ is given by 
\begin{equation} \label{z1}
| 0^{n}_{1,2j+1} \rangle  =  
\widehat{{\rm{A}}}_{n}| 0^{n-1}_{1,2j+1} \rangle.
\end{equation}
We use the relation 
\begin{equation}
{\rm R}_n \circ \phi^{-1}_{n, {\rm A}} = 
\phi^{-1}_{n, {\rm B}} \circ {\rm R}_{n-1}
\end{equation}
which implies that
\begin{equation} \label{myst1}
U_{{\rm R}_n}\widehat{{\rm{A}}}_{n} =  
\widehat{{\rm{B}}}_{n}U_{{\rm R}_{n-1}}.
\end{equation}
Unlike (\ref{myst}), equation (\ref{myst1}) 
is true only at the band-splitting points. Equation (\ref{myst1}) can be used to
verify that if we act on both sides of (\ref{z1}) by $U_{{\rm R}_n}$ its rhs reduces to zero.

The right state $| 0^{n}_{1,2j+1} \rangle_{{\rm G}_{n,{\rm B}}}$ 
is a good candidate for the Jordan state $| 0^{n}_{2,2j+1} \rangle$ since under one iteration by 
$U_{{\rm R}_n}$ it will have support only in $X_{n, \rm{A}}$ and under 
two iterations of $U_{{\rm R}_n}$ it will vanish. Tentatively, from 
(\ref{jzerob}) we write 
\begin{equation} \label{z2}
| 0^{n}_{2,2j+1} \rangle  =  
\widehat{{\rm{B}}}_{n} | 0^{n-1}_{1,2j+1} \rangle.
\end{equation}
This guess for the Jordan state can be verified by using 
relation (\ref{myst}).  Similarly (\ref{myst1})
can be used to show that 
\begin{equation} \label{z3}
| 0^{n}_{3,2j+1} \rangle  = 
\widehat{{\rm{A}}}_{n}| 0^{n-1}_{2,2j+1} \rangle.
\end{equation}
and equation (\ref{myst}) can be used to show that 
\begin{equation} \label{z4}
| 0^{n}_{4,2j+1} \rangle  =  
\widehat{{\rm{B}}}_{n}| 0^{n-1}_{2,2j+1} \rangle.
\end{equation}
In general, we find 
\begin{eqnarray} \label{r1}
| 0^{n}_{k,2j+1} \rangle & = & 
\widehat{{\rm{A}}}_{n} | 0^{n-1}_{\lceil k/2 \rceil,2j+1} \rangle \;\;\;\;
\mbox{for} \; k \; \mbox{odd} \\ \nonumber
| 0^{n}_{k,2j+1} \rangle & = & 
\widehat{{\rm{B}}}_{n} | 0^{n-1}_{k/2 ,2j+1} \rangle \;\;\;\;\;\;
\mbox{for} \; k \; \mbox{even}
\end{eqnarray}
where $k=1,2,\dots,2^n$ and $\lceil q \rceil$ is the ceiling function ($q$ if it is an integer or
else the next greatest integer). 
These recursions are illustrated in Figure~6 up to $n=2$.
\setlength{\unitlength}{1mm}
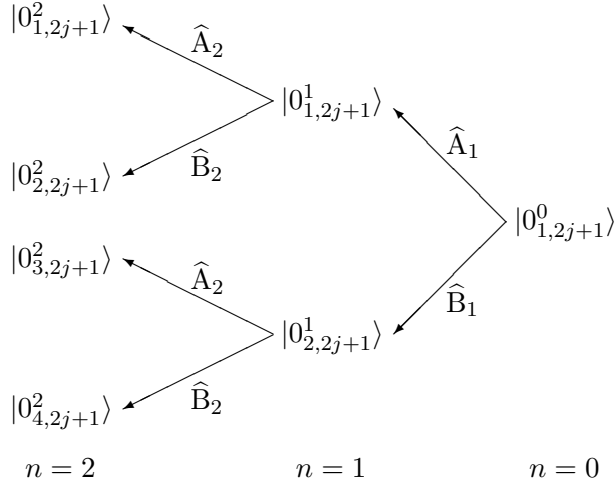
\begin{figure}
\begin{center}
\begin{picture}(67,60)
\put(67,33){$|0^{0}_{1,2j+1}\rangle$}
\put(66,34){\vector(-1,1){15}}
\put(66,34){\vector(-1,-1){15}}
\put(58,43){${\widehat{\rm A}}_{1}$}
\put(58,22){${\widehat{\rm B}}_{1}$}
\put(36,49){$|0^{1}_{1,2j+1}\rangle$}
\put(36,18){$|0^{1}_{2,2j+1}\rangle$}
\put(35,50){\vector(-2,1){20}}
\put(35,50){\vector(-2,-1){20}}
\put(35,19){\vector(-2,1){20}}
\put(35,19){\vector(-2,-1){20}}
\put(24,56.5){${\widehat{\rm A}}_{2}$}
\put(24,9){${\widehat{\rm B}}_{2}$}
\put(24,40){${\widehat{\rm B}}_{2}$}
\put(24,25){${\widehat{\rm A}}_{2}$}
\put(0,60){$|0^{2}_{1,2j+1}\rangle$}
\put(0,39){$|0^{2}_{2,2j+1}\rangle$}
\put(0,28){$|0^{2}_{3,2j+1}\rangle$}
\put(0,8){$|0^{2}_{4,2j+1}\rangle$}
\put(2,0){$n=2$}
\put(38,0){$n=1$}
\put(69,0){$n=0$}
\end{picture}
\parbox{5in}{\caption{\small States associated with eigenvalue zero obtained from the
action of $\widehat{{\rm{A}}}_{n}$ and $\widehat{{\rm{B}}}_{n}$ on $|0^{n-1}_{k,2j+1}\rangle$. }}
\end{center}
\end{figure}
This recursion relation can then be solved to write the eigen/Jordan
vectors at the $n^{\rm{th}}$ bsp in terms of the null vectors at the $0^{\rm{th}}$ bsp.
 To write down a compact solution we define $\Pi_{\sigma_n\sigma_{n-1}\dots\sigma_1}$, which is
similar to the $\widehat {\Pi}_{\sigma_n\sigma_{n-1}\dots\sigma_1}$ previously defined. 
 We define $\Pi_{\sigma_n\sigma_{n-1}...\sigma_1}$ to be an ordered 
$n$-product of $\widehat{{\rm{A}}}_{i}$'s and $\widehat{{\rm{B}}}_{i}$'s.
If $\sigma_i = 1$ then the $i^{\rm {th}}$ location in the $n$-product
(counting from the right) will be taken by  $\widehat{{\rm{B}}}_{i}$  and if
$\sigma_i = 0$ then the $i^{\rm {th}}$ location will be taken by 
$\widehat{{\rm{A}}}_{i}$. With each $\Pi_{\sigma_n\sigma_{n-1}\dots\sigma_1}$
we associate a binary
number formed from the string of $1$'s and $0$'s as  $\kappa = \sigma_n\sigma_{n-1}\dots\sigma_1 +1$.
Solving the recursion relation (\ref{r1}) we get
\begin{equation} \label{jord}
| 0^{n}_{\kappa,2j+1} \rangle = \Pi_{\sigma_n\sigma_{n-1}...\sigma_1}
| 0^{0}_{1,2j+1} \rangle,
\end{equation}
where $| 0^{0}_{1,2j+1} \rangle$ is the null vector of degree $2j+1$ of the
tent map with full height, and $\kappa$ here is the decimal equivalent of the binary $\kappa$,
which ranges from $1$ to $2^n$. 

\subsubsection{Left states on the attractor}
We obtain the left states, $ \langle \lambda^{n}_{k,j}
|$, which are orthonormal to the right states given by
(\ref{five}) and (\ref{jord})
by taking the duals of those expressions.
The dual expression of (\ref{five}) gives the left 
states corresponding to the non-zero eigenvalues as 
\begin{equation}
\langle \lambda^{n}_{k,2j} | =
\sum_{\{\sigma\}}\widehat{\Pi}^{\dag}_{\sigma_n\sigma_{n-1}\dots\sigma_1}
\frac{\langle
\lambda^{0}_{1,2j}|}{2^n},
\end{equation}
where the $n$-product here is of $(\widehat{{\rm{A}}}_{i}^{-1})^{\dag}$'s and 
$((\widehat{{\rm B}}^{n}_{k,2j,i})^{-1})^{\dag}$'s and the factor of $1/2^n$ is put for
normalization. The left states corresponding to the Jordan vectors associated with the
eigenvalue $0$ are 
\begin{equation} \label{leftjord}
\langle 0^{n}_{\kappa,2j+1} | =
\Pi^{\dag}_{\sigma_n\sigma_{n-1}\dots\sigma_1} \langle 0^{0}_{1,2j+1}|,
\end{equation}
where ${\Pi^{\dag}_{\sigma_n\sigma_{n-1}...\sigma_1}}$ is an 
ordered $n$-product of $(\widehat{{\rm{A}}}_{i}^{-1})^{\dag}$'s and 
$(\widehat{{\rm B}}_{i}^{-1})^{\dag}$'s.

\subsection{Decay onto the attractor of the rescaled map}

We saw in section 2.1 that $\rm{R}_{1}$ has no transient bands. 
The band structure of $\rm{G}_2$
consists of two scaled copies of that of $\rm{R}_{1}$ separated by the central interval  
$X_{2,{\rm C}}$. This central interval is a transient band of $\rm{R}_{2}$. Since
${\rm{G}}_{3,{\rm A}}$ and ${\rm{G}}_{3,{\rm B}}$ have a band structure similar to
that of ${\rm{R}}_{2}$ both of them have a transient interval also. We refer
to these two transient bands as peripheral transients since in
addition there is a central transient in the interval $X_{3,{\rm C}}$. 
In general, as discussed below (\ref{bandno}),
at the $n^{\rm{th}}$ bsp ${\rm{R}}_{n}$ has $2^{n-1} - 1$ transient bands, of which 
$2^{n-1} - 2$ are peripheral transient bands. At each bsp
all transient bands except the central one are rescaled versions of
the central transient bands at previous band-splitting points.
\begin{figure}[htb]
\begin{center}
\scalebox{.5}[.5]{\includegraphics{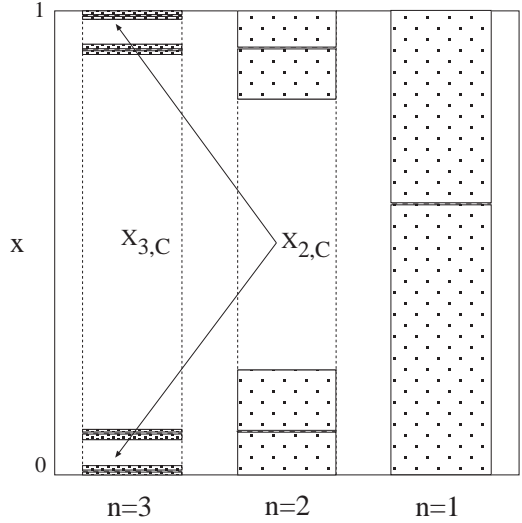}}
\parbox{5in}{\caption{\small Band structure at the $1^{\rm{st}}$, $2^{\rm{nd}}$ and 
$3^{\rm{rd}}$ band-splitting points. The central transient $X_{2,{\rm C}}$ at the $2^{\rm{nd}}$
bsp transforms into two peripheral transients at the $3^{\rm{rd}}$ bsp.}}
\end{center}
\end{figure} 

Under the map ${\rm{R}}_{n}$ the inverse image of any point in the central
interval $X_{n,{\rm C}}$ is contained within the interval itself. This implies that if
a function initially has no support in $X_{n,{\rm C}}$, it will continue to have no
support in $X_{n,{\rm C}}$ under repeated iterations by $U_{{\rm R}_n}$.
The spectrum and the form of the eigen/Jordan vector in
the central interval at all the band-splitting points can be obtained by
inspection. We notice that 
\begin{equation} \label{trans1}
U_{{\rm R}_n}\left[ \left(x-x^{*}_{n}\right)^{j}\chi_{n,{\rm C}}\right] =
(-1)^{j}\left(\frac{1}{\alpha_{n}}\right)^{j+1}\left\{\left(x-x^{*}_{n}\right)^{j}
\chi_{n,{\rm C}} +
\left(x-x^{*}_{n}\right)^{j}\chi_{n,b} \right\},
\end{equation}
where $\chi_{n,{\rm C}}$ is the indicator function on 
$X_{n,{\rm C}}$ and $\chi_{n,b} =1$ if 
$x \in [{\rm{R}}^{(3)}_{n}(x_c), {\rm{R}}^{(5)}_{n}(x_c) )$ and $0$ otherwise. The
form of the eigen/Jordan vector in $X_{n,{\rm C}}$ 
associated with decay out of the central interval is thus
$\left(x-x^{*}_{n}\right)^{j}$. The complete form will be determined below.
The eigenvalues associated with decay out of
the central transient at the $n^{\rm{th}}$ bsp are seen by (\ref{trans1}) to be
\begin{equation} \label{trans2}
\phi^{n,0}_{1,j} = (-1)^{j}\left(\frac{1}{\alpha_{n}}\right)^{j+1}
\end{equation}
 
Next we obtain the complete spectrum at the $n^{\rm{th}}$ bsp associated with
the decay out of all the transient regions onto $\Omega_{n}$. This is done
by transforming all the central transients at band-splitting points of
order less than $n$.  Of all the peripheral transient bands at the $n^{\rm th}$
bsp, $2^{n-2}$ of these are
$(n-2)$ times rescaled versions of the central transient at  ${\rm{R}}_{2}$,  $2^{n-3}$
of these are $(n-3)$ times rescaled versions of the central transient at 
${\rm{R}}_{3}$ and so on up to $2$ of these peripheral transients being rescaled
versions of the central transient at the $(n-1)^{\rm{th}}$ bsp.
This is shown in Figure~7 up to $n=3$.
We denote the transient eigenvalues at the $n^{\rm{th}}$ bsp by
$\phi^{n,l}_{k,j}$, where $l$ indicates that it was obtained from the
central transient at the $(n-l)^{\rm th}$ bsp, where 
$l=0,1,2,\dots,n-2$, $k$ is
an integer from $1$ to $2^l$, $j=0,1,2,3,\dots,$ and as before the right state 
$|\phi^{n,l}_{k,j} \rangle$ is piecewise-polynomial of degree $j$ in each of
the pieces. The eigenvalues are obtained in a similar fashion to (\ref{eig1}) as  
\begin{equation} \label{eig2} \phi^{n,l}_{k,j} =
\left(\phi^{n-l,0}_{1,j}\right)^{{1}/{2^l}}\mbox{exp}\left(
\frac{2\pi ik}{2^l}\right),
\end{equation}
where the $\phi^{n,0}_{1,j}$ are given by (\ref{trans2}). 
For even values of $j$ there are degeneracies in the spectrum while for odd
values of $j$ there are no degeneracies.  We consider first the states associated
with the degenerate eigenvalues.

\subsubsection{Transient right states of even degree}
As seen in (\ref{eig2}), for a given $l$ (with $n$ and $2j$ fixed) there are $2^l$ distinct
eigenvalues indexed by $k$.  At each integer step ($l$) all
the eigenvalues of the previous step ($l-1$) are present and $2^{l-1}$ new eigenvalues 
appear.  But identical eigenvalues from different steps have disparate $k$ values.
It is convenient to rearrange the $k$ index so that degenerate eigenvalues share
the same $k$.\footnote{This can be accomplished by choosing the new $k$'s asssociated
with the $l^{\rm th}$ step as $k_{2i-1}^l = k_i^{l-1}$ and 
$k_{2i}^l = k_{2i-1}^l + 2^{l-1}$, where $i=1,\dots,2^{l-1}$ denotes the order of its appearance and
$k_1^0=1$.} Table 2 contains $\phi^{5,l}_{k,0}$ for all possible values of 
$l$ and $k$, arranged to illustrate the reordering in $k$.

\begin{table}[htbp!]
\[
\begin{array}{|l|cccccccc|} \hline
 & k=1 & k=2 & k=3 & k=4 & k=5 & k=6 & k=7 & k=8 \\ \hline
& \vspace{-0.4mm} & \vspace{-0.4mm} & \vspace{-0.4mm} & \vspace{-0.4mm} & \vspace{-0.4mm} 
& \vspace{-0.4mm} & \vspace{-0.4mm} &  \\
\phi^{5,0}_{k,0} & \varphi & & & & & & &  \\
& \vspace{-0.4mm} & \vspace{-0.4mm} & \vspace{-0.4mm} & \vspace{-0.4mm} & \vspace{-0.4mm} 
& \vspace{-0.4mm} & \vspace{-0.4mm} & \\
\phi^{5,1}_{k,0} & \varphi & -\varphi & & & & & &  \\ 
& \vspace{-0.4mm} & \vspace{-0.4mm} & \vspace{-0.4mm} & \vspace{-0.4mm} & \vspace{-0.4mm} 
& \vspace{-0.4mm} & \vspace{-0.4mm} &\\
\phi^{5,2}_{k,0} & \varphi & -\varphi & i\varphi & -i\varphi & & & &  \\ 
& \vspace{-0.4mm} & \vspace{-0.4mm} & \vspace{-0.4mm} & \vspace{-0.4mm} & \vspace{-0.4mm} 
& \vspace{-0.4mm} & \vspace{-0.4mm} & \\
\phi^{5,3}_{k,0} & \varphi & -\varphi & i\varphi & -i\varphi & 
\varphi\exp\big( \frac{\pi i}{4} \big) &
\varphi\exp\big( \frac{3\pi i}{4} \big) &
\varphi\exp\big( \frac{5\pi i}{4} \big)& 
\varphi\exp\big( \frac{7\pi i}{4} \big) \\
\hline
\end{array}
\]
\caption{Transient spectrum at the $5^{\rm{th}}$ bsp with $j=0$. The
constant $\varphi \equiv 1/\alpha_{5}$.} \end{table}

Next we ask if there are independent eigenvectors
associated with the degenerate eigenvalues. 
Using the same procedure used to find the right eigenvectors
decaying out of region $\rm{I}$ it can be shown that in the rescaled map,
for $n > 2$ there is no eigenpolynomial of even degree with support in the central
transient. This means that the right states associated with $\phi^{n,0}_{1,2j}$
are Jordan vectors, for $n > 2$.  For $n = 2$, decay out of the central
transient is described by eigenvectors for all values of $j$. Since the right states associated
with 
$\phi^{3,1}_{k,j}$, $\phi^{4,2}_{k,j}$,\dots,$\phi^{n,n-2}_{k,j}$ are transformed
versions of $| \phi^{2,0}_{1,j} \rangle$, these are also eigenvectors. For
$n \geq 3$ decay out of the central transient is described by Jordan 
vectors for even values
of $j$. Therefore all the peripheral transients which are related by conjugacies
(\ref{imp1}) and (\ref{imp2}) to these central transients are also Jordan
vectors. Hence associated with the transient spectrum are Jordan blocks whose sizes correspond to the
algebraic multiplicity of the eigenvalues. Thus from Table 2 we see that at the $5^{\rm{th}}$ bsp
there is a $4 \times 4$ Jordan block associated with
$\phi^{5,0}_{1,2j}$, a $3 \times 3$ Jordan block associated with $\phi^{5,1}_{2,2j}$ and so on. At
the $n^{\rm{th}}$ bsp the largest Jordan block in the transient spectrum is
associated with $\phi^{n,0}_{1,2j}$ and is of size $n-1 \times n-1$. Since 
$\phi^{n,0}_{1,0}$ is the largest eigenvalue with modulus less than 1 in the
spectrum of $U_{\rm R_n}$, it corresponds to the slowest decay mode. Since there is a Jordan block
associated with $\phi^{n,0}_{1,0}$ this decay is modified exponential (polynomial factors in $t$
times exponential decay).

In general (for even $j$) we have Jordan vectors for $l\leq n-3$ and eigenvectors for $l=n-2$ as
\begin{eqnarray} \label{t5}
U_{{\rm R}_n}|\phi^{n,l}_{k,2j} \rangle & = & \phi^{n,l}_{k,2j}
|\phi^{n,l}_{k,2j} \rangle + |\phi^{n,l+1}_{k,2j}
\rangle \;\;\;\;\;\;\;\; n \geq 3, \; l \leq n-3 \nonumber \\ \noalign{\vskip4pt}
U_{{\rm R}_n}|\phi^{n,n-2}_{k,2j} \rangle & = & \phi^{n,n-2}_{k,2j}
|\phi^{n,n-2}_{k,2j} \rangle \;\;\;\;\;\;\;\;\;\;\;\;\;\;\;\; n \geq 2. 
\end{eqnarray}
By inspection we have
already obtained the form of the Jordan state in the central region in
(\ref{trans1}).  We have
\begin{equation} \label{t6} | \phi^{n,0}_{1,2j} \rangle =
a_{n,2j}\left(x-x^*_n\right)^{2j}\chi_{n,{\rm C}} + f_{n,2j}(x)   \end{equation}
where $f_{n,2j}(x)$ is piecewise polynomial with degree $2j$ over each of the
$S_{n}$ intervals (excluding the central interval
$X_{n, {\rm C}}$) at the $n^{\rm{th}}$ bsp.  Since a Jordan vector multiplied by a
scalar doesn't remain a Jordan vector (with respect to the same eigenvector in the block)
we do not have the freedom to choose
the $a_{n,2j}$'s to be $1$ as we did in (\ref{haha}).   Using (\ref{trans1})
and (\ref{t5}) we know that if (\ref{t6}) is to be a Jordan vector the
arbitrary functions $f_{n,2j}(x)$ must satisfy 
\begin{equation} \label{t1} U_{{\rm R}_n}
f_{n,2j}(x) = \phi^{n,0}_{1,2j}\left[ f_{n,2j}(x) -
a_{n,2j}\left(x-x^{*}_{n}\right)^{2j}\chi_{n, b}
\right] + | \phi^{n,1}_{1,2j} \rangle
\end{equation}
A formal approach to determine the $a_{n, 2j}$'s and $f_{n, 2j}(x)$ is 
described in Appendix D.

To find the eigenstates corresponding to the peripheral transients
we transform the central transient eigenstates at the $2^{\rm{nd}}$ bsp
to all the higher band-splitting points. We define 
\begin{equation}
\check{\rm B}^{n,n-2}_{k,2j,i} = \left( \phi^{n,n-2}_{k,2j} \right)^{i}\widehat{\rm B}_{i},
\end{equation}
and $\check{\Pi}_{\sigma_n\sigma_{n-1}\dots\sigma_{3}}$ is an ordered product of 
$n-2$ operators $\widehat{\rm A}_{i}$ and $\check{\rm B}^{n,n-2}_{k,2j,i}$ ($i
= 3,4,\dots,n$). If $\sigma_i = 0$ the $i^{\rm th}$ place in the
product is taken by  $\widehat{\rm A}_{i}$ and if $\sigma_i = 1$ the $i^{\rm
th}$ place in the  product is taken by  $\check{\rm B}^{n,n-2}_{k,2j,i}$.
Following the same procedure  used to obtain (\ref{five}) we find that the transient
eigenvectors, $| \phi^{n,n-2}_{k,2j}\rangle$, $(n\geq 3)$, are given by
\begin{equation} \label{transeig}
| \phi^{n,n-2}_{k,2j} \rangle = \sum_{\{ \sigma\}}
\check{\Pi}_{\sigma_n \sigma_{n-1}\dots\sigma_3}
| \phi^{2,0}_{1,j} \rangle
\end{equation}

To find the Jordan states corresponding to the peripheral transients, 
$| \phi^{n,l}_{k,2j} \rangle$ for $1 \leq l \leq n-3$, 
we use an approach similar to that used to find the
Jordan vectors  describing decay out of the central transient, 
$| \phi^{n,0}_{1,2j} \rangle $. To clarify the notation we 
note that $| \phi^{n,1}_{k,2j} \rangle $ does not
have support in the central transient band, but has 
support over all the other transient and attracting bands. Similarly
$| \phi^{n,2}_{k,2j} \rangle $ has support on all bands except the
central transient band and the two transient bands that are related by a single
transformation to the central transient at the $(n-1)^{\rm{th}}$ bsp. This
pattern continues and $| \phi^{n,n-2}_{k,2j} \rangle$ has support
on all the attracting bands and the $2^{n-2}$ transient bands that are related
by transformations to the central transient at the $2^{\rm{nd}}$ bsp. Following
(\ref{t6}) and using the fact that the peripheral transients at the
$n^{\rm{th}}$ bsp are transformed versions of the transients at the
$(n-1)^{\rm{th}}$ bsp we write
\begin{equation} \label{jord10}
| \phi^{n,l}_{k,2j} \rangle  = a^{n,l}_{k,2j}\left( 
\widehat{\rm A}_n| \phi^{n-1,l-1}_{\lceil k/2 \rceil,2j} \rangle +
\phi^{n,l}_{k,2j} \widehat{\rm B}_n | \phi^{n-1,l-1}_{\lceil k/2 \rceil,2j}
\rangle \right) + g^{n,l}_{k,2j}(x)
\end{equation}
where $g^{n,l}_{k,2j}(x)$ is piecewise polynomial of degree $2j$. The
polynomial  $g^{n,l}_{k,2j}(x)$ can be obtained using the procedure similar to
the one described in Appendix D. We present here the
solution only for the simplest case,
\begin{equation} \label{jord14}
| \phi^{n,n-3}_{k,2j} \rangle = 2 \, \phi^{n,n-3}_{k,2j} \left(
\widehat{\rm A}_n| \phi^{n-1,n-4}_{\lceil k/2 \rceil,2j} \rangle
 + \phi^{n,n-3}_{k,2j}
\widehat{\rm B}_n| \phi^{n-1,n-4}_{\lceil k/2 \rceil,2j} \rangle
\right) - \frac{1}{2 \, \phi^{n,n-3}_{k,2j}}
| \phi^{n,n-2}_{k',2j} \rangle
\end{equation}
where $k'=k+1$ if $k$ is odd and $k'=k-1$ if $k$ is even. The recursion
relations for the other Jordan vectors involve more terms, and in general
to find all the Jordan vectors at the $n^{\rm{th}}$ bsp one must know
all the Jordan vectors at the $(n-1)^{\rm{th}}$ bsp. Also at a particular
bsp to determine $| \phi^{n,l}_{k,2j} \rangle $ one must know 
all the $| \phi^{n,l'}_{k,2j} \rangle $ for $l' < l$. So
generally one proceeds in the following order
$| \phi^{2,0}_{1,2j} \rangle $, $|
\phi^{3,1}_{k,2j}\rangle $, $| \phi^{3,0}_{1,2j}\rangle $,
$| \phi^{4,2}_{k,2j} \rangle $, $|
\phi^{4,1}_{k,2j}\rangle \dots $. The transient eigenvectors can be
obtained directly from (\ref{transeig}) without regard to this order.

\subsubsection{Transient Right states of odd degree}

For odd values of $j$ the transient spectrum at the $n^{\rm th}$ bsp, 
given by equation (\ref{eig2}), is nondegenerate. We first find the 
eigenvectors with support in the central transient bands, at all the
band-splitting points. By inspection we have already obtained the form 
of the eigenvector in the central transient in (\ref{trans1}) so that
\begin{equation} \label{oddj}
| \phi^{n,0}_{1,2j+1} \rangle = \left( x-x^{*}_{n} \right)^{2j+1} \chi_{n,C} 
+ f_{n,2j+1}(x),
\end{equation}
where $f_{n,2j+1}(x)$ is a polynomial of degree $2j+1$ in each of the 
$S_n$ intervals excluding the central interval.  For this expression is to be an 
eigenvector $f_{n,2j+1}(x)$  must satisfy
\begin{equation} \label{oddj1}
U_{{\rm R}_n}
f_{n,2j+1}(x) = \phi^{n,0}_{1,2j+1}\left[ f_{n,2j+1}(x) -
\left(x-x^{*}_{n}\right)^{2j+1}\chi_{n, b}
\right]. 
\end{equation}
This equation for $f_{n,2j+1}(x)$ is similar to equation (\ref{t1}), except
that it has one term less on the rhs, and may be solved using the procedure
outlined in Appendix~D.

Once the $| \phi^{n,0}_{1,2j+1} \rangle$ are known the eigenvector
corresponding to any $| \phi^{n,l}_{k,2j+1} \rangle$ can be written as 
\begin{equation}
| \phi^{n,l}_{k,2j+1} \rangle = \sum_{\{ \sigma \}
}\check{\Pi}_{\sigma_n\sigma_{n-1}\dots\sigma_{n-l+1}}
|\phi^{n-l,0}_{1,2j+1} \rangle,
\end{equation}
where $\check{\Pi}_{\sigma_n\sigma_{n-1}\dots\sigma_{n-l+1}}$ 
 is an ordered product of 
$l$ operators $\widehat{\rm A}_{i}$ and $\check{\rm B}^{n,l}_{k,2j,i}$ ($i =
n-l+1,\dots,n$). The  $\check{\rm B}^{n,l}_{k,2j,i}$ operators are defined
here as 
\begin{equation}
\check{\rm B}^{n,l}_{k,2j,i} \equiv  \left(\phi^{n,l}_{k,2j+1}\right)^i
 \widehat{\rm B}_i
\end{equation}
If $\sigma_i = 0$ the
$i^{\rm th}$ place in the product is taken by 
$\widehat{\rm A}_{i}$ and if $\sigma_i = 1$ the $i^{\rm th}$ place in the 
product is taken by  $\check{\rm B}^{n,l}_{k,2j,i}$. The sum is over all
possible strings of $0$'s and $1$'s of length $l$.

\subsubsection{Left States}
 
The left states  $\langle \phi^{n,l}_{k,j} |$ form a  bi-orthonormal set with all
the previously obtained right states at the $n^{\rm{th}}$ bsp as
\alpheqn
\begin{eqnarray}
\langle \phi^{n,l}_{k,j} | \lambda^{n}_{k'j'}  
\rangle & = & 0 \\ 
\langle \phi^{n,l}_{k,j} | \phi^{n,l'}_{k'j'}  
\rangle & = & \delta_{ll'}\delta_{kk'}\delta_{jj'}.
\end{eqnarray}
\reseteqn
Among the right states only the states with $l=0$ have support in the central transient band.
From (\ref{t6}) we see that the associated left states are 
\alpheqn
\begin{eqnarray}
\langle \phi^{n,0}_{1,2j} | & = & \frac{1}
{a_{n,2j}}\delta^{(2j)}\left( x - x^{*}_{n} \right) \\ 
\langle \phi^{n,0}_{1,2j+1} | & = & 
-\delta^{(2j+1)}\left( x - x^{*}_{n} \right).
\end{eqnarray}
\reseteqn
The left states corresponding to the peripheral transients at 
the $n^{\rm{th}}$ bsp, $\langle \phi^{n,l}_{k,j} |$,
are found by transforming the left states at the 
$(n-1)^{\rm{th}}$ bsp, 
$\langle \phi^{n-1,l-1}_{\lceil k/2 \rceil,j} |$.
Then the entire set at the $n^{\rm{th}}$ bsp has to be orthonormalized
using a Gram--Schmidt procedure.

\subsection{Back to ${\rm T}_n(x)$}

Going back to the tent map ${\rm{T}}_{n}(x)$, we transform
all the right states of ${\rm{R}}_{n}(x)$ by $U_{\phi^{-1}_{n}}$ and the left states by
$K_{\phi_{n}}$. The states describing decay out of 
$[ {\rm{T}}_{n}^{(3)}(x_{c}),1 ]$ are null states. 
The eigenvalues describing decay out of 
$[ 0,{\rm{T}}_{n}^{(4)}(x_{c})]$ are 
$\phi_j^n \equiv \left( \frac{1}{\alpha_{n}}\right)^{j+1}$ and have
associated polynomial eigenvectors of degree $j$.  For $j$ even this part of the spectrum overlaps
with the spectrum describing decay of transients of ${\rm R}_n$. There are Jordan vectors associated
with this part of the spectrum; denoting them as
$| \phi^{n}_{2j} \rangle$ we have
\alpheqn
\begin{eqnarray}
U_{{\rm T}_n}| \phi^{n}_{2j} \rangle = \phi^{n}_{2j}
| \phi^{n}_{2j} \rangle +| \phi^{n,0}_{1,2j} 
\rangle \\ 
U_{{\rm T}_n}| \phi^{n}_{2j+1}\rangle = \phi^{n}_{2j+1}
| \phi^{n}_{2j+1} \rangle .
\end{eqnarray}
\reseteqn
These right states can be determined by an extension of the 
methods used in
Section 2.2 to determine the eigenstates describing 
decay out of interval $\rm{I}$ for the tent map at the first bsp.

\section{Conclusion}

We have presented the generalized spectral
decomoposition of the Frobenius--Perron
operator of the tent map at the
band-splitting points. The right eigenstates are
piecewise-polynomial functions and the left eigenstates
are generalized functions. The 
spectrum is discrete and gives the
characteristic decay times of the map.  From the decomposition
one can calculate correlations of arbitrary polynomials (as
well as functions expandable in terms of the polynomial 
eigenstates).  Furthermore, since the modes corresponding
to transient decay onto the attractor have been obtained, the
full nonequilibrium dynamics of initial probability densities
is accessible.

The slowest decay mode, corresponding to the
eigenvalue $\alpha_{n}^{-1}$ at the
$n^{\rm th}$ bsp, describes decay onto the
attractor. At the $n^{\rm th}$ bsp there is an 
$n \times n$ Jordan block associated with this
eigenvalue and therefore the decay is modified
exponential. The asymptotic periodicity
of the map is clearly reflected in the
spectrum as at the $n^{\rm th}$ bsp, all the 
$n^{\rm th}$ roots of unity are part of the
spectrum.  Our analytic solution of density 
evolution in this system may be useful for 
comparision with the behavior of systems governed by the Ginzburg--Landau
equation since a component of its dynamics~\cite{Moon} can be reduced to the tent map.

\section*{Acknowledgements}

We thank I.~Prigogine for his support and encouragement and G.E.~Ord\'{o}\~{n}ez
for several useful discussions and his comments on the manuscript.
We acknowledge US
Department of Energy grant no. FG03-94ER14465,  
the Welch Foundation
grant no. F-0365 and the European Communities Commision (contract no.
27155.1/BAS) for support of this work.
\section*{Appendix A: Topological conjugacy}

In this appendix we review the spectral decompositions 
of maps related by a coordinate 
transformation~\cite{Deanbook}. Let ${\rm{T}}: X \rightarrow X$ be a map defined on the
interval
$X$. Transforming the interval $X$ by the one-to-one, 
onto, continuous function $\phi : X \rightarrow Y$ gives a new map, ${\rm{S}}:Y
\rightarrow Y$.  This map is determined as
\begin{equation}
y_{t+1} = \phi(x_{t+1}) = \phi({\rm T}(x_t)) \equiv {\rm S}(y_t).
\end{equation}
Using that $\phi$ has an inverse gives
\begin{equation}
\phi({\rm T}(x_t)) = \phi({\rm T}(\phi^{-1}(\phi(x_t)))) 
= \phi \circ {\rm T} \circ \phi^{-1}(y_t),
\end{equation}
so that
\begin{equation} \label{strel}
{\rm S} = \phi \circ {\rm T} \circ \phi^{-1}.
\end{equation}
The maps ${\rm{T}}$ and ${\rm{S}}$ are said to be topologically 
conjugate to each other. 

The Koopman operator, $K_{\rm{S}}$,
corresponding to ${\rm{S}}$ is given from (\ref{strel}) by 
\begin{equation} \label{kooprel}
 K_{\rm{S}} = K^{-1}_{\phi}K_{\rm{T}}K_{\phi},
\end{equation}
where we have used the fact that $K_{\phi^{-1}} = K^{-1}_{\phi}$.
The Frobenius--Perron operator, $U_{\rm{S}}$, coresponding to ${\rm{S}}$
is the adjoint of $K_{\rm{S}}$. Taking the adjoint of (\ref{kooprel})
and using $(K_{\phi}^{-1})^{\dagger} = (K_{\phi}^{\dagger})^{-1}$ gives
\begin{equation} \label{frobrel}
U_{\rm{S}} = U_{\phi} U_{\rm{T}} U^{-1}_{\phi}.
\end{equation}
Since $U_{\rm S}$ and $U_{\rm T}$ are related by the similarity (\ref{frobrel})
the spectrum of $U_{\rm S}$ is identical to that of $U_{\rm T}$ and eigenstates
transform as
\begin{equation} \label{app2}
\left| \lambda_{n} \right\rangle_{\rm{S}} = U_{\phi}\left| \lambda_{n}
\right\rangle_{\rm{T}},
\end{equation}
where we use a Dirac-style bra-ket notation for the states.
From (\ref{kooprel}) the left states transform as 
\begin{equation} \label{app3}
\left\langle \lambda_{n} \right|_{\rm{S}} = 
K_{\phi^{-1}} \left\langle \lambda_{n} \right|_{\rm{T}}.
\end{equation}
Jordan states of the maps are also related as in (\ref{app2}) and (\ref{app3}) and
both the algebraic and geometric multiplicities of the eigenvalues are preserved 
under conjugacy.

\section*{Appendix B: The tent map with unit height}

The Frobenius--Perron operator of the tent map with unit height is given by 
\begin{equation}
U_{\rm{T_{0}}}\rho(x) = \frac{1}{2}\left[ \rho \left( \frac{x}{2} \right)
 + \rho\left( \frac{2-x}{2} \right) \right].
\end{equation}
The operator $U_{\rm T_0}$ admits polynomial eigenstates with support on the
whole unit interval. Associated with eigenpolynomials of order
$2j$ are the nonzero eigenvalues  $2^{-2j}$. There
is an infinite degeneracy of the eigenvalue $0$ with an independent
odd-order eigenpolynomial associated with each occurence of the eigenvalue. 
Thus, the odd-order eigenpolynomials are not unique but we choose them as Euler
polynomials so that the associated left eigendistributions take a simple form.
The right eigenvectors of $U_{\rm T_0}$ are~\cite{Gonzalo,fox}
\alpheqn 
\begin{eqnarray}
|2^{-2j}\rangle_{{\rm{T}}_{0}} & = & 
B_{2j}(x/2)  \\
| 0_{2j+1} \rangle_{\rm T_0} & = &  E_{2j+1}(x), 
\end{eqnarray}
\reseteqn
where $B_{j}(x)$ is the Bernoulli polynomial of order $j$ and 
$E_{j}(x)$ is the Euler polynomial of order $j$~\cite{Absteg}.
The corresponding left states are
\alpheqn
\begin{eqnarray}
\langle {2^{-2j}}|_{\rm{T}_{0}} & = & 
2^{2j}\widetilde B_{2j} (x)  \\
\left\langle 0_{2j+1} \right|_{\rm{T}_{0}} & = & 
\frac{-1}{(2j+1)!}\delta^{(2j+1)}_{-}(x-1),
\end{eqnarray}
\reseteqn
where $\tilde B_0(x) = 1$ and for $j\geq 1$
\begin{equation} 
\widetilde B_{2j} (x) \equiv \frac{(-1)^{2j-1}}{(2j)!}\left[ 
\delta^{(2j-1)}_{-}(x-1) - \delta^{(2j-1)}_{+}(x) \right],  
\end{equation}
where the action of $\delta^{(m)}_\pm (x-c)$ on a sufficently differentiable
function $f(x)$ is given by 
\begin{equation}
\int_a^b dx \, \delta^{(m)}_{\pm}(x-c) f(x) =
\lim_{\epsilon \rightarrow 0} (-1)^{m} f^{(m)}(c \pm \epsilon),
\end{equation}
for $a \leq c \leq b$ and $\epsilon$ is a positive infinitesimal.

The time evolution of a density is expressed in terms of the spectral decomposition of
$U_{\rm T_0}$ as
\begin{equation}
U_{\rm T_0}^t \, \rho(x) = \sum_{j=0}^\infty
\left[ (2^{-2j})^t |2^{-2j}\rangle \langle 2^{-2j}| \rho \rangle
+ \delta_{t,0} | 0_{2j+1} \rangle \langle 0_{2j+1}| \rho \rangle \right],
\end{equation}
where the bilinear form is defined by
\begin{equation}
\langle f | g \rangle \equiv \int_0^1 dx \, f^*(x) g(x).
\end{equation}

\section*{Appendix C: Calculation of transient right states}

To determine the functions 
$g_{{\rm II},j}(x)$ and $g_{{\rm III},j}(x)$, which appear
on the rhs of (\ref{form1}) we expand 
$g_{{\rm II},j}(x)\chi_{\rm{II}} + g_{{\rm III},j}(x)\chi_{\rm{III}}$ in terms of the
eigenstates given in Table 1 of $U_{\rm T_1}$ on the attractor as 
\begin{eqnarray} \label{expand1}
g_{{\rm II},j}(x)\chi_{\rm{II}} + g_{{\rm III},j}(x)\chi_{\rm{III}} & = &
\sum_{i=1}^{\lfloor j/2
\rfloor} a_{i,j}|{+2^{-i}} \rangle_{\Omega} + 
\sum_{i=1}^{\lfloor j/2\rfloor}
 b_{i,j}|{-2^{-i}} \rangle_{\Omega} \nonumber \\
 & & \mbox{} + \sum_{i=1}^{\lfloor
\frac{j-1}{2} \rfloor} c_{i,j}| 0_{2i+1} \rangle_{\Omega} +
\sum_{i=1}^{\lfloor
\frac{j-1}{2} \rfloor} d_{i,j}| 0_{J_{2i+1}} \rangle_{\Omega},  
\end{eqnarray}
where $\lfloor x \rfloor$ denotes the integer
part (floor) of the real number $x$.
Then acting with $U_{\rm T_1}$ gives
\begin{eqnarray} \label{expand2}
U_{{\rm{T_1}}} \left( g_{{\rm II},j}(x)\chi_{\rm{II}} + g_{{\rm
III},j}(x)\chi_{\rm{III}}
\right) & = &
\sum_{i=1}^{\lfloor j/2
\rfloor} \frac{a_{i,j}}{2^i}|{+2^{-i}} \rangle_{\Omega} - 
\sum_{i=1}^{\lfloor j/2\rfloor}
 \frac{b_{i,j}}{2^i}|{-2^{-i}} \rangle_{\Omega} \nonumber \\
& & \mbox{} + \sum_{i=1}^{\lfloor
\frac{j-1}{2} \rfloor} d_{i,j}| 0_{2i+1} \rangle_{\Omega}.  
\end{eqnarray} 
We substitute (\ref{expand1}) 
and (\ref{expand2}) into (\ref{form1}) and act on (\ref{form1}) by all the left states
on the attractor in succession. Using orthonormality we obtain the
following equations for the expansion coefficients:
\alpheqn
\begin{eqnarray} \label{form2}
\frac{a_{i,j}}{2^i} + \alpha_{i,j} & = & 2^{-(j+1)/2} \, a_{i,j} \\
\frac{b_{i,j}}{2^i} - \beta_{i,j} & = & -2^{-(j+1)/2} \, b_{i,j} \\ 
d_{i,j} + \gamma_{i,j} & = &  2^{-(j+1)/2} \, c_{i,j} \\
d_{i,j} & = &  0,
\end{eqnarray}
\reseteqn 
where
\alpheqn
\begin{eqnarray} \label{def}
\alpha_{i,j} & \equiv & 2^{-(j+1)/2} 
\langle{+2^{-i}}| x^j\chi_{\rm{II}} \rangle  \\ 
\beta_{i,j} & \equiv & 2^{-(j+1)/2}
\langle{-2^{-i}}| x^j\chi_{\rm{II}} \rangle  = 
\alpha_{i,j}  \\ 
\gamma_{i,j} & \equiv & 2^{-(j+1)/2}
\langle 0_i |  x^j\chi_{\rm{II}} \rangle,
\end{eqnarray}
\reseteqn
and we used $\langle 0_{J_i}| x^j\chi_{\rm{II}} \rangle = 0$ because 
$\langle 0_{J_i}|$ has support only in $\chi_{\rm III}$.  
Explicit evaluation of $\langle +\frac{1}{2^i}|
x^j\chi_{\rm{II}} \rangle$ gives
\begin{equation}
\langle +\frac{1}{2^i}| x^j\chi_{\rm{II}} \rangle  =  
\left\{ \begin{array}{lc} \frac{\sqrt{2}}{2(\sqrt{2}-1)(j+1)} 
((2 - \sqrt{2})^j -1 )
& i=0  \\ \noalign{\vskip4pt}
0  & 2i-1 \geq j  \\ \noalign{\vskip4pt}
-\frac{j!(2 - \sqrt{2})^{j+2i-1}}{(2i)!(j-2i+1)!}(2^{(j-2i+1)/2} -1) & 2i-1 < j.  
\end{array} \right.
\end{equation}
Evaluation of $\langle 0_i |  x^j\chi_{\rm{II}} \rangle$ gives
\begin{equation}
\langle 0_i |  x^j\chi_{\rm{II}} \rangle = 
\left\{ \begin{array}{lc} \frac{j!}{(2i+1)!(j-2i-1)!}(
\sqrt{2} -1 )^{j+2i+1} & 2i+1 \leq j \\ \noalign{\vskip4pt}
0 & \mbox{otherwise} .
\end{array} \right.
\end{equation}
These results are then
used in (109) to determine the expansion coefficients for the transient
eigenstates with support in region $\rm I$.

\section*{Appendix D: Transient right states at higher bsps}

We expand the arbitrary functions $f_{n,2j}(x)$ in terms of the
transient and non-transient eigenvectors of $U_{{\rm R}_n}$. The
non-transient eigenvectors are given in (\ref{five}) and (\ref{jord}). The
transient eigenvectors will be transformed versions of the central
eigenvectors at the previous band-splitting points. The expansion is
\begin{eqnarray} \label{t2}
f_{n,2j}(x) & = &
\sum_{k=1}^{2^n}\sum_{j^{'}=1}^{j} b^{n,2j}_{k,2j^{'}}
| \lambda^{n}_{k,2j^{'}} \rangle +
\sum_{k=1}^{2^n}\sum_{j^{'}=1}^{j-1} c^{n,2j}_{k,j^{'}}
| 0^{n}_{k,2j^{'}+1} \rangle \nonumber \\
 & & + \sum_{l=1}^{n-2}\sum_{k=1}^{2^l}\sum_{j^{'}=1}^{j}
d^{n,l,2j}_{k,j^{'}}
| \phi^{n,l}_{k,2j^{'}} \rangle + \sum_{l=1}^{n-2}\sum_{k=1}^{2^l}
\sum_{j^{'}=1}^{j-1} e^{n,l,2j}_{k,j^{'}}
| \phi^{n,l}_{k,2j^{'}+1} \rangle.
\end{eqnarray}
Since $d^{n,n-2,2j}_{1,j}$ is the coefficent of the eigenvector of the
Jordan block it can be set to zero.  
Applying $U_{{\rm R}_n}$ to the function $f_{n,2j}(x)$ we get
\begin{eqnarray} \label{t3}
\lefteqn{U_{{\rm R}_n} f_{n,2j}(x)  =  
\sum_{k=1}^{2^n}\sum_{j^{'}=1}^{j}b^{n,2j}_{k,j^{'}}
\lambda^{n}_{k,2j^{'}}| \lambda^{n}_{k,2j^{'}}\rangle 
+\sum_{k=1}^{2^{n}-1}\sum_{j^{'}=1}^{j-1}
c^{n,2j}_{k+1,j^{'}}
| 0^{n}_{k,2j^{'}+1} \rangle } \hspace{20pt} \nonumber \\
& & \mbox{} 
  + \sum_{l=1}^{n-2}\sum_{k=1}^{2^l}\sum_{j^{'}=1}^{j}
d^{n,l,2j}_{k,j^{'}}
\phi^{n,l}_{k,2j^{'}}| \phi^{n,l}_{k,2j^{'}} \rangle + \sum_{l=2}^{n-2}\sum_{k=1}^{2^{l-1}}\sum_{j^{'}=1}^{j}
d^{n,l-1,2j}_{k,j^{'}}
\phi^{n,l}_{k,2j^{'}}| \phi^{n,l}_{k,2j^{'}} \rangle \nonumber \\
 & &\mbox{}  
+ \sum_{l=1}^{n-2}\sum_{k=1}^{2^l}\sum_{j^{'}=1}^{j}
e^{n,l,2j}_{k,j^{'}}
\phi^{n,l}_{k,2j^{'}+1}| \phi^{n,l}_{k,2j^{'}+1} \rangle
\end{eqnarray} 
We substitute (\ref{t3}) and (\ref{t2}) into (\ref{t1}) and
hit both sides of the equation with $\langle \lambda^{n}_{k,2j^{'}}|$,
$\langle 0^{n}_{k,2j^{'}+1} |$, $\langle
\phi^{n,l}_{k,2j^{'}+1} |$ and $\langle
\phi^{n,l}_{k,2j^{'}} |$ successively.  Letting
\alpheqn
\begin{eqnarray}
\alpha^{n,2j}_{k,j'} & \equiv & \langle \lambda^{n}_{k,2j'} |
\left(x-x^*_n \right)^2j\chi_{b,n} \rangle \\
\beta^{n,2j}_{k,j'} & \equiv & \langle 0^{n}_{k,2j'+1} |
\left(x-x^*_n \right)^{2j}\chi_{b,n} \rangle \\
\gamma^{n,l,2j}_{k,j'} & \equiv & \langle \phi^{n,l}_{k,2j'}|
\left(x-x^*_n \right)^{2j}\chi_{b,n} \rangle 
\end{eqnarray}
\reseteqn 
we obtain the following equations for the expansion coefficents
$a_{n,j}$, $b^{n,2j}_{k,j^{'}}$, $c^{n,2j}_{k,j^{'}}$,
$d^{n,l,2j}_{k,j^{'}}$ and $e^{n,l,2j}_{k,j'}$
\begin{eqnarray} \label{t4}
b^{n,2j}_{k,j'}\lambda^{n}_{k,2j'} & = & \phi^{n,0}_{1,j}\left(
b^{n,2j}_{k,j'} - \alpha^{n,2j}_{k,j'} \right) \nonumber \\
c^{n,2j}_{k+1,j^{'}} & = & \phi^{n,0}_{1,j}\left(
c^{n,2j}_{k,j'} - \beta^{n,2j}_{k,j'} \right) \nonumber \\
a_{n,2j}\phi^{n,0}_{1,2j}\gamma^{n,1,2j}_{1,j} & = & 1 \nonumber \\
e^{n,l,2j}_{k,j'}\phi^{n,l}_{k,2j'+1} & = & \phi^{n,0}_{1,j}\left(
e^{n,l,2j}_{k,j'} - \delta^{n,l,2j}_{k,j'} \right).
\end{eqnarray}
The equations for $d^{n,l,2j}_{k,j'}$ differ depending on the values of
$l,k$ and $j'$. For $k=1$, $j'=j$ and $l=2,3,\dots,n-2$
\begin{eqnarray} 
  d^{n,l-1,2j}_{1,j} & = & a_{n,2j}\phi^{n,0}_{1,2j}\gamma^{n,l,2j}_{1,j} \nonumber \\
  d^{n,n-2,2j}_{1,j} & = & 0.
\end{eqnarray}
When $j \neq j'$, $k \neq 1$ and $l=1$ we have
\begin{equation} 
d^{n,1,2j}_{k,j'}\left(
\phi^{n,0}_{1,2j} - \phi^{n,1}_{k,2j'} \right)
= a_{n,j}\phi^{n,0}_{1,2j}\gamma^{n,1,2j}_{k,j'}.  
\end{equation}
When $j \neq j'$, $k \neq 1$ and $l=2,3,\dots,n-2$ we have
\begin{equation} \label{t7}
 d^{n,l,2j}_{k,j'}\left(
\phi^{n,0}_{1,2j} - \phi^{n,l}_{k,2j'} \right) - d^{n,l-1,2j}_{k,j'} =
a_{n,2j}\phi^{n,0}_{1,2j}\gamma^{n,l,2j}_{k,j'}. 
\end{equation}

The equations (\ref{t4}) -- (\ref{t7}) for $a_{n,2j}$, 
$b^{n,2j}_{k,j'}$, $c^{n,2j}_{k,j'}$ 
$d^{n,l,2j}_{k,j'}$ and $e^{n,l,2j}_{k,j'}$ are either uncoupled or are coupled in a simple
manner and can be solved explicitly to find the expansion coeffients. Plugging
these coefficents into (\ref{t2}) and (\ref{t6}) we get all the Jordan
states with support in the central transient, for all $n > 2$. For $n = 2$
set all the $d^{l,j}_{2,k,j'} = 0$, $e^{n,l,2j}_{k,j'} = 0$ and $a_{2,j} = 1$ and solve for
$b^{j}_{n,k,j'}$, $c^{j}_{n,k,j'}$ from (\ref{t7}) to obtain the
eigenvectors with support in the central transient at the second bsp.

\end{document}